\documentclass[traditabstract]{aa}
\usepackage{graphicx}
\usepackage{txfonts}
\usepackage{color}
\usepackage{natbib}
\usepackage{epsfig}

\usepackage[usenames,dvipsnames,svgnames,table]{xcolor}
\usepackage[colorlinks=true,
            linkcolor=blue,
            urlcolor=blue,
            citecolor=blue]{hyperref}

\begin{document}

\title{Multiwavelength observation of a large-scale flux rope eruption above kinked mini-filament}

\author{Pankaj Kumar, Kyung-Suk Cho} 
\institute{Korea Astronomy and Space Science Institute (KASI), Daejeon, 305-348, Republic of Korea \\
\email{pankaj@kasi.re.kr}}
\abstract
{We analyse multiwavelength observations of a western limb flare (C3.9) occurred in AR NOAA 111465 on 30 April 2012. The high resolution images recorded by SDO/AIA 304, 1600 \AA~ and Hinode/SOT H$\alpha$ show the activation of a mini-filament (rising speed$\sim$40 km s$^{-1}$) associated with kink instability and the onset of a C-class flare near the southern leg of the filament. The first magnetic reconnection occurred at one of the footpoints of the filament causing the breaking of its southern leg. The filament shows unwinding motion of the northern leg and apex in the counterclockwise direction and failed to erupt. A flux-rope (visible only in hot channels, i.e., AIA 131 and 94 \AA~ channels and Hinode/SXT) structure was appeared along the neutral line during the second magnetic reconnection taking place above the kinked filament. Formation of the RHESSI hard X-ray source (12-25 keV) above the kinked filament and simultaneous appearance of the hot 131 \AA~ loops associated with photospheric brightenings (AIA 1700 \AA) suggest the particle acceleration along these loops from the top of the filament. In addition, EUV disturbances/waves observed above the filament in 171 \AA~ also show a close association with magnetic reconnection. The flux rope rises slowly ($\sim$100 km s$^{-1}$) producing a rather big twisted structure possibly by reconnection with the surrounding sheared magnetic fields within $\sim$15-20 minutes, and showed an impulsive acceleration reaching a height of about 80--100 Mm. AIA 171 and SWAP 174 \AA~ images reveal a cool compression front (or CME frontal loop) surrounding the hot flux rope structure.
}
 
 \keywords{Sun: flares---Sun: magnetic topology---Sun: filaments, prominences---Sun: coronal mass ejections (CMEs)}
\maketitle
\section{Introduction}
 A magnetic flux rope is a coherent helical structure that contains twisted field lines wrapped around its guiding axis. Flux rope is a crucial part of a CME and plays an important role in the onset of solar eruptions and associated flares. The CME models involves the eruption of a flux rope either by magnetic reconnection or by MHD instabilities.
For example, emerging flux trigger \citep{chen2000}, flux cancellation \citep{amari2003a,amari2003b,amari2010}, tether cutting \citep{moore2001}, and breakout models \citep{antiochos1999,lynch2004,karpen2012} require magnetic reconnection below or above the flux rope. On the other hand, numerical MHD simulations of the kink instability suggest that if the twist of the flux rope exceeds a critical value ($\sim$1.75 field line turns), then it becomes unstable \citep{fan2003,fan2004,kliem2004,torok2003,torok2004}. Alternatively, the decrease of overlying magnetic field (B) with height (H) (decay index n=-${d (log B)}/{d (log H)}>1.5$, condition for torus instability) above the eruption site alone can decide whether the eruption of a flux rope is successful or failed \citep{kliem2006,aulanier2010,demoulin2010,olmedo2010}. In addition, the interaction of filaments or flux ropes caused by the rotation of sunspots is also found to be associated with the CME onset \citep{kumar2010a,torok2011,yan2012,kumar2013a,torok2013}. For the detail mechanisms and instabilities involved in the solar eruptions, see the recent reviews by \citet{chen2011}, \citet{webb2012} and \citet{aulanier2014}.
 
 The helical structure of the flux rope has been inferred from SOHO/LASCO coronagraph and IPS (interplanetary scintillation) observations (e.g., \citealt{dere1999}, \citealt{manoharan2010}). The cylindrical or circular features observed in the coronagraph images support the flux rope topology in projection (see, \citealt{vourlidas2013,vourlidas2014}). The flux rope is recognized as a magnetic cloud in the solar wind observations at 1 AU by low plasma beta ($\beta$) and smooth rotation of the magnetic field vector (B$_z$) \citep{burlaga1982,bothmer1998,kumar2011a,marubasi2012}.
In the solar atmosphere, the filament plasma is considered to be supported in the dips of a helical flux rope \citep{priest1989,su2011}. In addition, twisted flux rope structure is sometime observed co-spatially in Soft X-ray and EUV channels (hot and cool channels) above the neutral line \citep{kumar2011b}. Recently, flux ropes have been observed in the AIA hot channels (131 and 94 \AA, T$\sim$8 MK), which are elongated S shaped structures \citep{zhang2012,pats2013,cheng2014}. Expansion and rising of these structures are associated with the formation of CMEs in the low corona.

Most of the CME models require the presence of the flux rope prior to the eruption (e.g, flux emergence, flux cancellation, MHD instabilities related models) whereas other models (including breakout and tether-cutting models) consider that the flux rope is formed (above the neutral line) during magnetic reconnection \citep{gosling1995,longcope2007}. 
The details of the flux rope formation in the low corona is not well understood. It is under debate whether the flux rope is present well before the eruption or formed by magnetic reconnection during the eruption.
A model of solar prominences indicates that the prominence plasma is supported inside the twisted flux tube \citep{priest1989,van2000}, and this twisted flux tube is formed by the flux cancellation in a sheared arcade \citep{van1989}. Alternatively, the helical magnetic field of the filament is explained by the twisted flux rope model \citep{rust1994}, in which  the flux tube emerges (already twisted) through the photosphere supporting the filament plasma. Numerical simulation by \citet{fan2001} suggests that the emergence process stops when the flux rope reaches the photosphere and forms a sheared arcade in the corona. \citet{gary2004} and \citet{kumar2012} have provided evidences of 3-4 helical turns within or below the prominence, which support the flux rope model. \citet{okamoto2008} showed the evidences of the emergence of a helical flux rope below an active region prominence. On the other hand, \citet{gosling1995} suggested that magnetic reconnection occurring within the leg of the outward moving magnetic loops causes the formation of helical flux rope structure. \citet{qiu2007} quantitatively demonstrated a comparison between the total magnetic reconnection flux in the low corona (measured from flare ribbons) in the wake of coronal mass ejections (CMEs) and the magnetic flux in magnetic clouds, favouring the flux rope formation during magnetic reconnection. \citet{longcope2007} model also explains the flux rope formation by a sequence of magnetic reconnection above the polarity inversion line in a two-ribbon flare. 
 
Soft X-ray images often show the sigmoid (S and `inverse S' shaped) structures \citep{canfield1999,pevtsov2002} exhibiting a twisted magnetic flux rope topology \citep{rust1996, manoharan1996}. Forward and reverse S shaped sigmoid are generally (not all) observed in southern and northern hemisphere, exhibiting the positive (right handed) and negative (left handed) helicity according to the hemispheric helicity rule \citep{pevtsov1995,rust1996}.  
\citet{gibson2006} explained the evolving sigmoid as an evidence of long-lived coronal flux rope before, during, and after a CME eruption. 
 The tether cutting model \citep{moore2001} explains the flux rope formation by magnetic reconnection in between two opposite `J' shaped elbow structures (arcade loops), which form a continuous S shaped loop and an underlying loop as a result of magnetic reconnection. In \citet{aulanier2010} model, the flux rope formation takes place slowly due to photospheric bald patch and coronal slip-running tether-cutting reconnection.
Furthermore, few studies (using Hinode SXT and SOHO/MDI data) suggest the gradual transformation of arcade field into a sigmoid/flux rope as a result of photospheric flux cancellation over extended periods prior to the onset of an eruption \citep{green2009,tripathi2009,green2011}.  

Using multiwavelength observations from TRACE, STEREO, and Hinode SXT,  \citet{kumar2011b} provided an example of the presence of a helical flux rope prior to and during the solar eruption at multitemperature plasma. Recent AIA observations have also revealed the formation of a hot flux rope (visible in AIA 94 and 131 \AA) during the flare impulsive phase on 3 November 2010 \citep{cheng2011,cheng2012}, which was followed by magnetic breakout \citep{kumar2013c}. Moreover, some of the AIA observations suggest the formation of hot flux rope well before the solar eruption \citep{zhang2012, pats2013, kumar2013b}. However, the exact mechanism of the flux rope formation, eruption (i.e., trigger), and their relationship with the filament structures is not well known yet.

At present, high resolution multiwavelngth data from SDO/AIA and Hinode provide an opportunity to investigate the onset mechanism of the solar eruptions and associated phenomena in great detail.
In this paper, we present multiwavelength observations of a flux rope eruption followed by a C-class flare occurred on 30 April 2012. AIA observations combined with RHESSI hard X-ray images reveal the appearance of a flux rope above the mini-filament during magnetic reconnection. So far, the appearance of a hot flux rope structure above a kinked mini-filament has not been reported in earlier data sets. In section 2, we present the multiwavelength observational data sets and in the last section, we discuss the results.

\section{Observations and Results}

The Atmospheric Image Assembly (AIA; \citealt{lemen2012}) onboard the Solar Dynamics Observatory (SDO; \citealt{pesnell2012}) obtains full disk images of the Sun (field-of-view $\sim$1.3 R$_\odot$) with a spatial resolution of 1.5$\arcsec$ (0.6$\arcsec$  pixel$^{-1}$) and a cadence of 12 sec in 10 extreme ultraviolet (EUV) and UV filters. For the present study, we utilized 171~\AA\ (Fe IX, T$\sim$0.7 MK), 131~\AA\ (Fe VIII/XXI/XXIII, T$\sim$0.4, 11 \& 16 MK), 94 \AA~(Fe XVIII, T$\sim$6.3 MK), 304~\AA\ (He II, T$\sim$0.05 MK), 1600 \AA~ (CIV \& continuum, T$\sim$0.1 MK), and 1700 (continuum, T$\sim$5000 K) \AA~ images.
\begin{figure}
\centering{
\includegraphics[width=9cm]{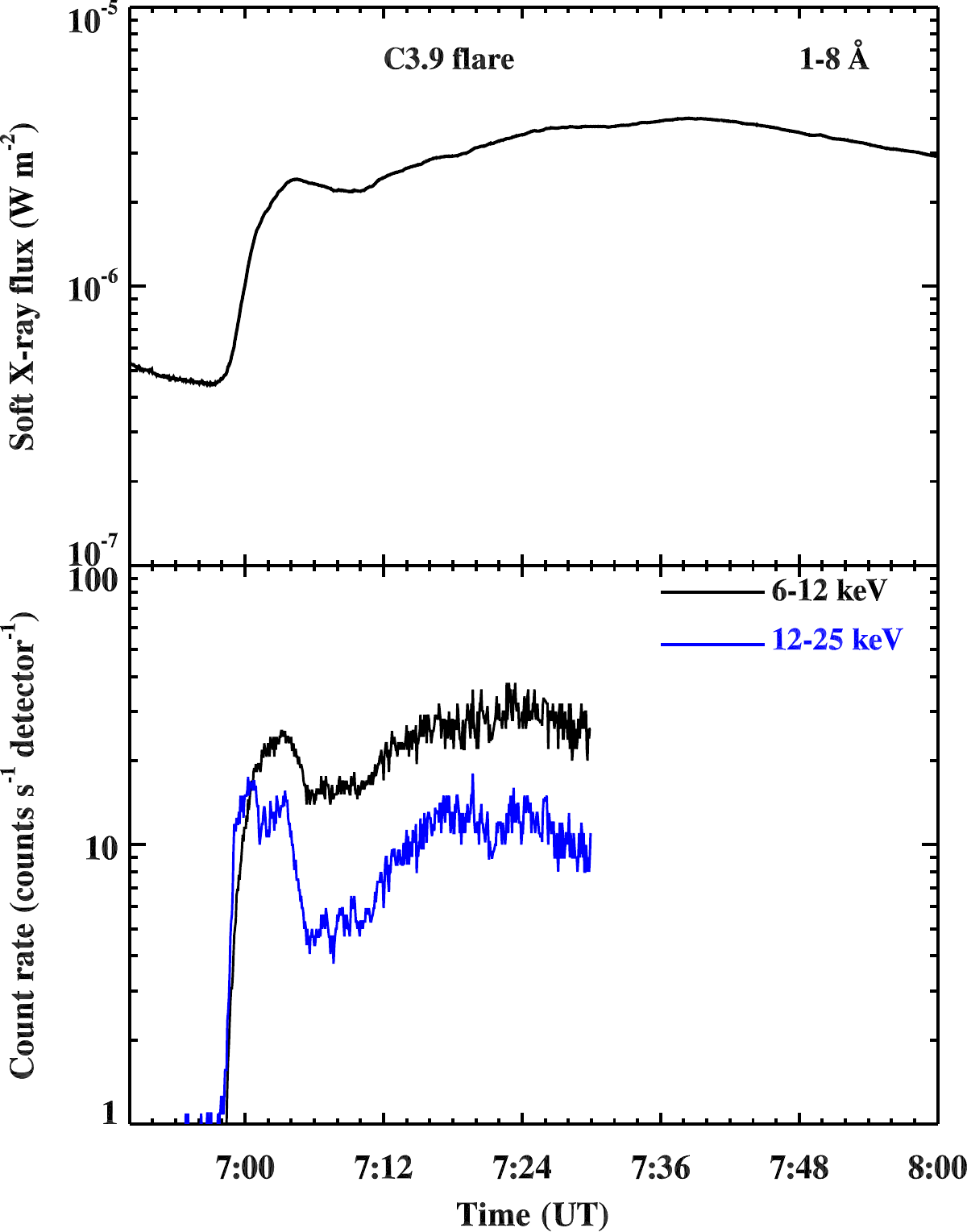}
}
\caption{GOES soft X-ray flux in 1-8 \AA~ channel (top) and RHESSI hard X-ray flux profiles in 6-12 and 12-25 keV channels (bottom) for the C3.9 flare on 30 April 2012.}
\label{flux}
\end{figure}
The active region NOAA 11465, located at the western limb, produced a C3.9 flare on 30 April 2012. This flare was a long duration event, which started at $\sim$06:56 UT, peaked at $\sim$07:38 UT and ended at $\sim$08:19 UT. Figure \ref{flux} displays the GOES soft X-ray flux profile in 1-8 \AA~ channel (top) and RHESSI hard X-ray flux profiles (bottom) in 6-12 (black) and 12-25 keV (blue) energy bands. The hard X-ray fluxes indicate two stages of energy release during (i) 06:58--07:05 UT, and (ii) 07:13--07:30 UT. 

\begin{figure*}
\centering{
\includegraphics[width=4cm]{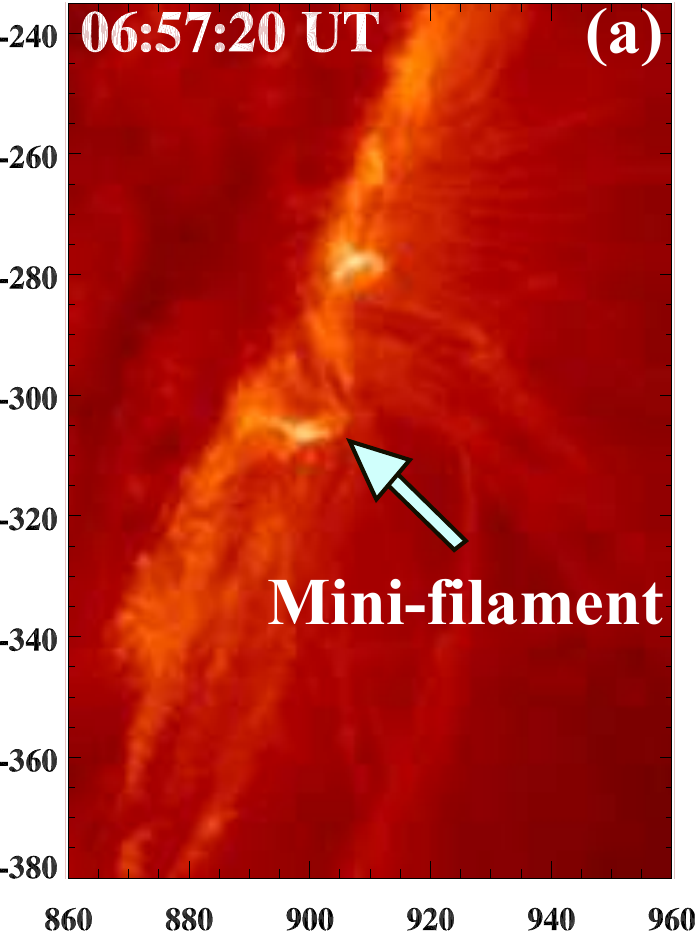}
\includegraphics[width=4cm]{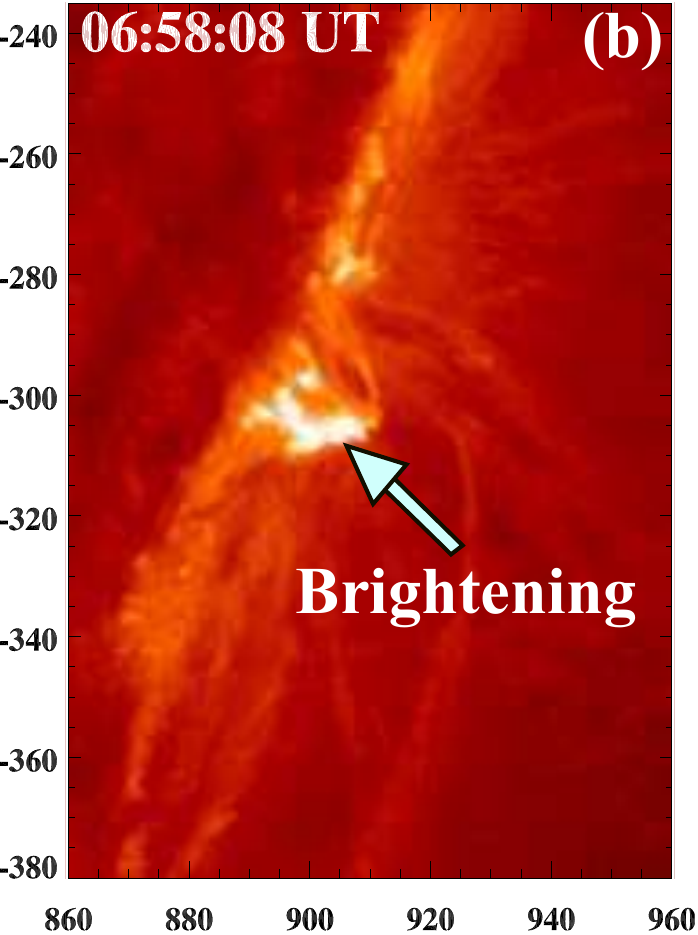}
\includegraphics[width=4cm]{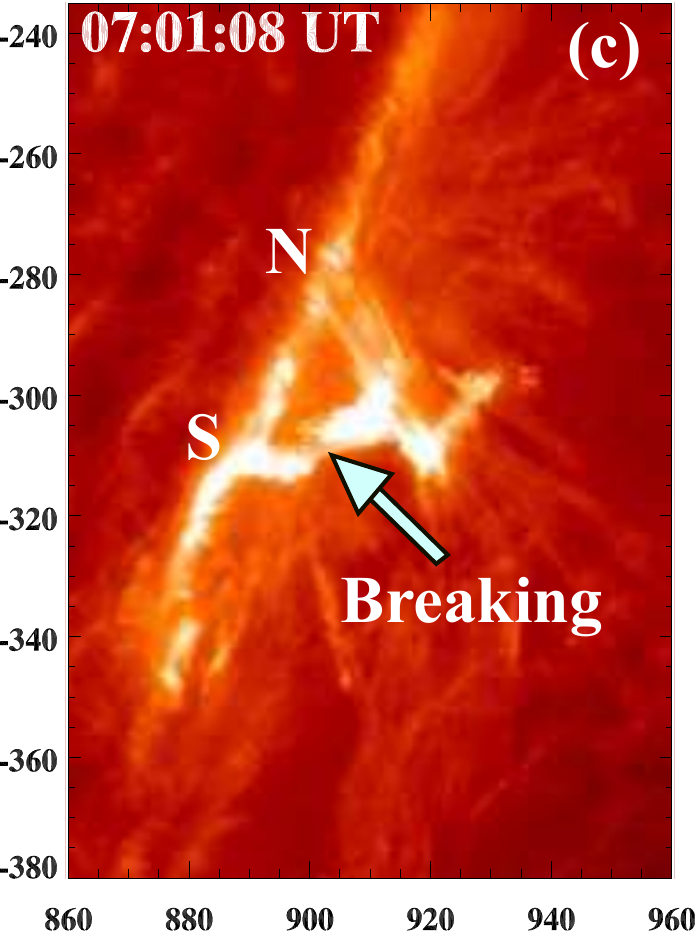}
\includegraphics[width=4cm]{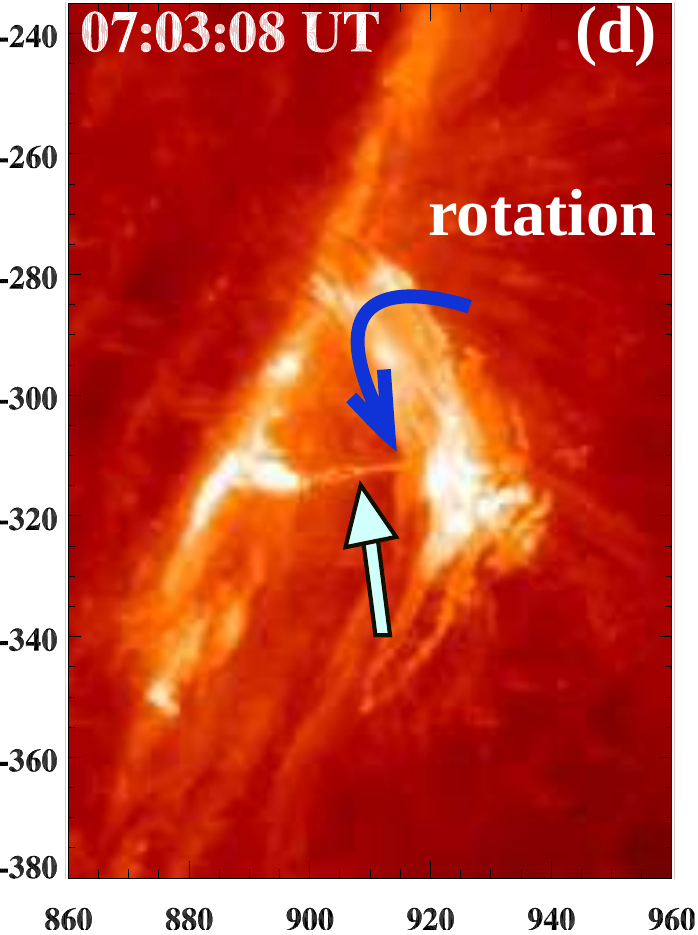}

\includegraphics[width=4cm]{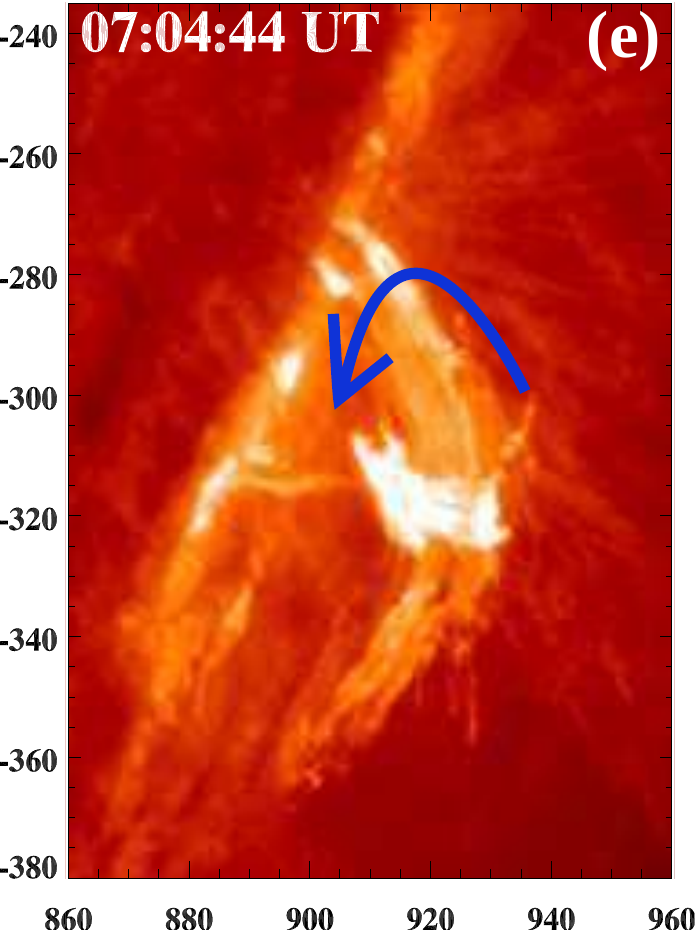}
\includegraphics[width=4cm]{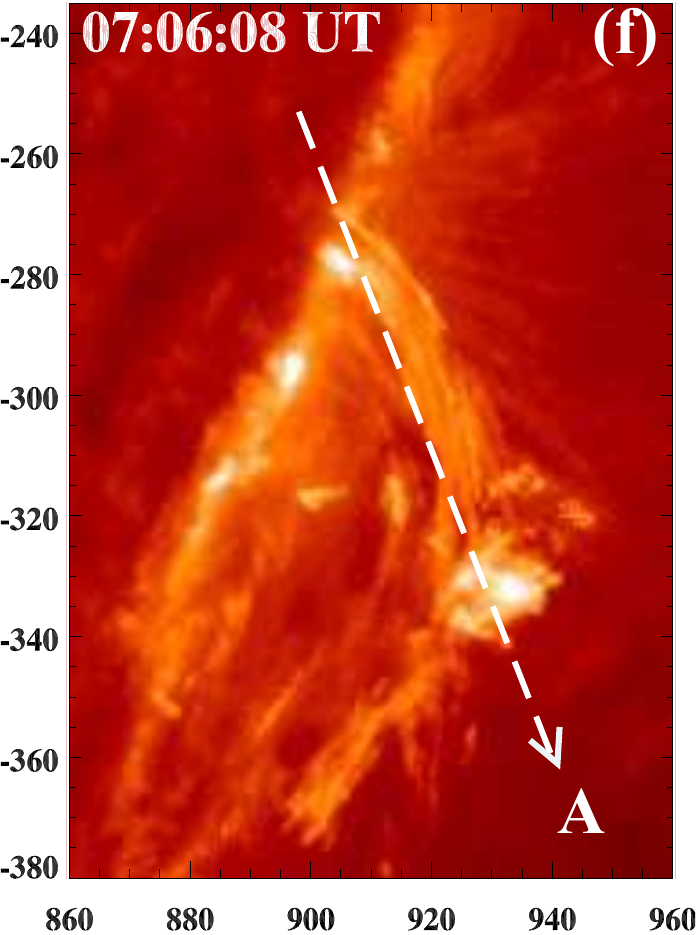}
\includegraphics[width=4cm]{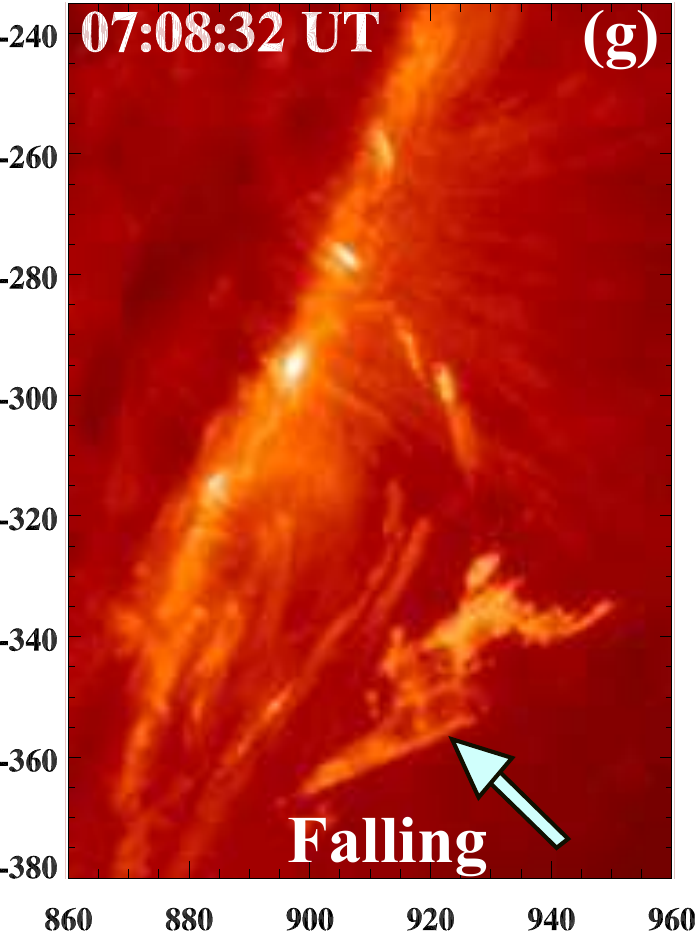}
\includegraphics[width=4.4cm]{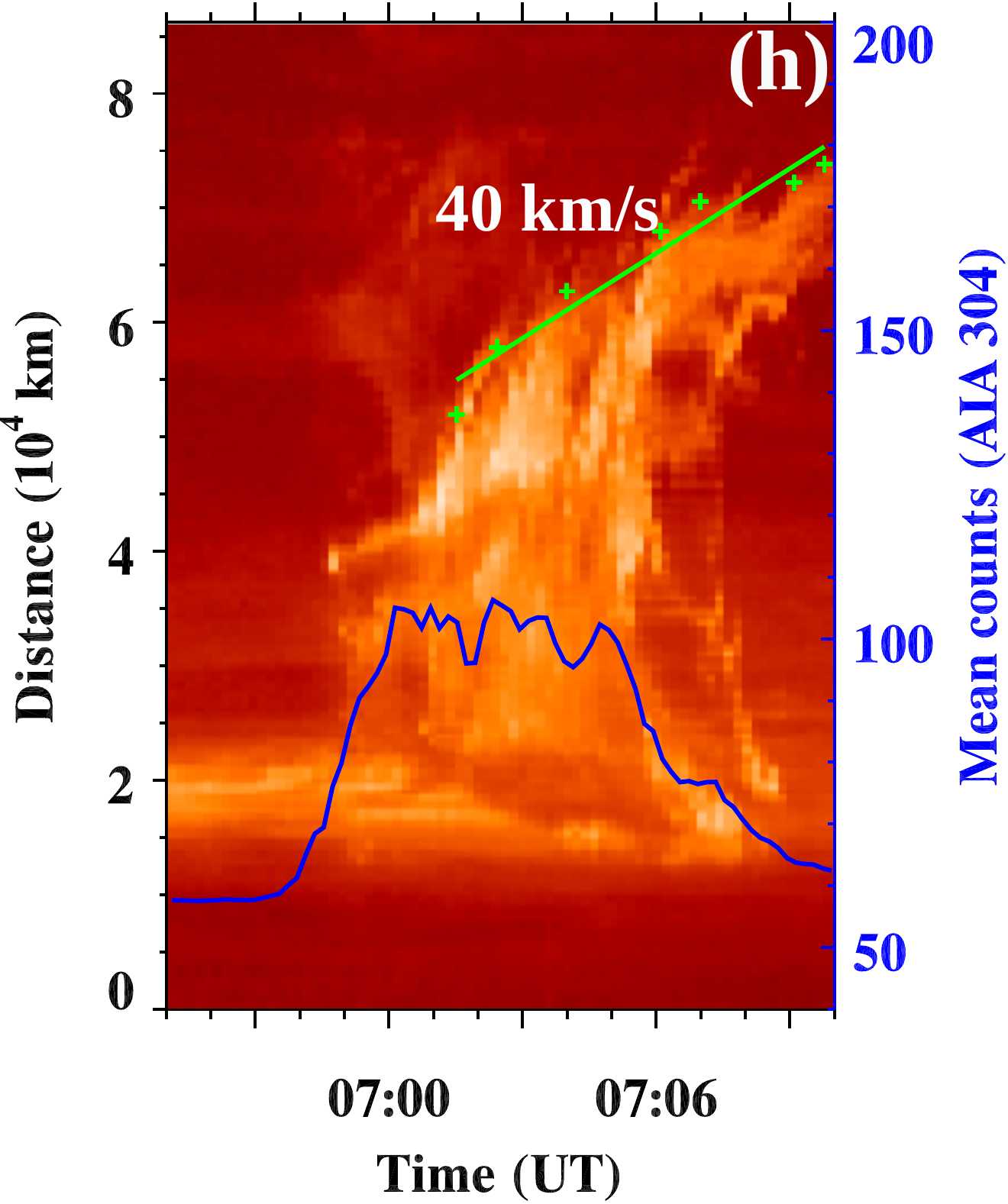}

\includegraphics[width=5.5cm]{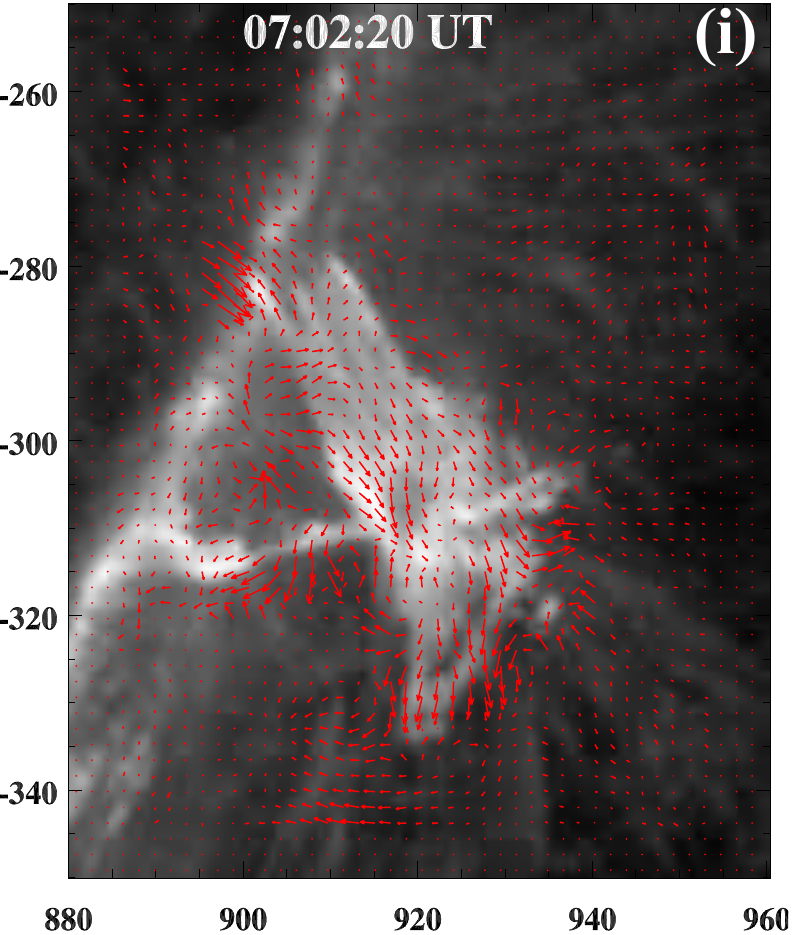}
\includegraphics[width=5.5cm]{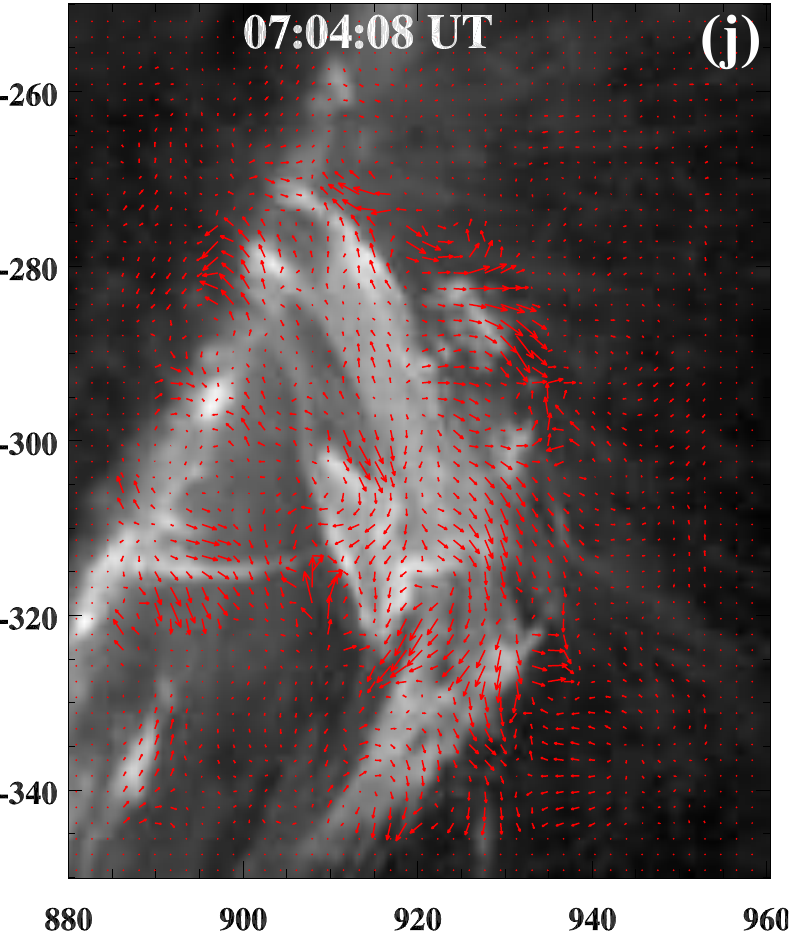}
\includegraphics[width=5.5cm]{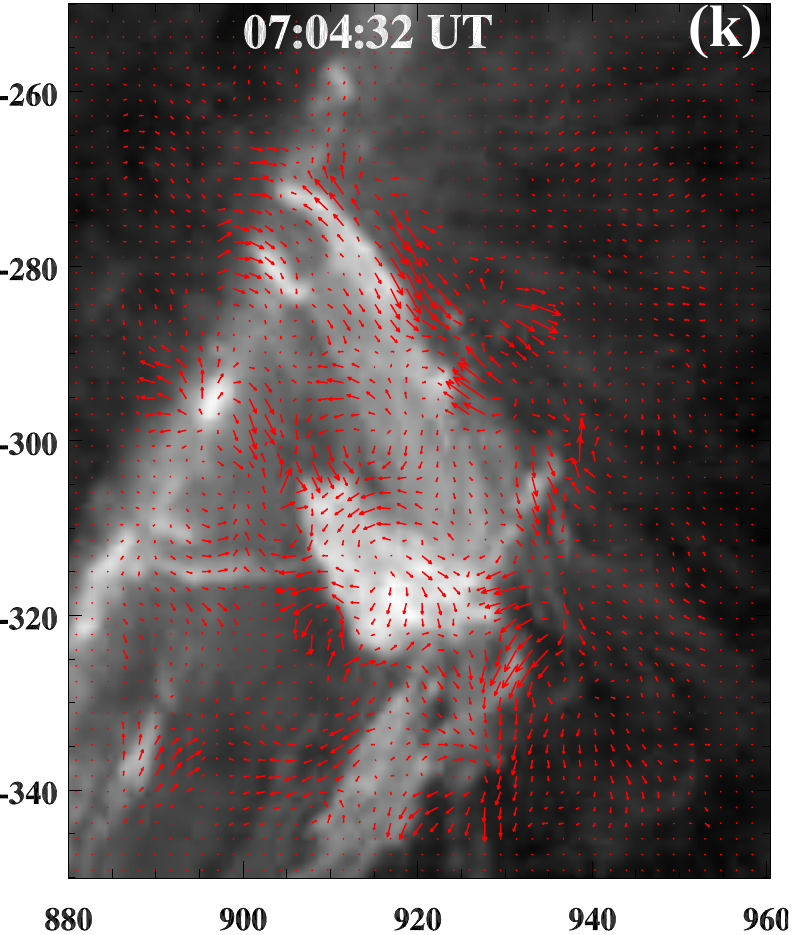}
}
\caption{(a-g) SDO/AIA EUV images in the 304 \AA~ channel (T$\sim$0.05 MK) showing the eruption of a kinked mini-filament associated with the C3.9 flare on 30 April 2012. N and S represent the northern and southern legs of the filament.  The blue arrows show the unwinding motion of the  northern leg and top portion of the filament. $X$ and $Y$ axes are in arcsecs. (h) Space-time plot along slice A (shown by dotted line in panel f) plotted with AIA 304 \AA~ mean counts in blue curve.  (i-k) Apparent velocity field computed with DAVE method using AIA 304 \AA~ images, showing the plasma flows in and around the prominence threads. The longest arrow (in the panels) corresponds to the speed of (i) 204 km s$^{-1}$ (j) 178 km s$^{-1}$ (k) 158 km s$^{-1}$. The temporal evolution of the kinked filament in 304 \AA~ can be found in a movie available in the online edition.}
\label{aia304}
\end{figure*}
\begin{figure*}
\centering{
\includegraphics[width=5cm]{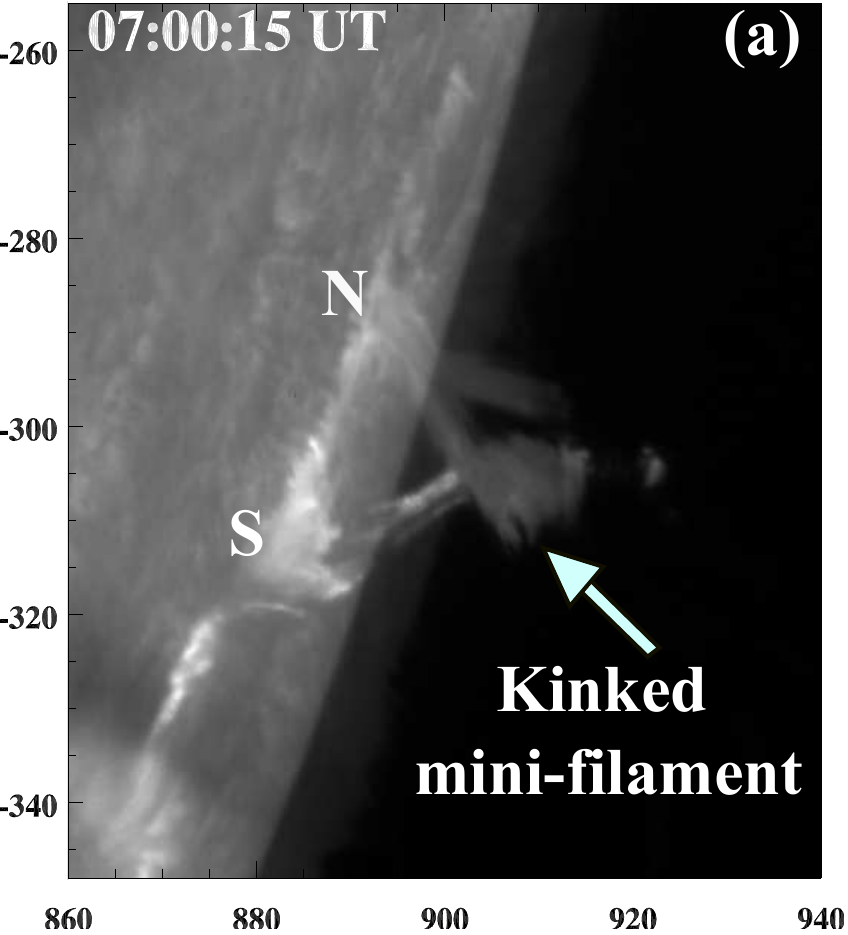}
\includegraphics[width=5cm]{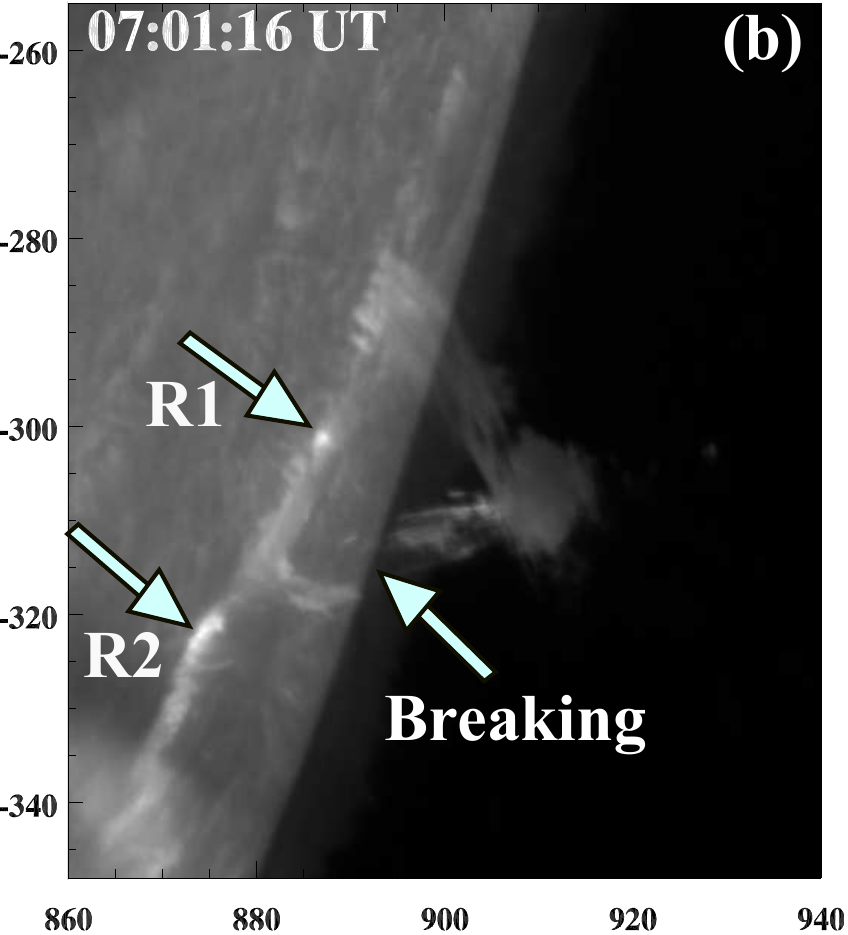}
\includegraphics[width=5cm]{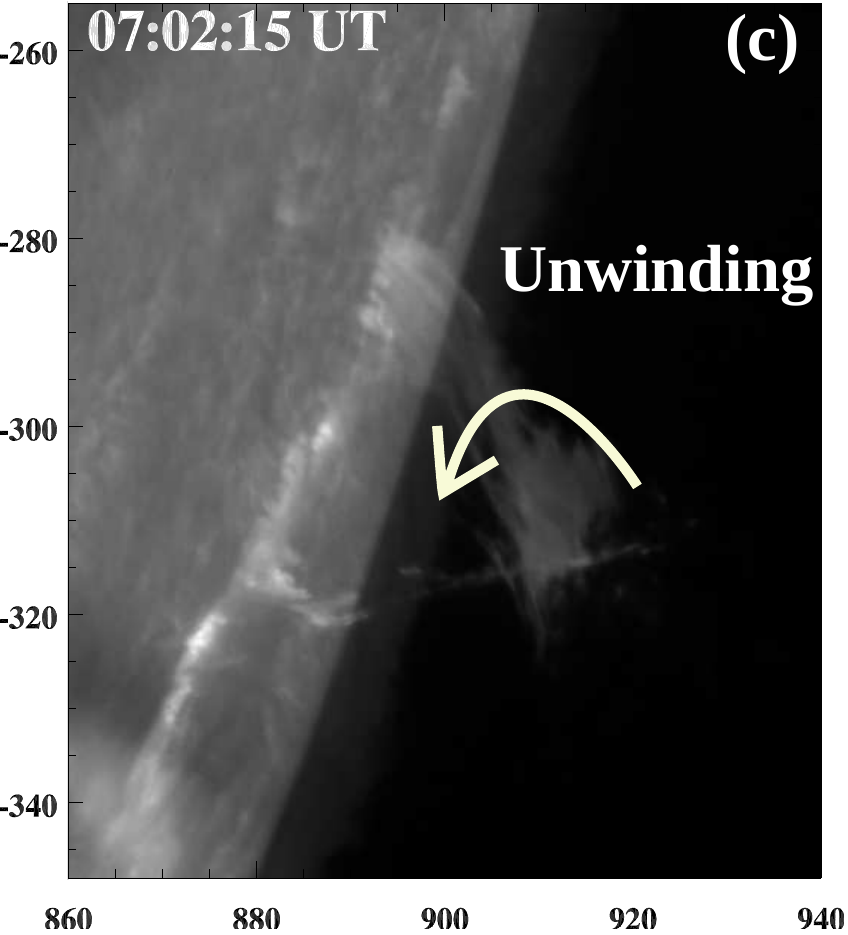}

\includegraphics[width=5cm]{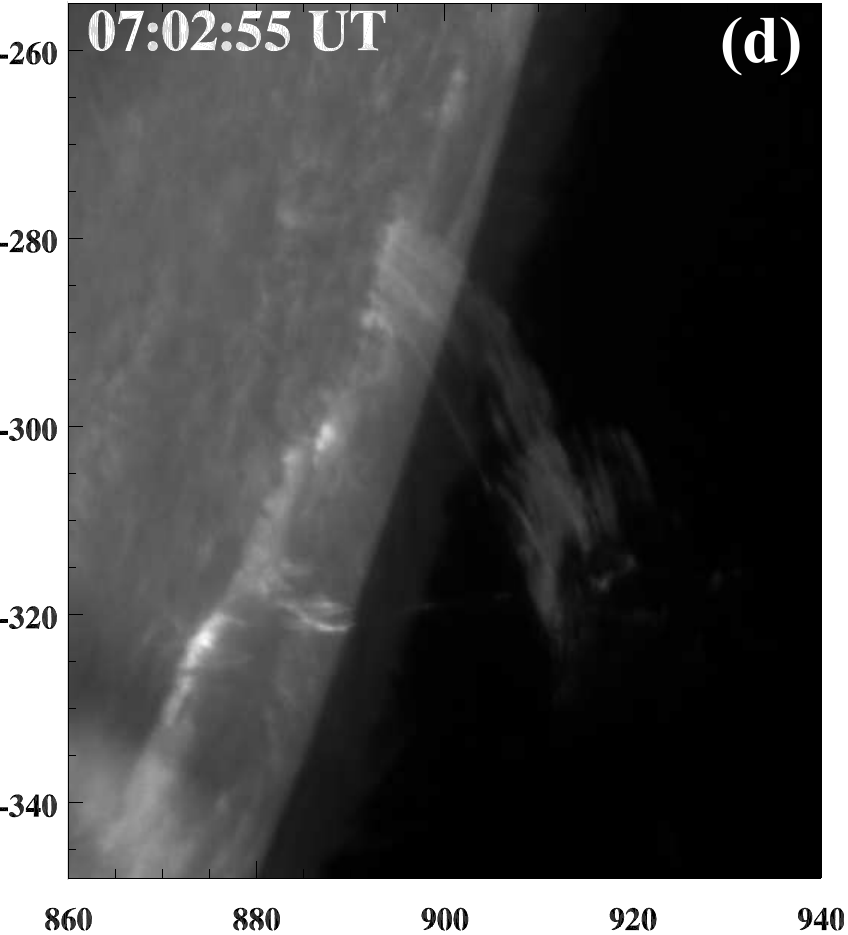}
\includegraphics[width=5cm]{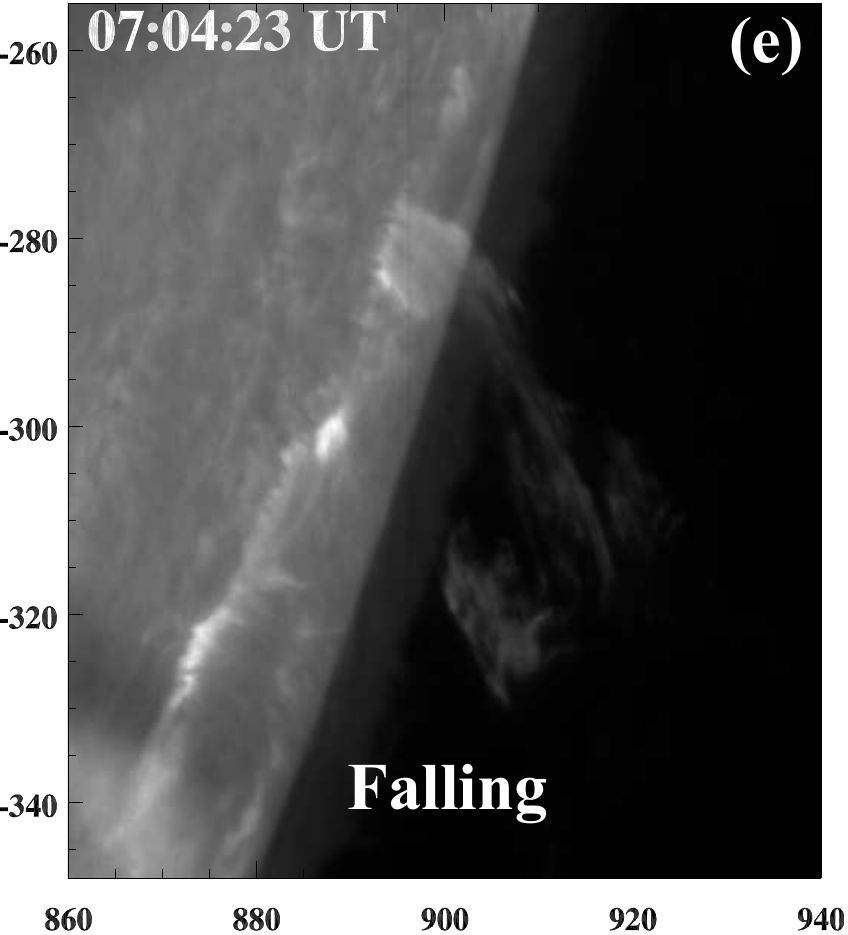}
\includegraphics[width=5cm]{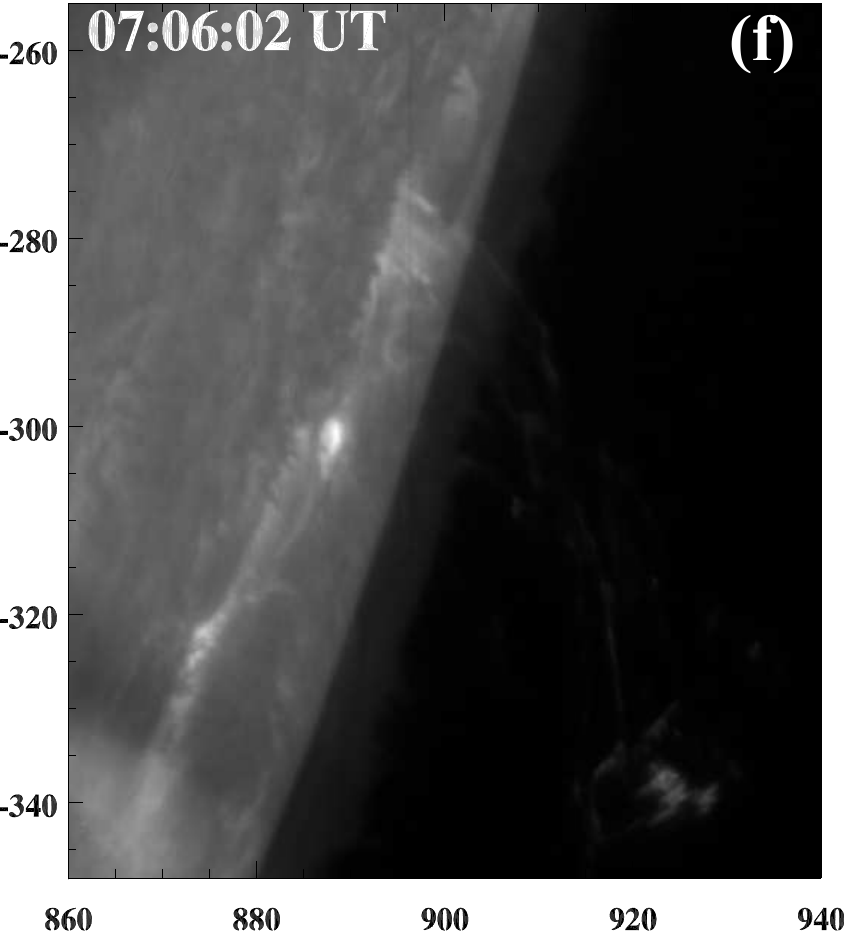}
}
\caption{Hinode/SOT H$\alpha$ images showing eruption of the kinked mini-filament associated with the C3.9 flare. R1 and R2 indicate the flare kernals.}
\label{sot}
\end{figure*}

\begin{figure*}
\centering{
\includegraphics[width=4cm]{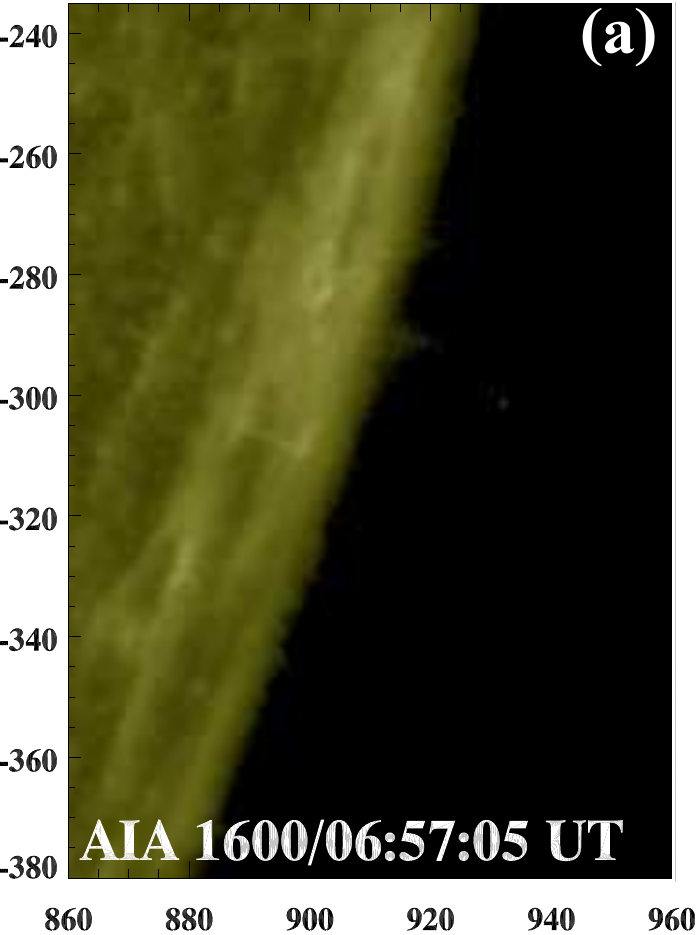}
\includegraphics[width=4cm]{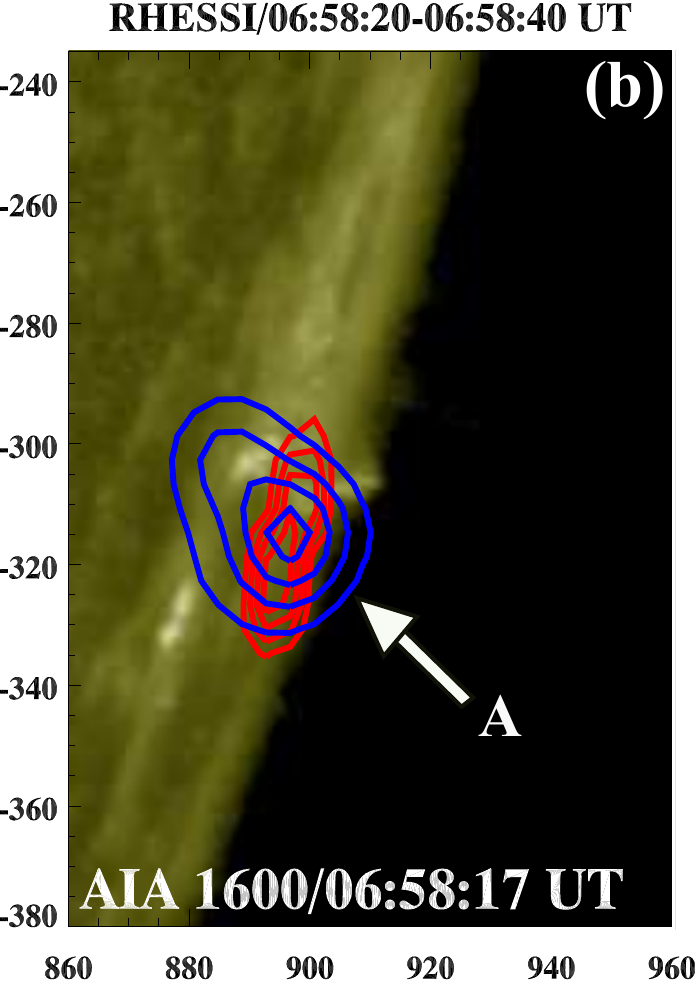}
\includegraphics[width=4cm]{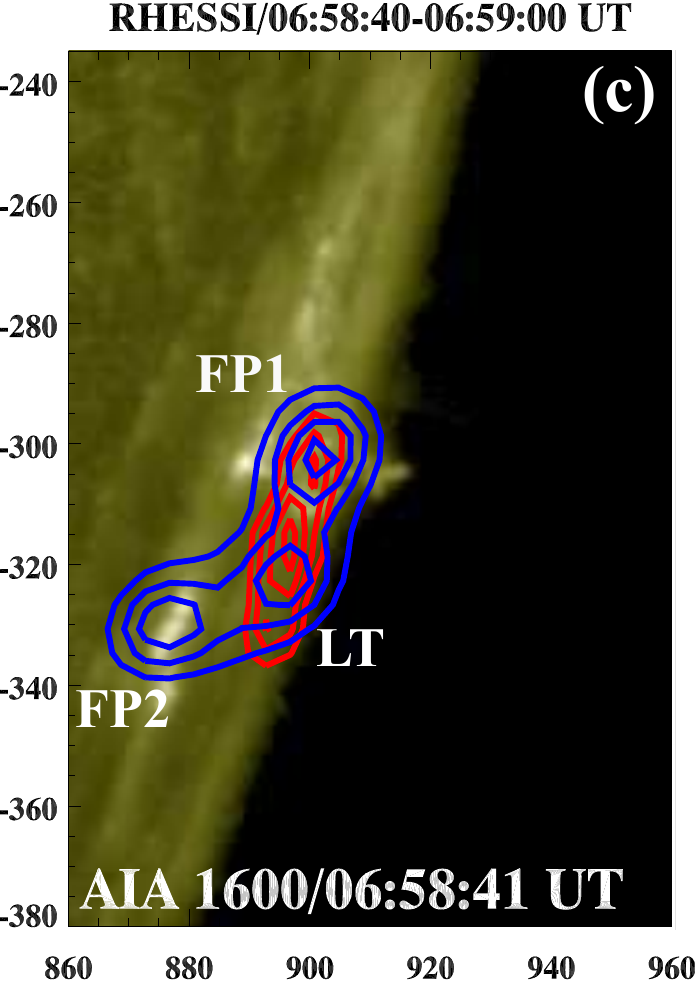}
\includegraphics[width=4cm]{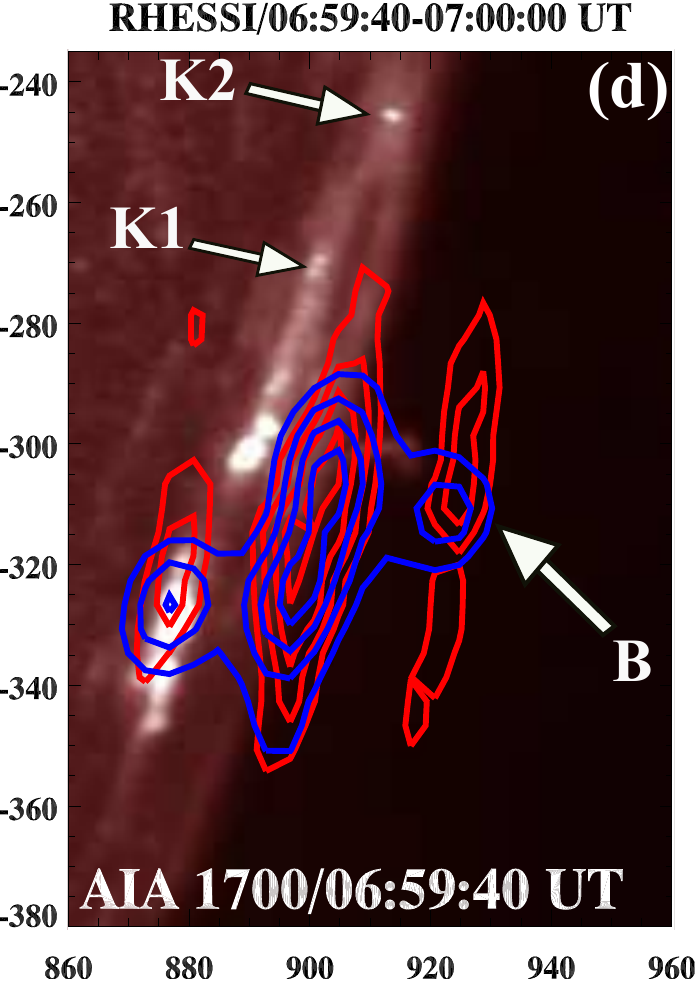}

\includegraphics[width=4cm]{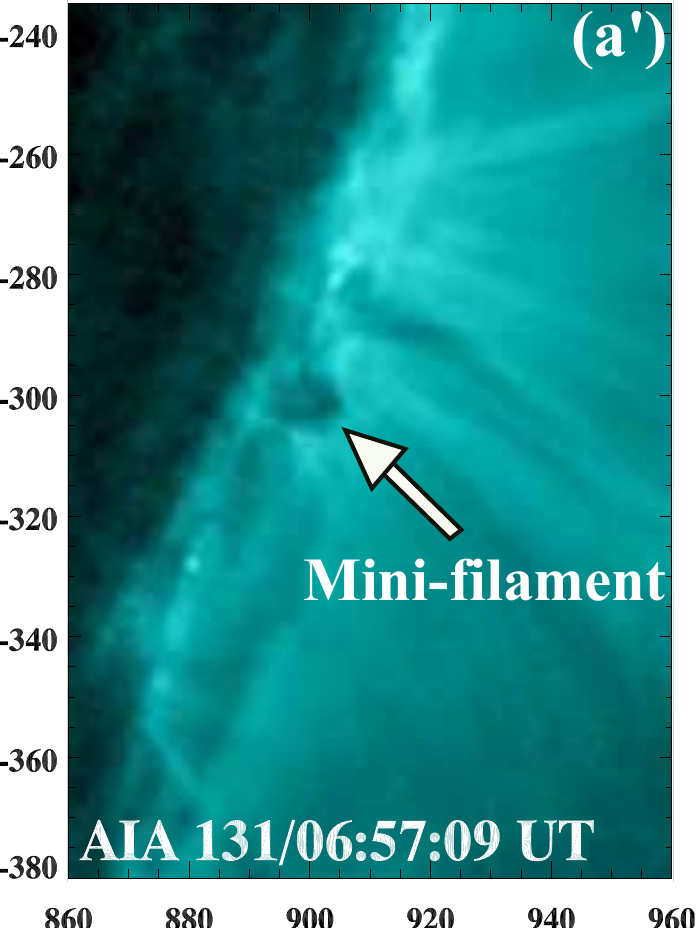}
\includegraphics[width=4cm]{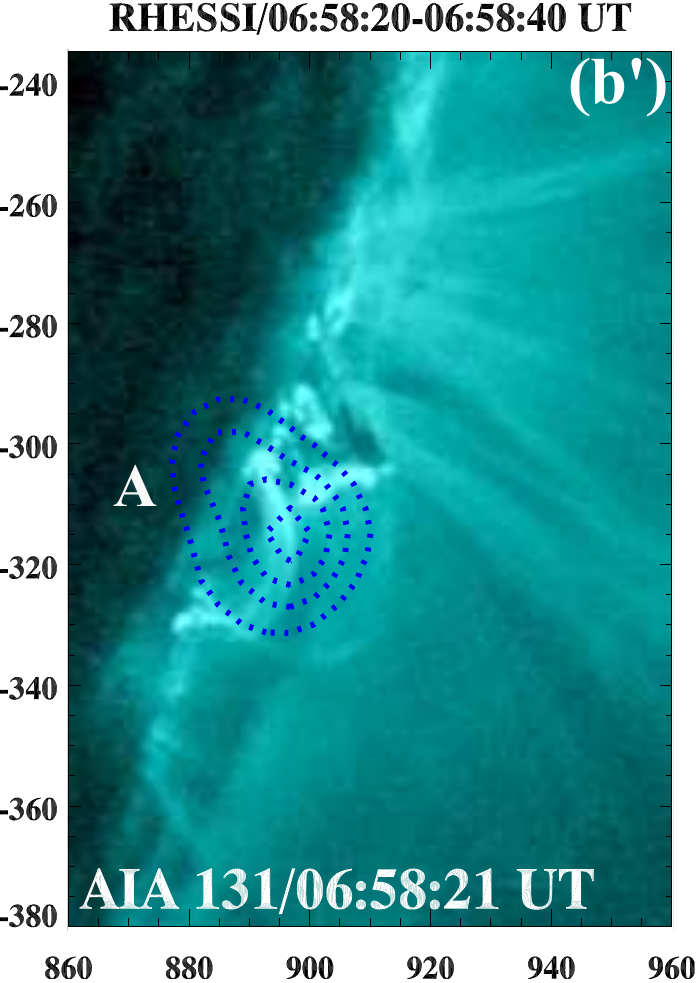}
\includegraphics[width=4cm]{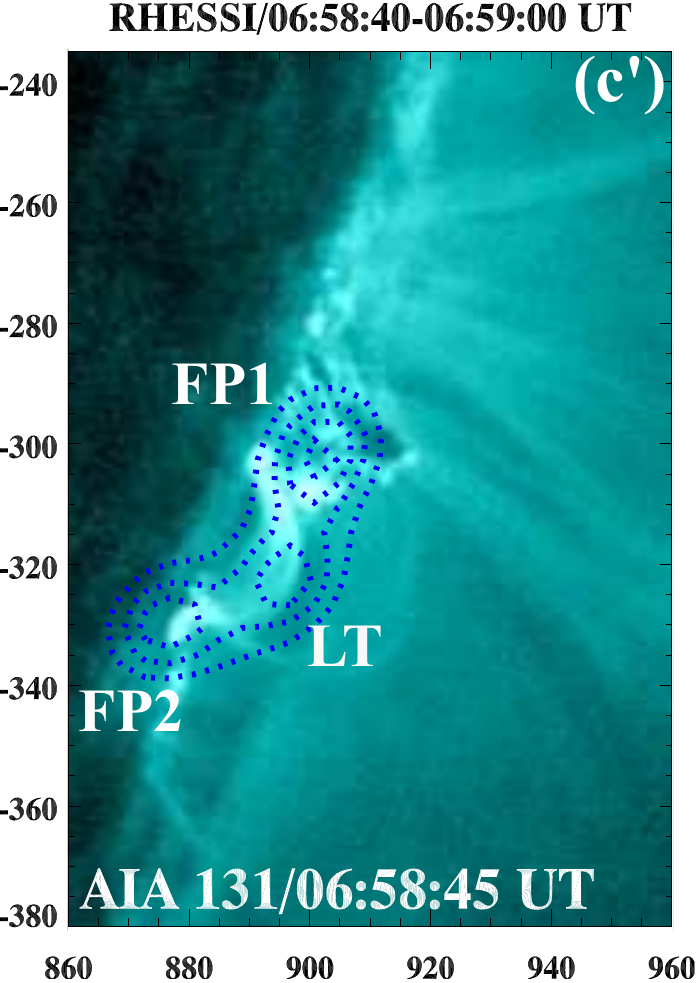}
\includegraphics[width=4cm]{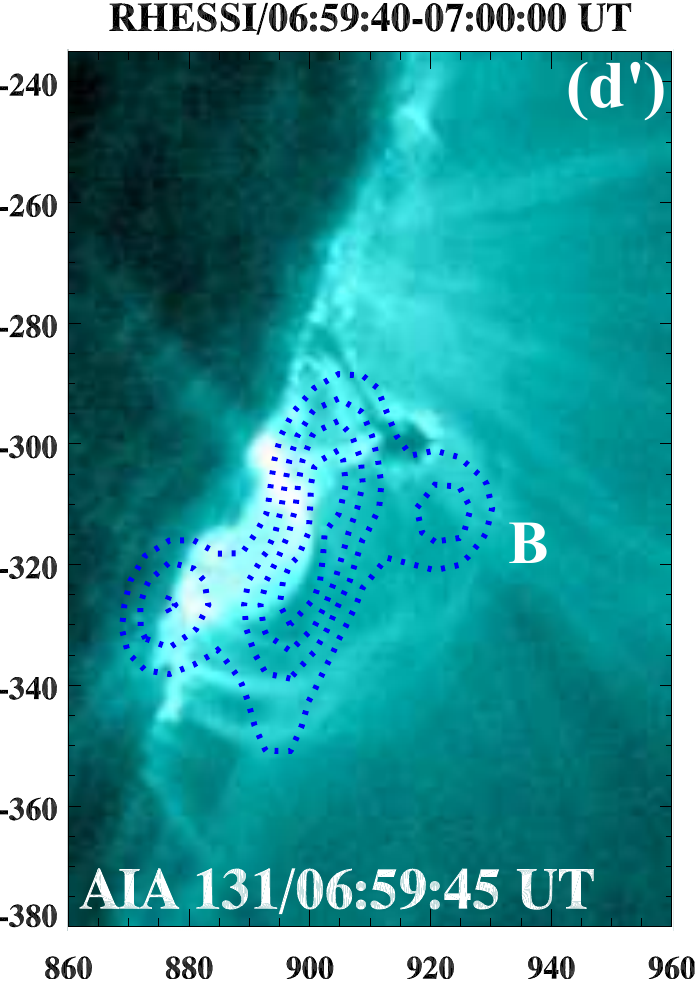}

\includegraphics[width=4cm]{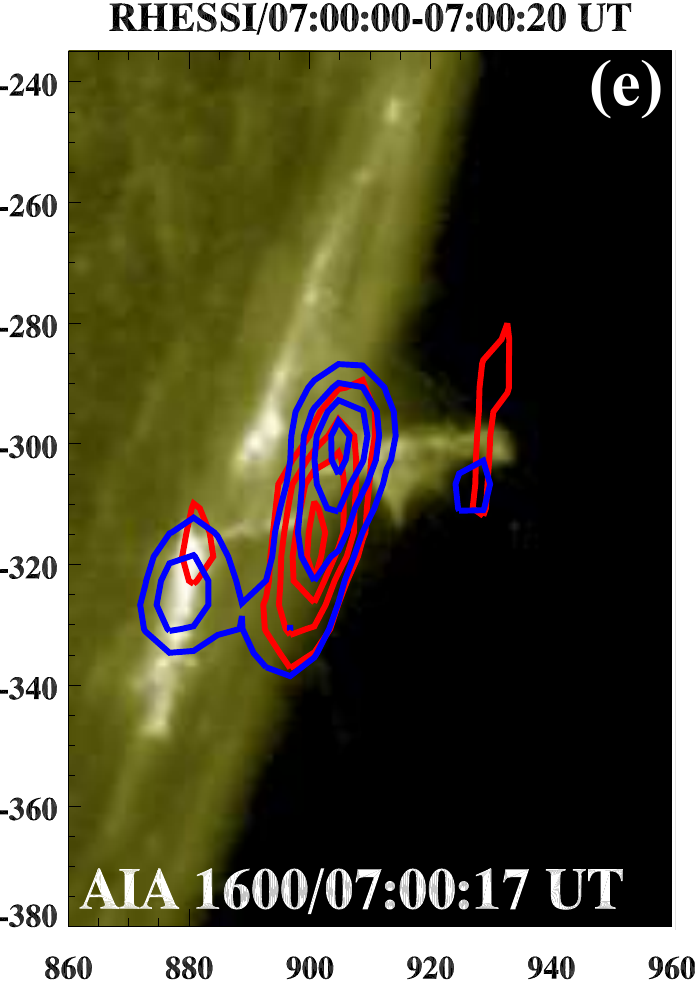}
\includegraphics[width=4cm]{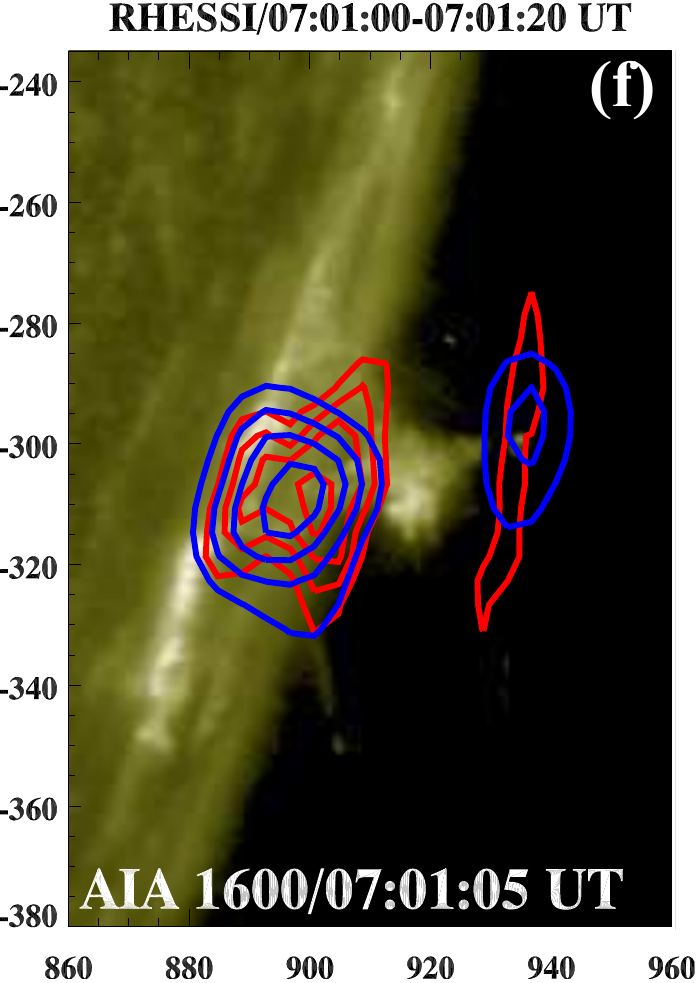}
\includegraphics[width=4cm]{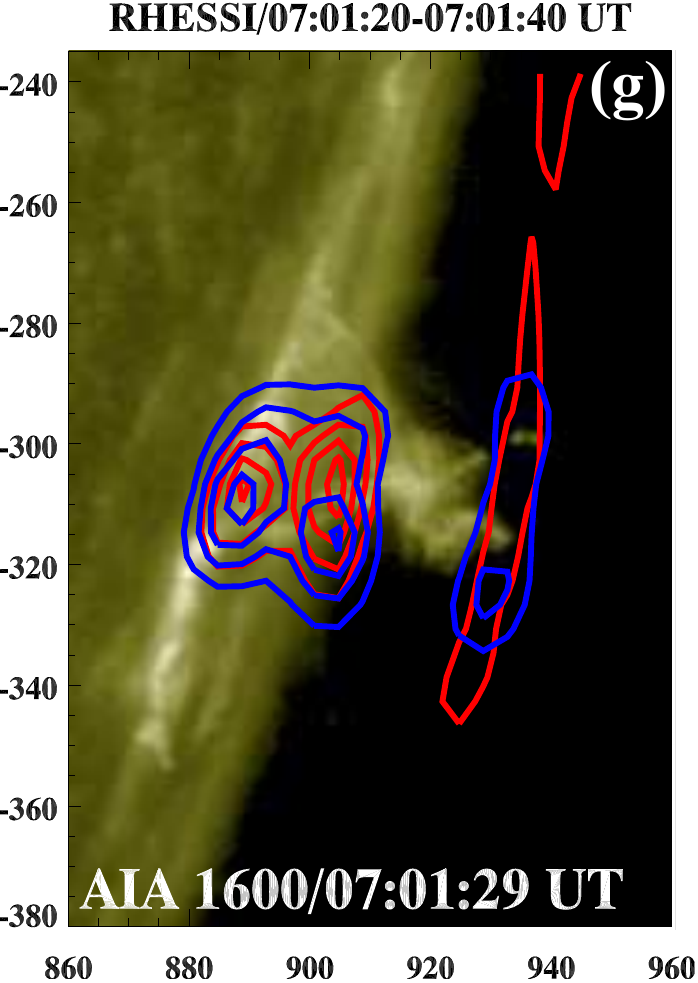}
\includegraphics[width=4cm]{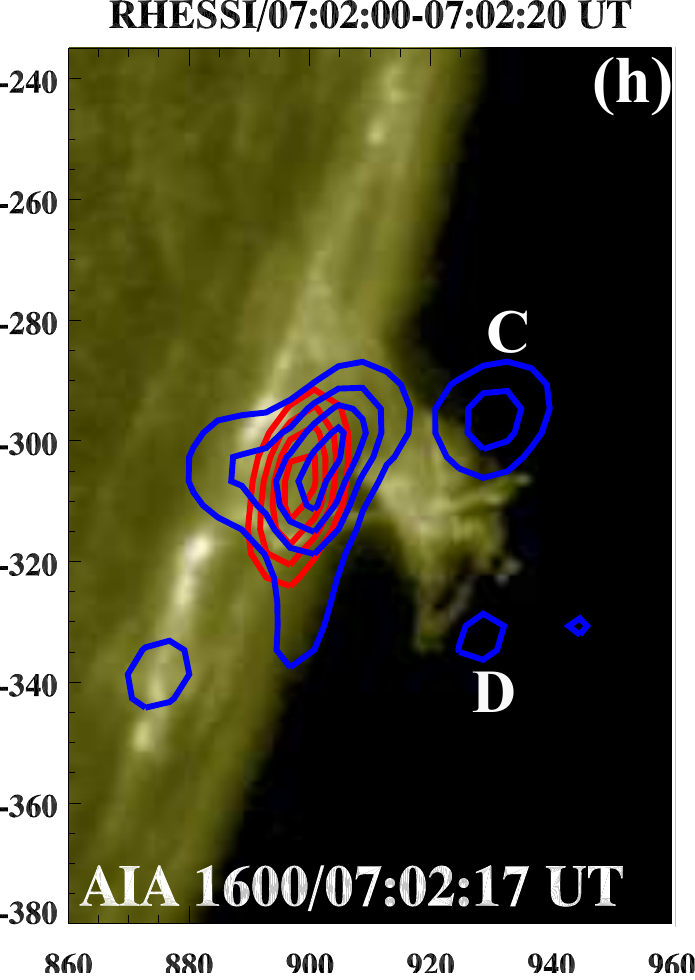}

\includegraphics[width=4cm]{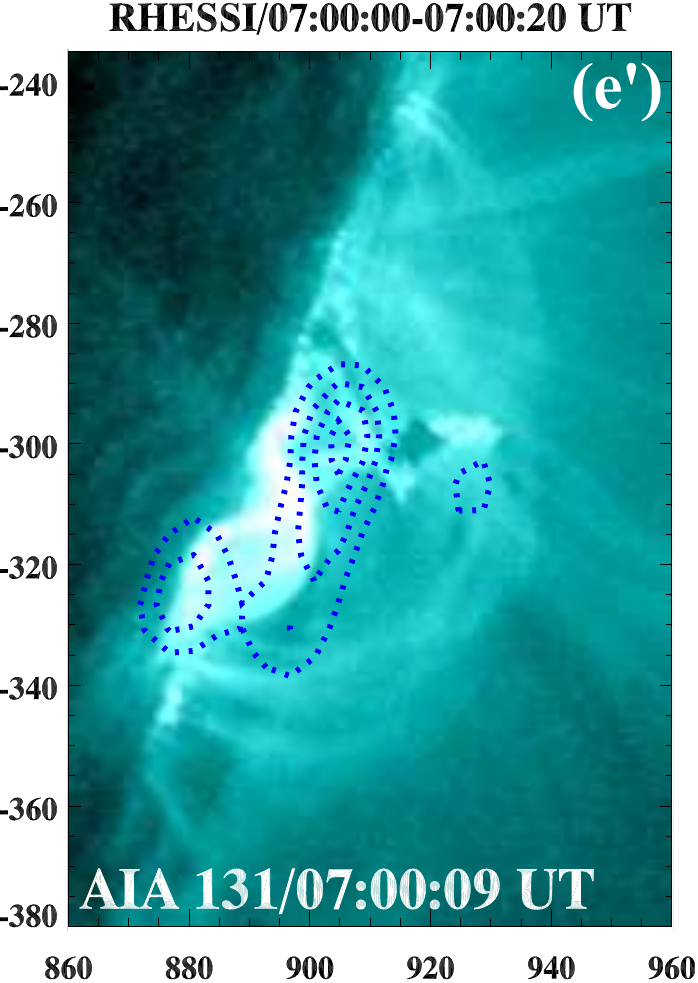}
\includegraphics[width=4cm]{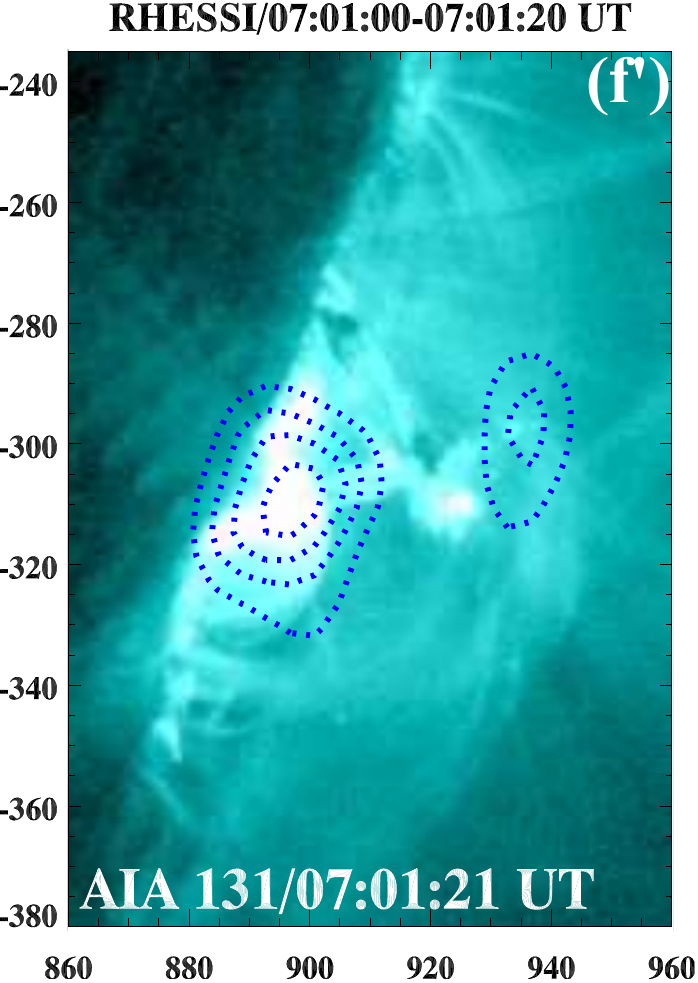}
\includegraphics[width=4cm]{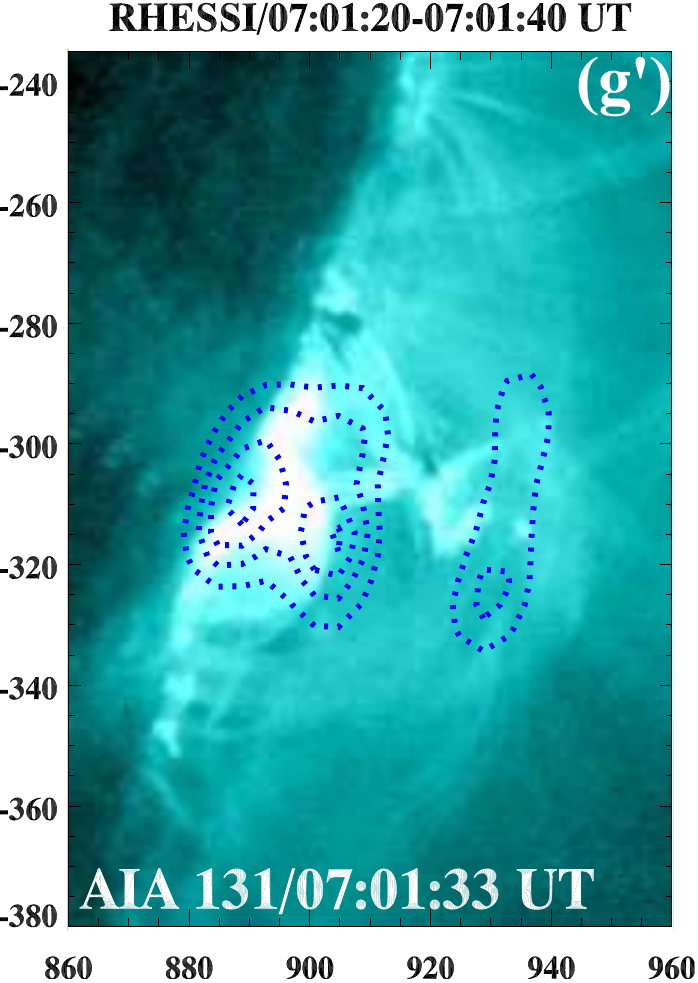}
\includegraphics[width=4cm]{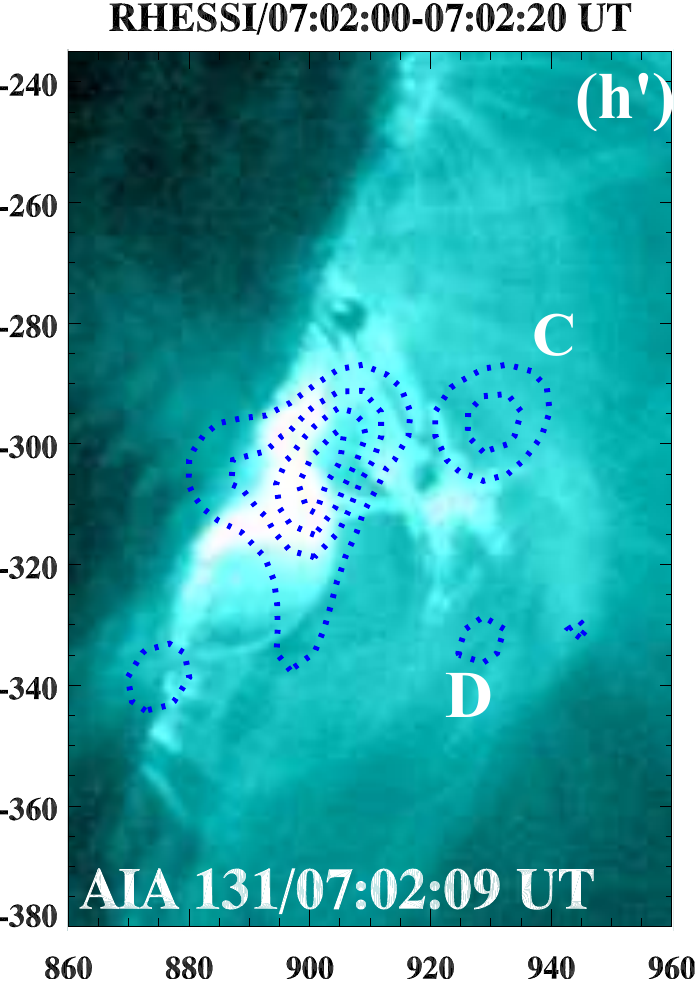}

}
\caption{SDO/AIA 1600, 1700, and 131 \AA~ images overlaid by the RHESSI hard X-ray image contours in 6-12 (red) and 12-25 (blue) keV energy channels. The contour levels are 30$\%$, 50$\%$, 70$\%$, and 90$\%$ of the peak intensity.}
\label{hessi}
\end{figure*}

\subsection{Mini-filament eruption and associated flare}
To see the evolution of the flare onset in the lower solar atmosphere (i.e., chromosphere and transition region), we used the AIA 304 \AA~ images. Figure \ref{aia304} shows some of the selected images of the flare site just before the impulsive phase and during the flare. We observe a mini-filament at the flare site at 06:57:20 UT (Figure \ref{aia304}a). The brightening starts below the southern leg of the filament at 06:58:08 UT, and the filament started to rise.  Figure \ref{aia304}c shows the kinked filament with two legs (marked by N and S). The southern (S) leg of the filament breaks possibly as a result of reconnection, and it shows unwinding motion at the apex and counterclockwise rotation of the northern leg (N) (see Figure \ref{aia304}d-f). The direction of the rotation is marked by blue arrows. These sense of rotation is detected by the bright and dark treads of the filament. The apex of the filament breaks and falls back (i.e., failed eruption) to the solar surface (Figure \ref{aia304}g). The temporal evolution of the filament (unwinding motion with failed eruption) in 304 \AA~ is shown in the animation attached to Figure \ref{aia304}.  
To determine the rising speed of the filament and its relation with the flare onset, we plotted the intensity distribution along the slice A, which is shown by the dotted line in Figure \ref{aia304}f. The stack plot is shown in Figure \ref{aia304}h. The average intensity of the eruption site (in AIA 304 \AA~) has also been plotted with blue curve. The rising of the filament and associated brightening in 304 \AA~ occurred simultaneously. From the linear fit, the estimated average speed of the filament was $\sim$40 km s$^{-1}$.

We used the Differential Affine Velocity Estimator (DAVE; \citealt{schuck2006}) method to track the apparent plasma flows (in the plane of sky) in and around the mini-filament. Coaligned pairs of the AIA images in 304 \AA~ (with 12 sec cadence) are utilized to deduce the plasma flows. Figures \ref{aia304}i-k display the AIA 304 \AA~ images overlaid by flow fields during the filament eruption. The flow vectors are in the upward direction along the northern leg, and in the southward direction at the apex. The direction of flow vectors agree with the expansion and rotation of the northern leg/apex of the filament in the counterclockwise direction. The longest arrows represent the flow speeds of 204, 178, and 158 km s$^{-1}$ respectively, which are larger than the flow speeds observed in the tornado like prominence (55-95 km s$^{-1}$), computed by \citet{li2012} using the local correlation tracking (LCT) method.

To see the chromospheric evolution of the filament in a better spatial resolution than AIA, we also utilized the high resolution (0.08$\arcsec$ pixel$^{-1}$) H$\alpha$ (6563 \AA) images observed by the Solar Optical Telescope (SOT; \citealt{tsuneta2008,suematsu2008}) onboard Hinode satellite \citep{kosugi2007}. Figure \ref{sot} displays a series of H$\alpha$ images showing the evolution of the mini-filament and associated flare. Figure \ref{sot}a shows the kinked mini-filament (07:00:15 UT) associated with the brightenings near its southern leg. The twisted (writhed) fields at the apex of the filament is much clear here. The southern leg of the filament detached from the surface as a result of magnetic reconnection and showed the unwinding motion of the apex and northern leg in the counterclockwise direction (Figure \ref{sot}e-f). The apex of the filament also detached and showed a failed eruption similar to the AIA 304 \AA~ images. Most of the plasma is drained back to the solar surface. These images clearly demonstrate the dynamical evolution of the filament associated with the kink instability.


\subsection{Hard X-ray emission}

To investigate the particle acceleration sites during the flare associated with the kinked mini-filament eruption, we utilized Reuven Ramaty High Energy Solar Spectroscopic Imager (RHESSI) hard X-ray images \citep{lin2002}. We used PIXON algorithm technique to reconstruct the RHESSI images because of its most accurate image photometry \citep{metcalf1996,aschwanden2004}. We adopted 20 sec integration time for the image reconstruction in 6-12 and 12-25 keV energy channels. The hard X-ray emission was limited to these energy channels only and was not detected in higher energy bands.

Figure \ref{hessi} displays the selected AIA 1600, 1700, and 131 \AA~ images overlaid by RHESSI hard X-ray images contours in 6-12 (red) and 12-25 keV (blue). We use the AIA 1600 and 1700 \AA~ images to explore the photospheric brightenings associated with accelerated particles from the corona (i.e., hard X-ray emission). Flare heated loops and associated (hot and cool) magnetic structures (0.4, 11, and 16 MK) are well observed in the AIA 131 \AA~ channel.  The co-temporal AIA images overlaid by the RHESSI images contours are shown in the first and second rows. A mini-filament (indicated by arrow,  Figure \ref{hessi}a$^\prime$) was observed prior to the flare onset and no brightening was observed in AIA 1600 \AA~ at 06:57:05 UT. The filament activation and subsequent rising motion was closely associated with the formation of a hot loop (Figure \ref{hessi}b and b$^\prime$) near the southern leg of the filament. A hard X-ray source (observed in both energy bands, marked by A) was found to be cospatial with the hot loop. Furthermore, we see a clear loop structure in 12-25 keV channel showing the two footpoints (FP1 and FP2) and a looptop source (LT) during 06:58:40-06:59:00 UT. As the filament rises up, we notice the formation of an another source B (Figure \ref{hessi}d and d$^\prime$) located above the filament apex. The underlying flare loop becomes more brighter as the filament moves up due to the progressive magnetic reconnection associated with the breaking of the southern leg.  The source B in 6-12 keV exhibits the elongated structure parallel to the flare ribbons. Simultaneously, we observed the brightenings along the northern side of the filament (two kernals K1 and K2 are marked by arrows, Figure \ref{hessi}d). The height of the source B increases with the rising motion of the filament (Figure \ref{hessi}e, e$^\prime$, f, f$^\prime$). Furthermore, the structure of the source B becomes elongated during 07:01:20--07:01:40 UT (Figure \ref{hessi}g and g$^\prime$). Later, we noticed two sources C and D along both sides of the filament apex (Figure \ref{hessi}h and h$^\prime$). 

Appearance of the source A at the beginning of the flare suggests the occurrence of magnetic reconnection above or near the southern leg of the filament, which causes the formation of a hot underlying loop, observed in AIA 131 \AA. Consequently, the filament moved up and breaking of the southern leg took place. Furthermore, source B above the filament apex indicate the particle acceleration (energy release site) probably caused by the magnetic reconnection above the filament, resulting the appearance of an overlying mini-flux rope like structure (which we will discuss in the next section). In addition, the elongation of the source B, and appearance of the source C and D confirm the particle acceleration from the filament apex to the north and south directions.

Using the RHESSI and TRACE observations, \citet{ji2003} have reported the hard X-ray source (12-25 keV) above a large-scale filament, which was failed to erupt associated with the kink instability. They have suggested that the reconnection (i.e., energy release) must have occurred above the filament. This result is not supported by the standard flare (i.e., CSHKP) model, because the reconnection takes place below the filament or flux rope in the CSHKP model. 
\citet{alex2006} also have studied the same event and confirmed the presence of a hard X-ray source (12-25 keV) in the corona above the filament prior to the main activation phase and identified a second coronal hard X-ray source under the apex of the strongly kinked filament during the eruption. They interpreted the second hard X-ray source as the energy release site at the vertical current sheet formed in between the crossing legs of the filament, as suggested in the numerical simulation of kink instability \citep{torok2004}. We also observed the hard X-ray source above the filament similar to these observational reports but no hard X-ray source in between the crossing legs of the filament.

\begin{figure*}
\centering{
\includegraphics[width=4.4cm]{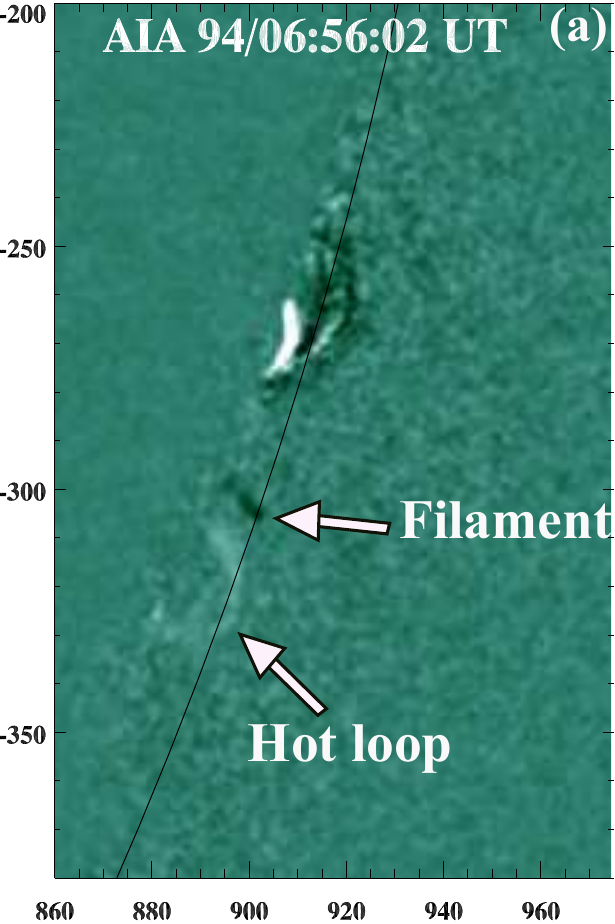}
\includegraphics[width=4.4cm]{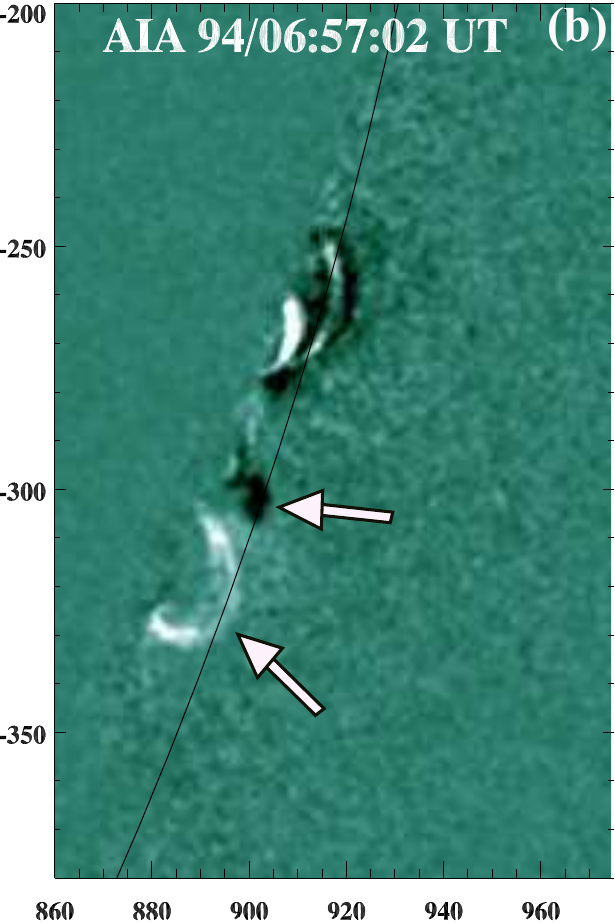}
\includegraphics[width=4.4cm]{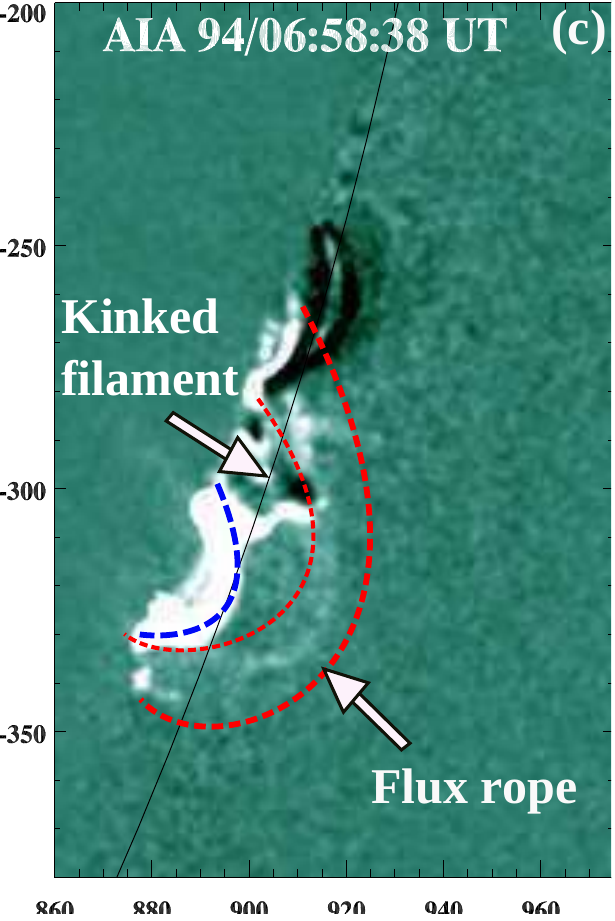}
\includegraphics[width=4.4cm]{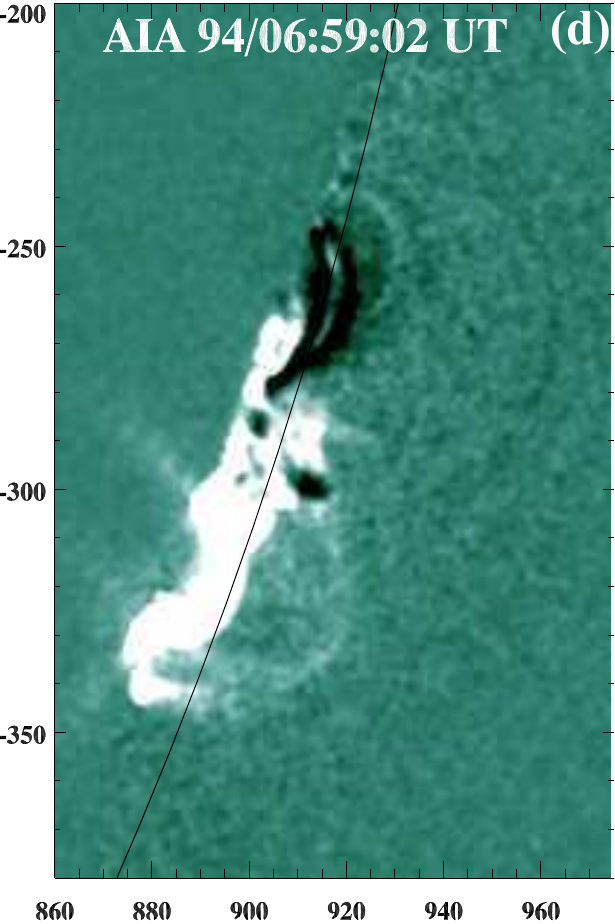}

\includegraphics[width=4.4cm]{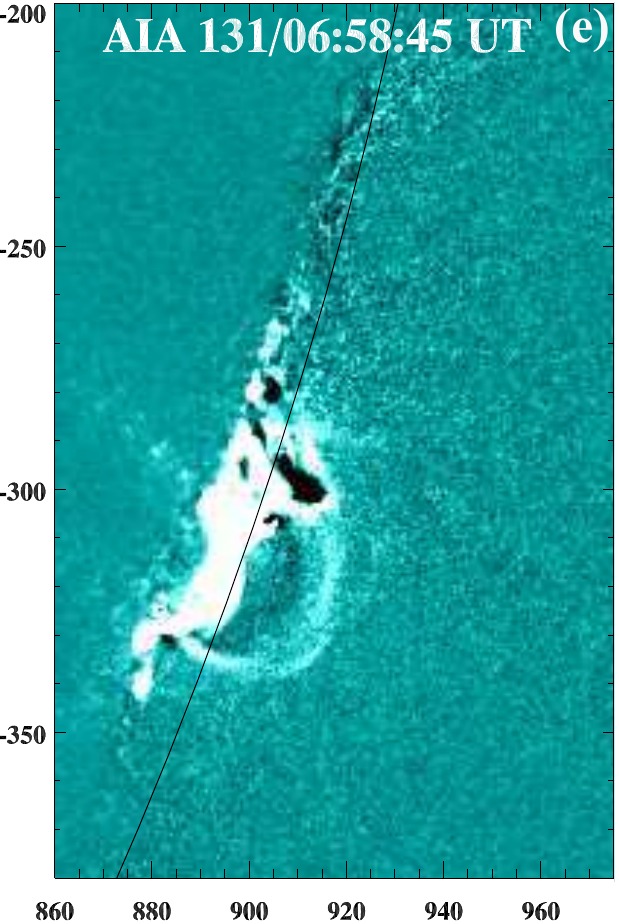}
\includegraphics[width=4.4cm]{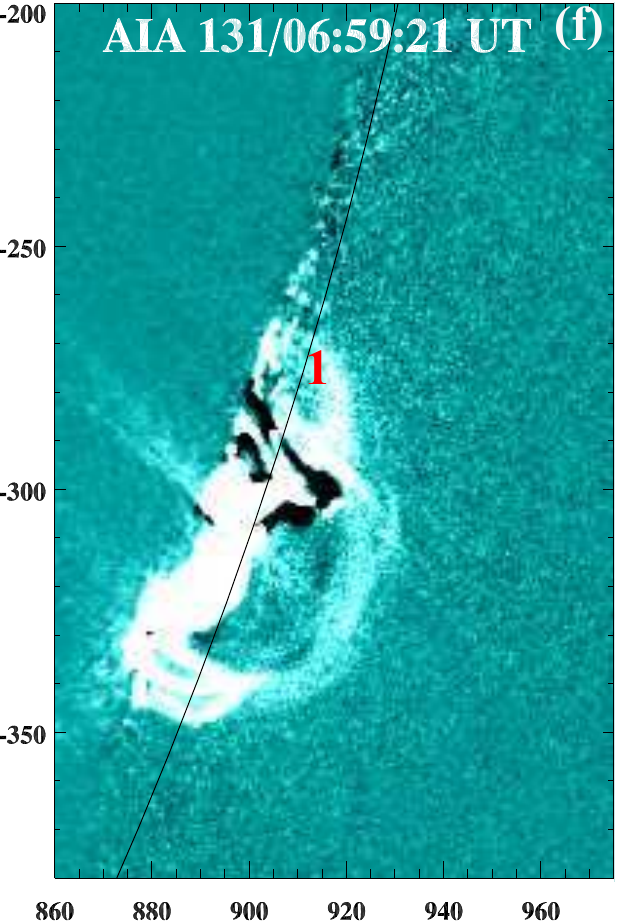}
\includegraphics[width=4.4cm]{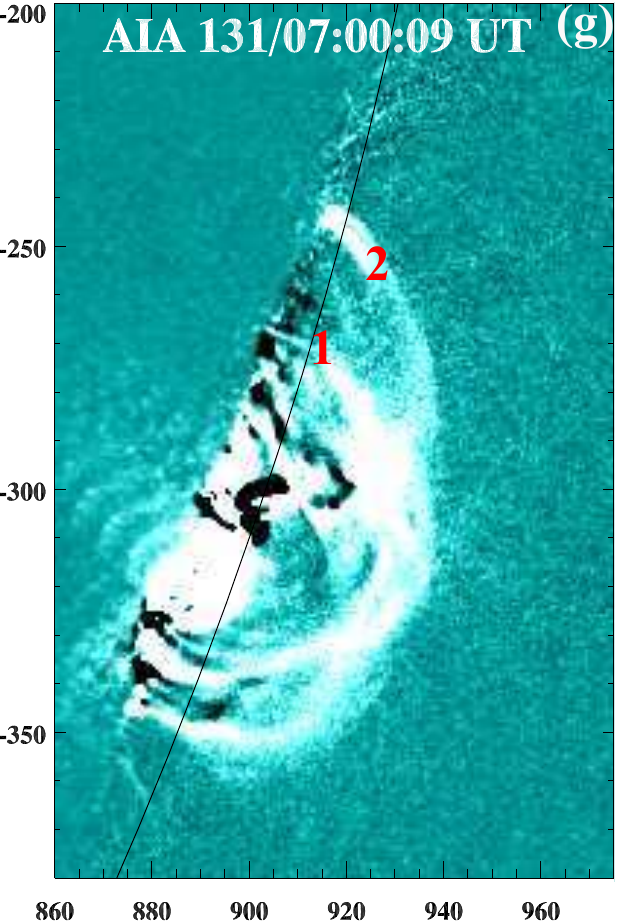}
\includegraphics[width=4.4cm]{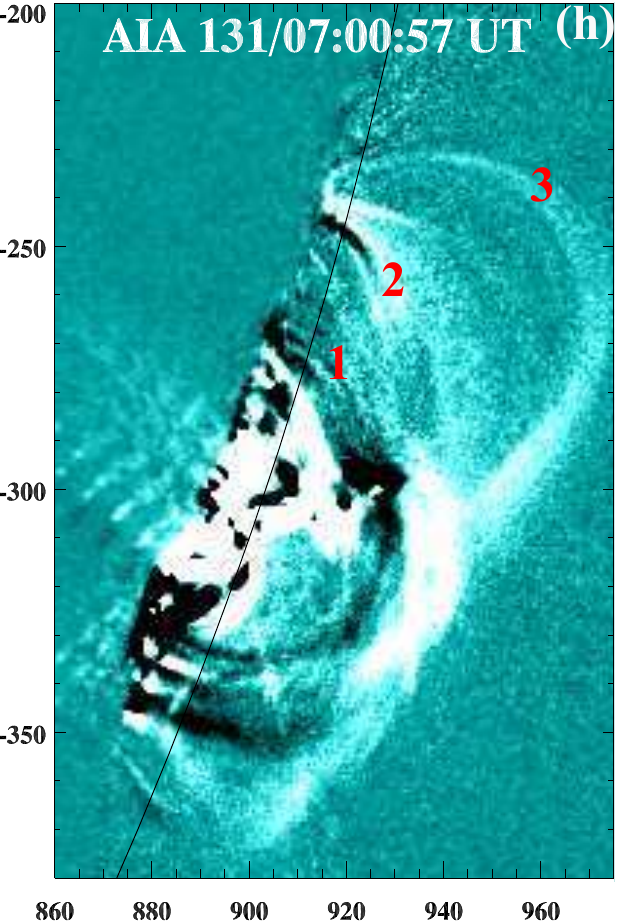}

\includegraphics[width=4.5cm]{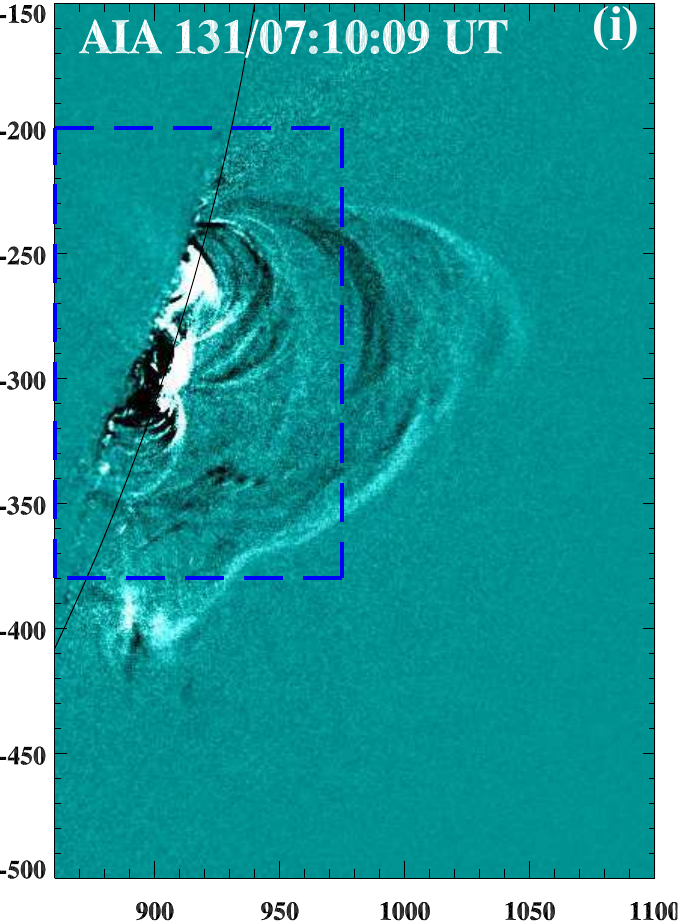}
\includegraphics[width=4.5cm]{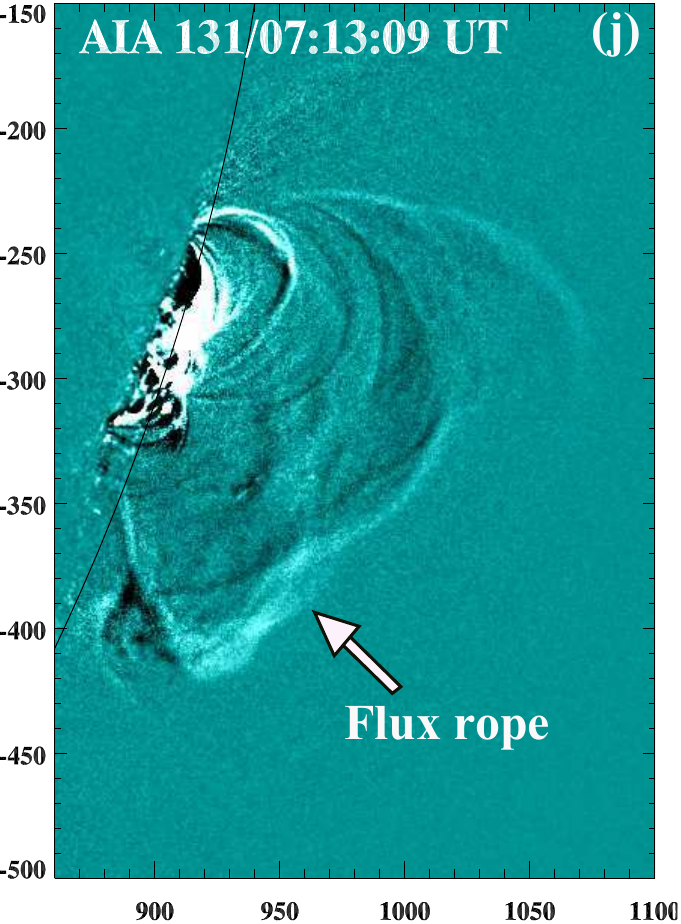}
\includegraphics[width=4.5cm]{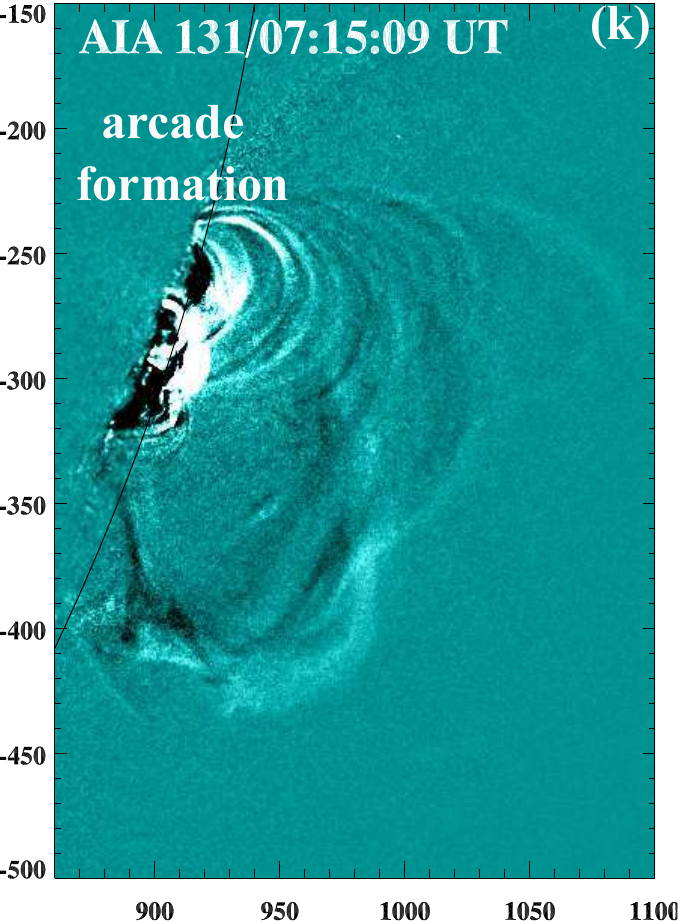}
\includegraphics[width=4.5cm]{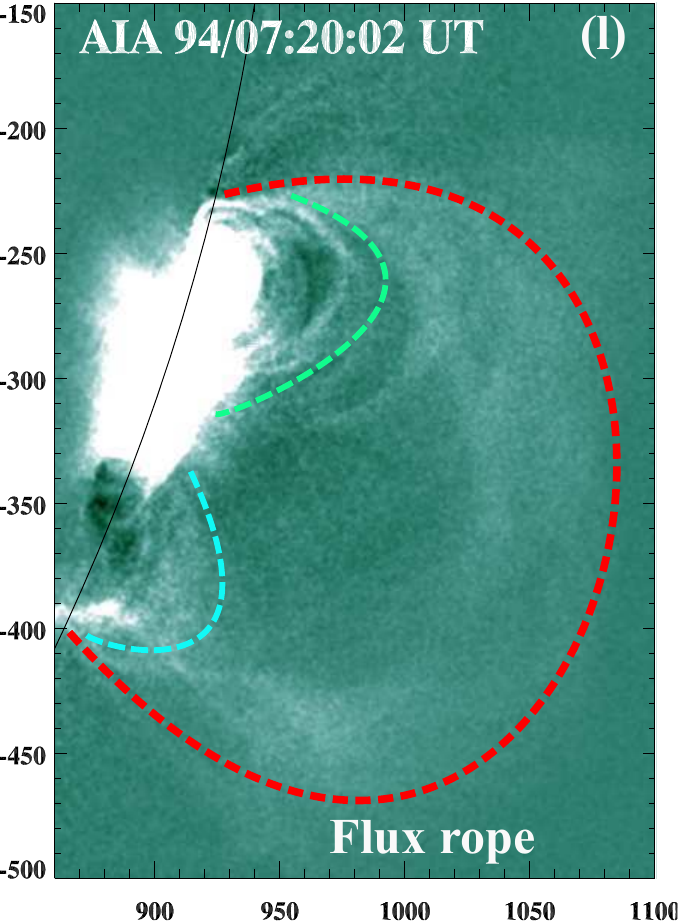}

}
\caption{SDO/AIA 94 \AA~ base difference and 131 \AA~ running difference images showing the formation and ejection of the twisted flux rope above the kinked filament during magnetic reconnection. Red dotted curve shows the flux rope and other curves indicate the underlying closed loops (panels c and l). Blue rectangle (panel i) represents the size of the upper panels.}
\label{aia131_94}
\end{figure*}
\begin{figure*}
\centering{
\includegraphics[width=4.5cm]{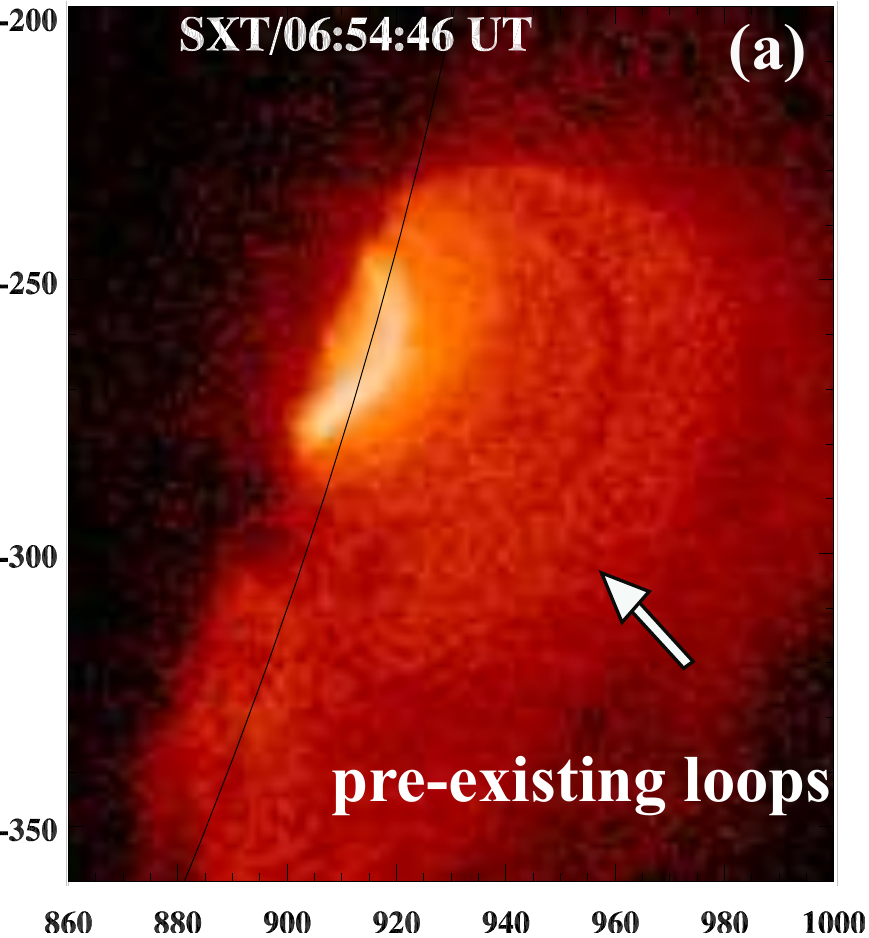}
\includegraphics[width=4.5cm]{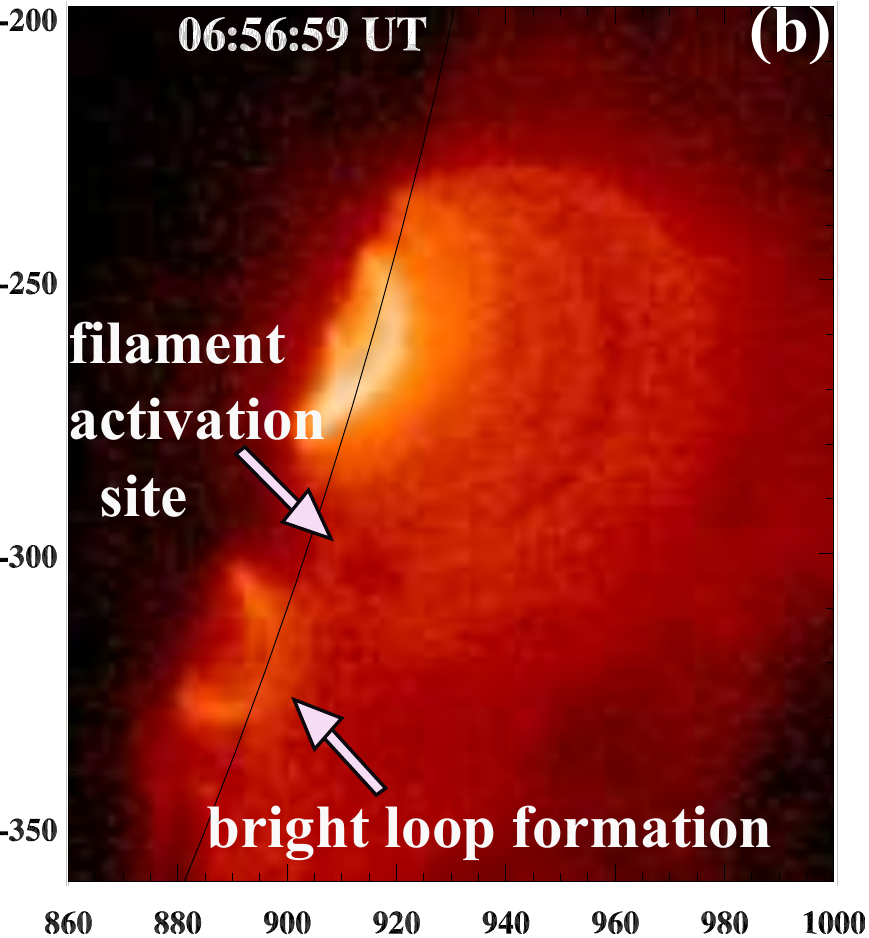}
\includegraphics[width=4.5cm]{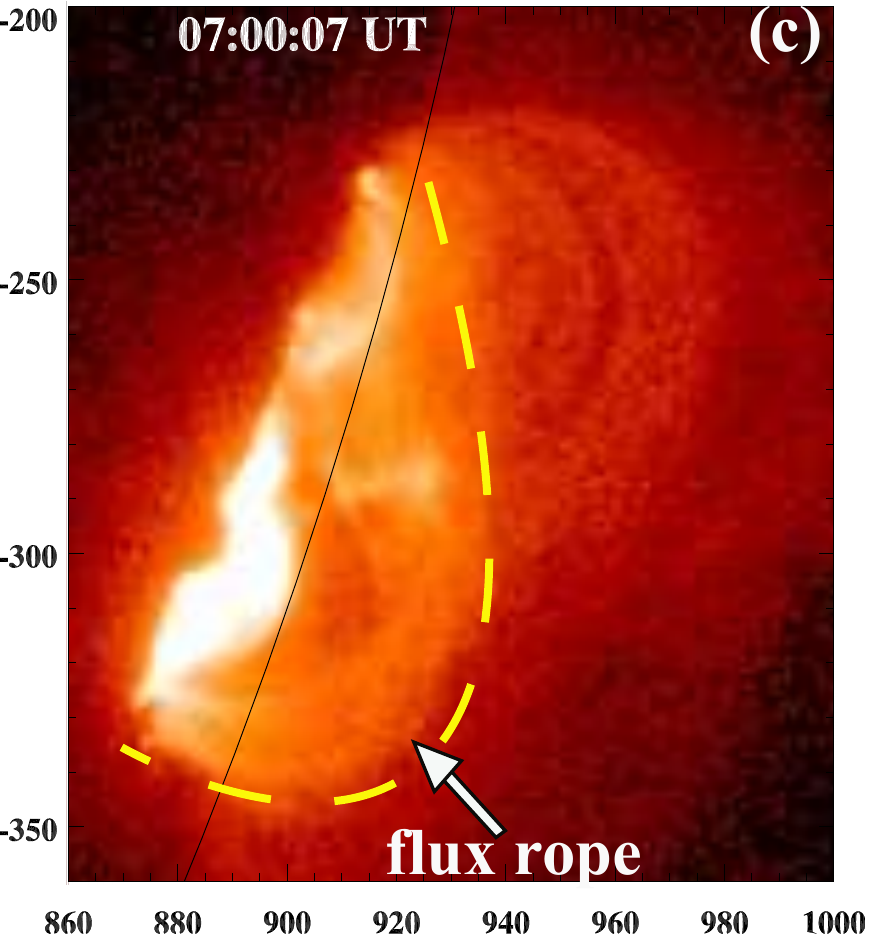}
\includegraphics[width=4.5cm]{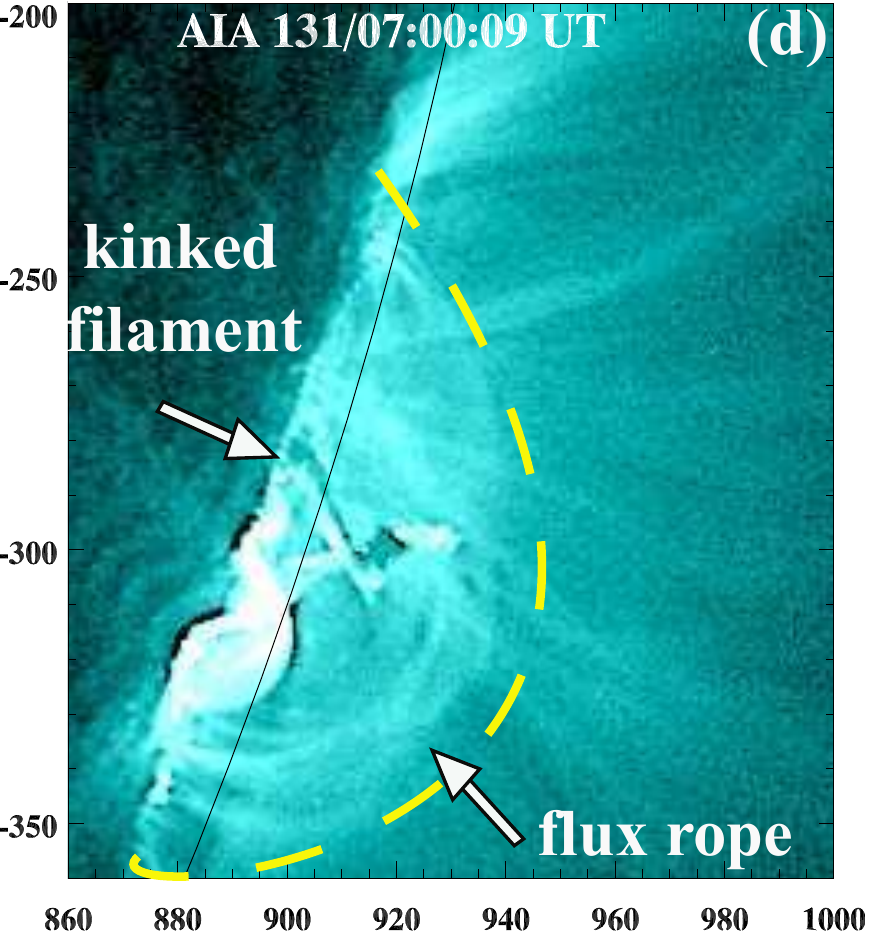}

\includegraphics[width=4.96cm]{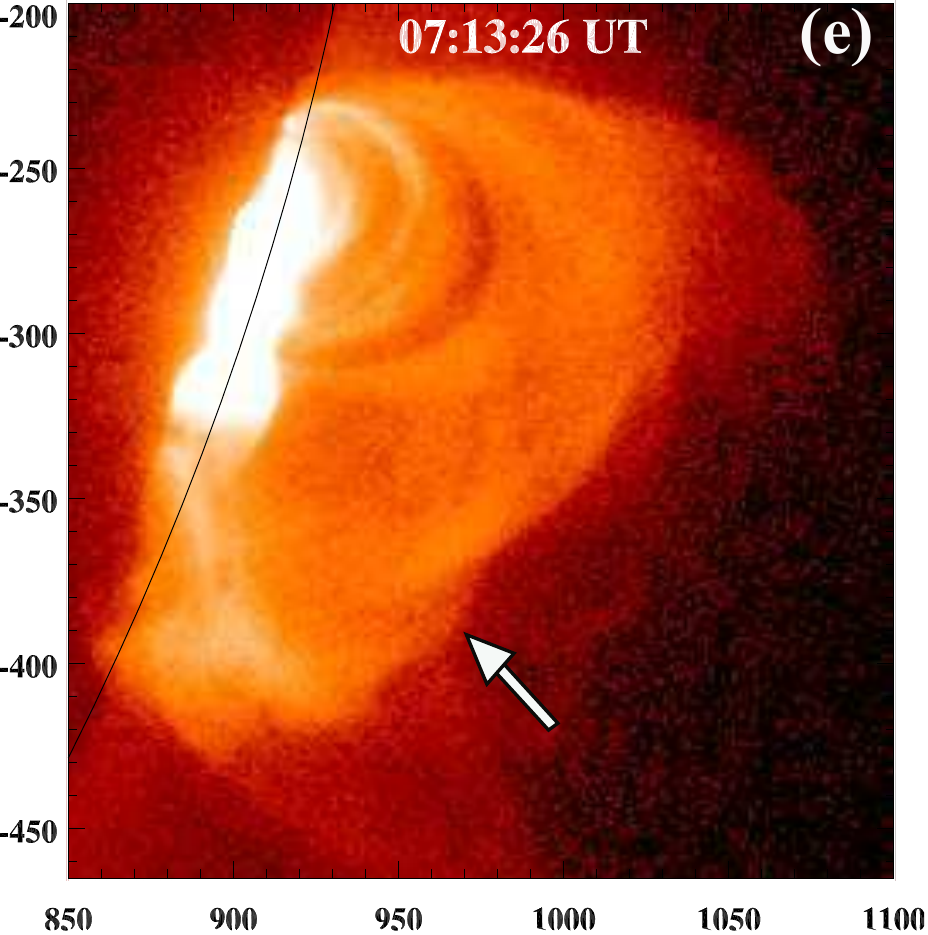}
\includegraphics[width=4.96cm]{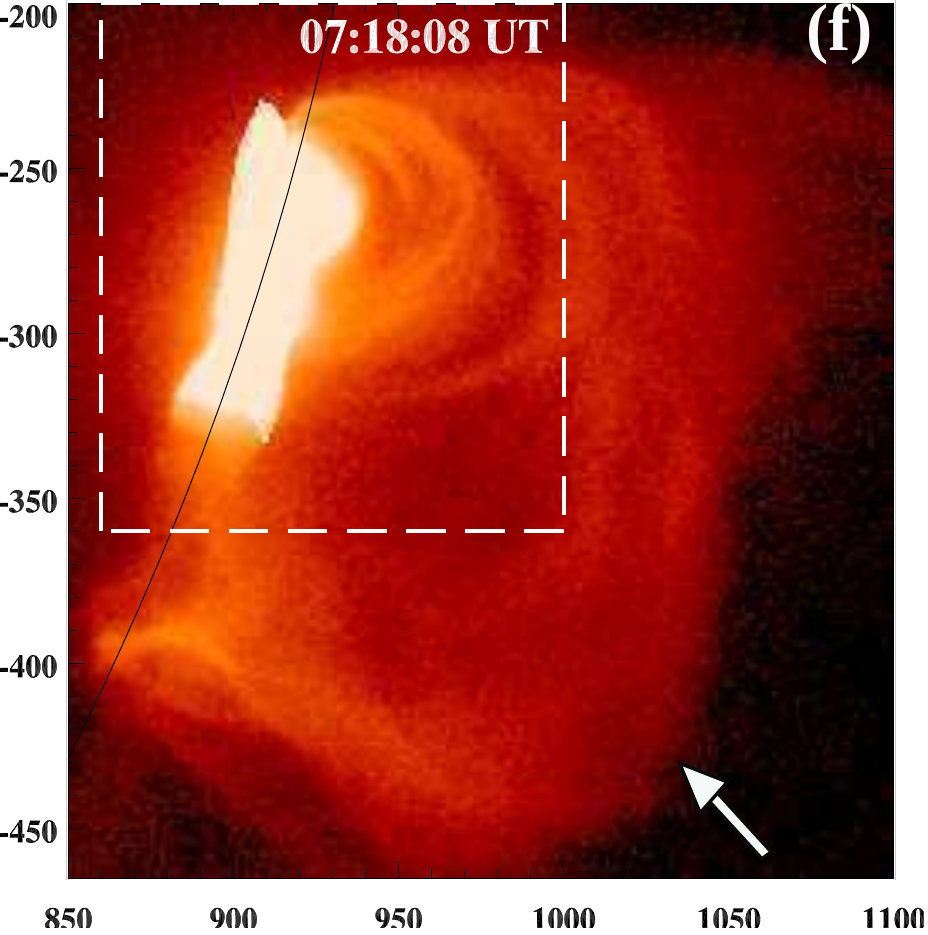}
\includegraphics[width=4.96cm]{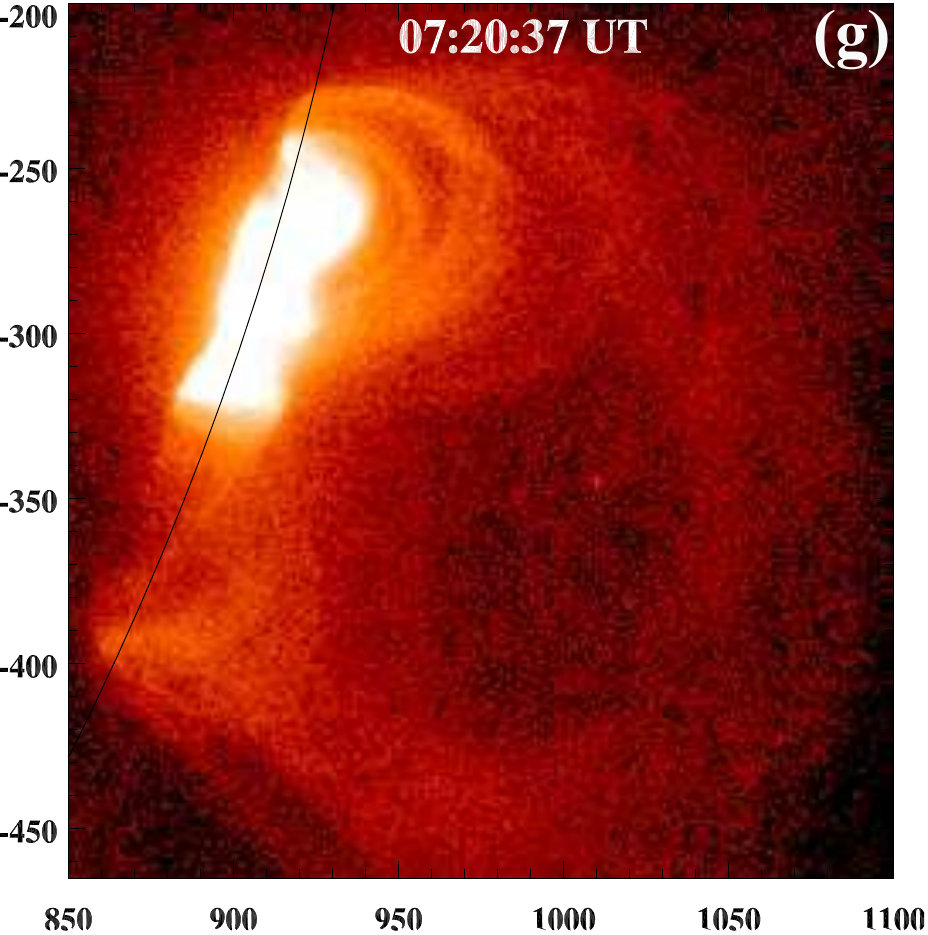}
\includegraphics[width=3.1cm]{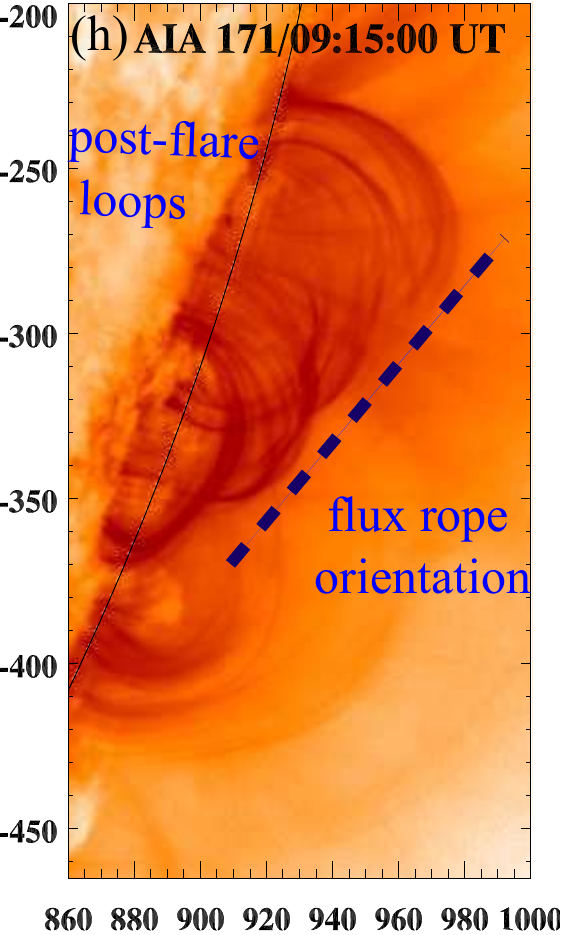}

\includegraphics[width=3.4cm]{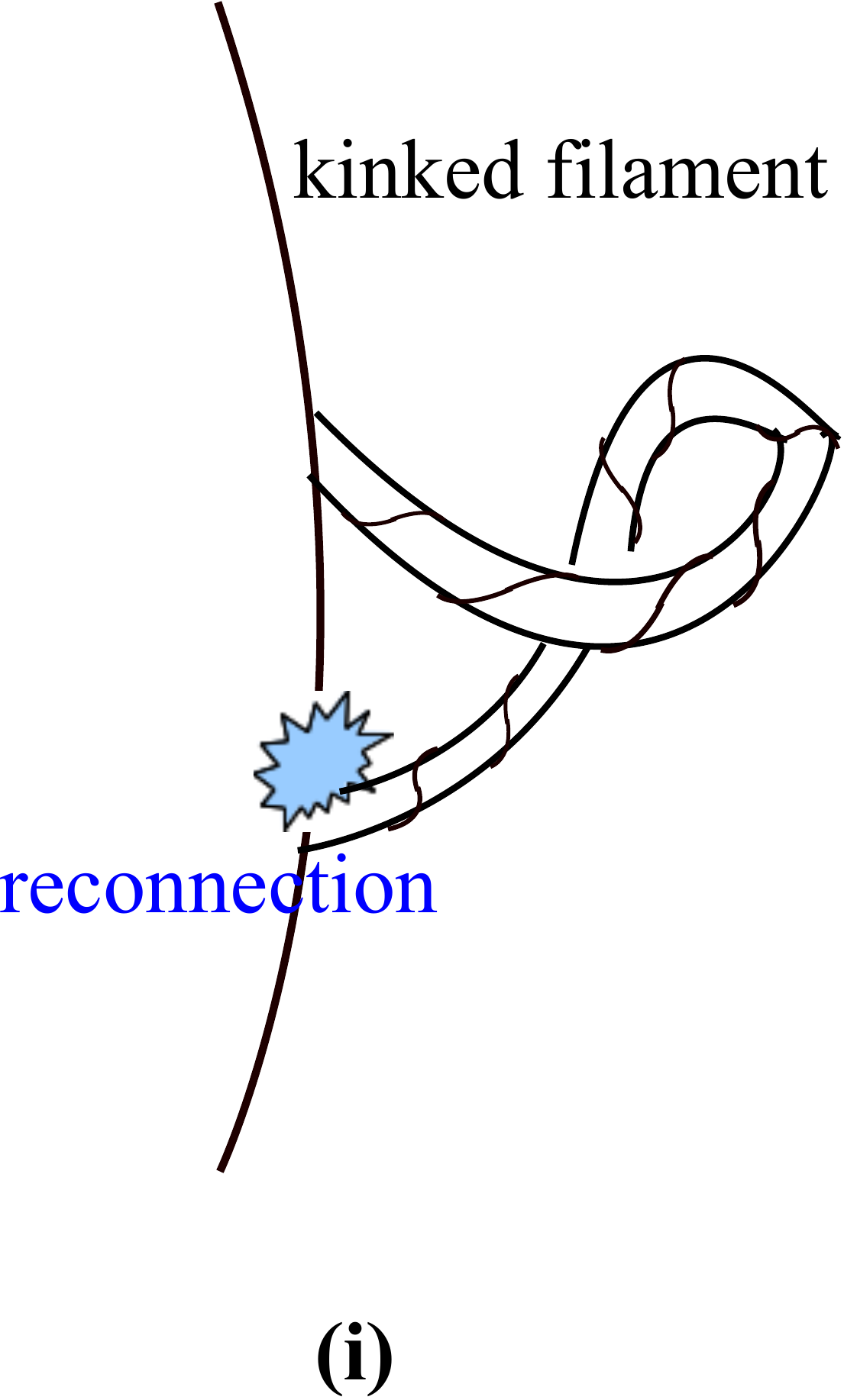}
\includegraphics[width=6.3cm]{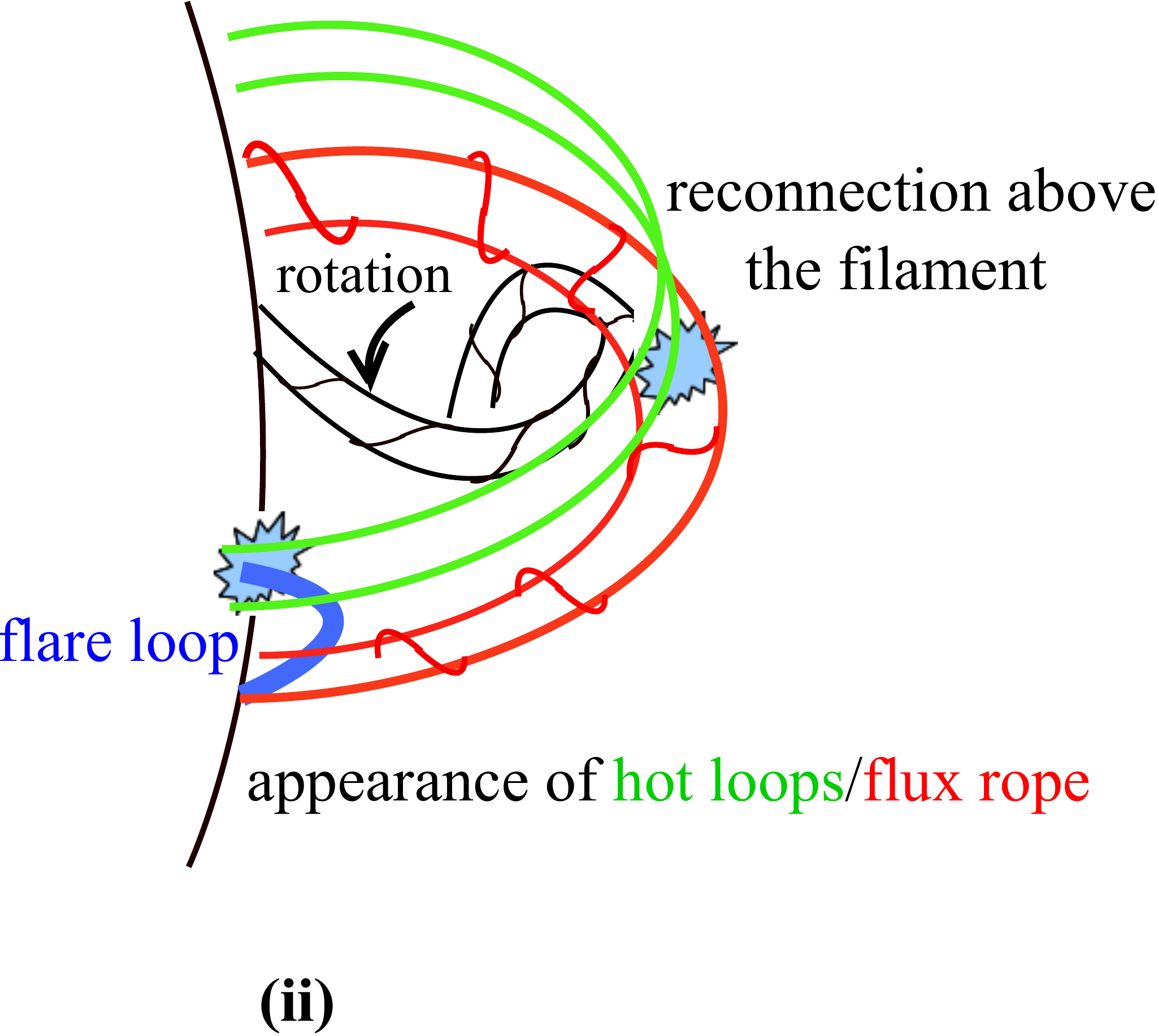}
\includegraphics[width=5.5cm]{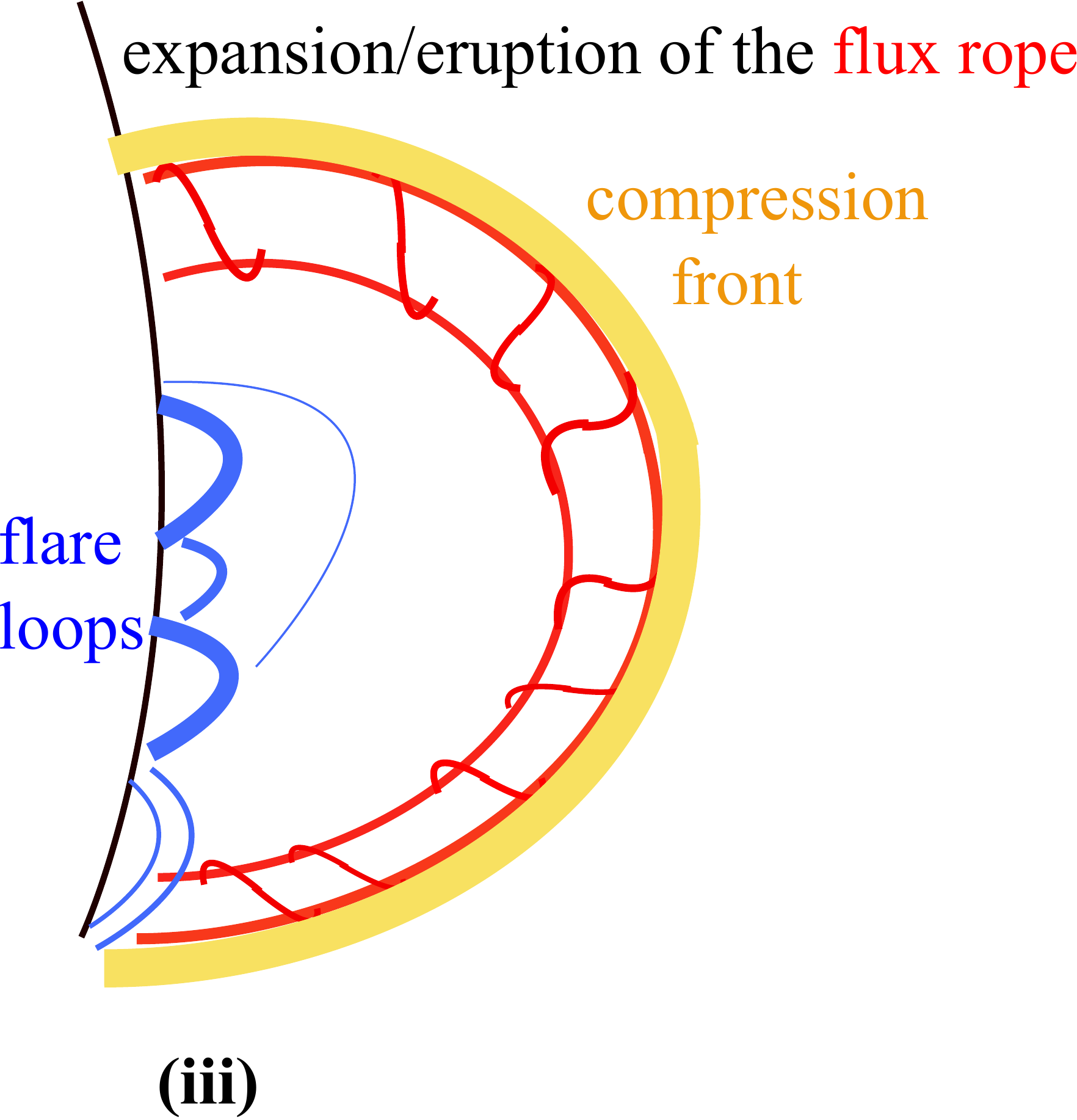}
}
\caption{Hinode SXT images of the eruption site showing the evolution of the flux rope during the flare. The appearance of the flux rope is shown by a yellow dotted curve (c and d). The dotted rectangular box in panel (f) represents the field of view of upper row panels. Panel (g) shows the post-flare loops observed in AIA 171 \AA~ channel, and orientation of the flux rope is marked by the dotted lines. Bottom: Schematic cartoon showing the different stages of the eruption.}
\label{sxt}
\end{figure*}

\begin{figure*}
\centering{
\includegraphics[width=6cm]{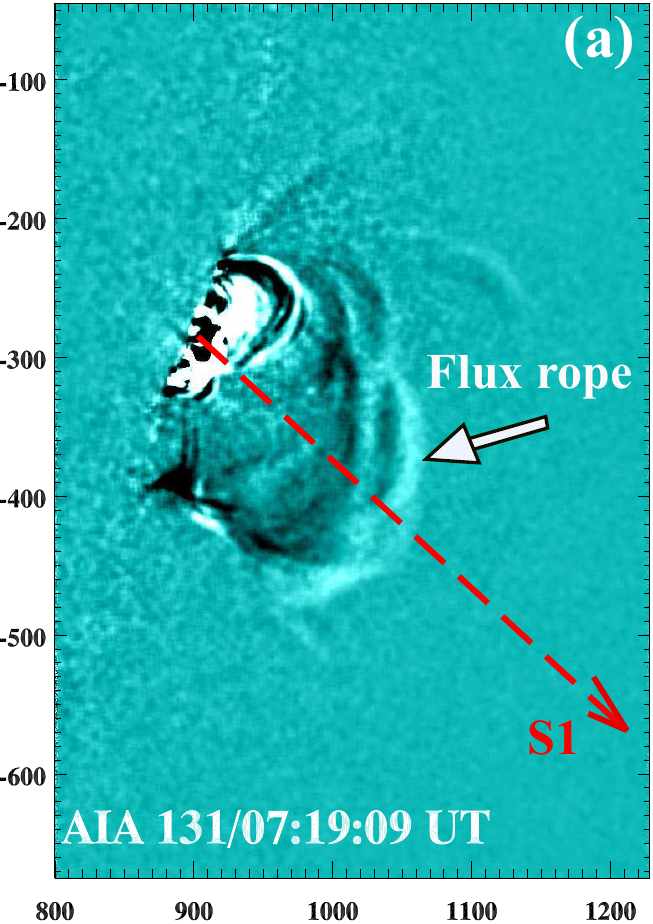}
\includegraphics[width=5.9cm]{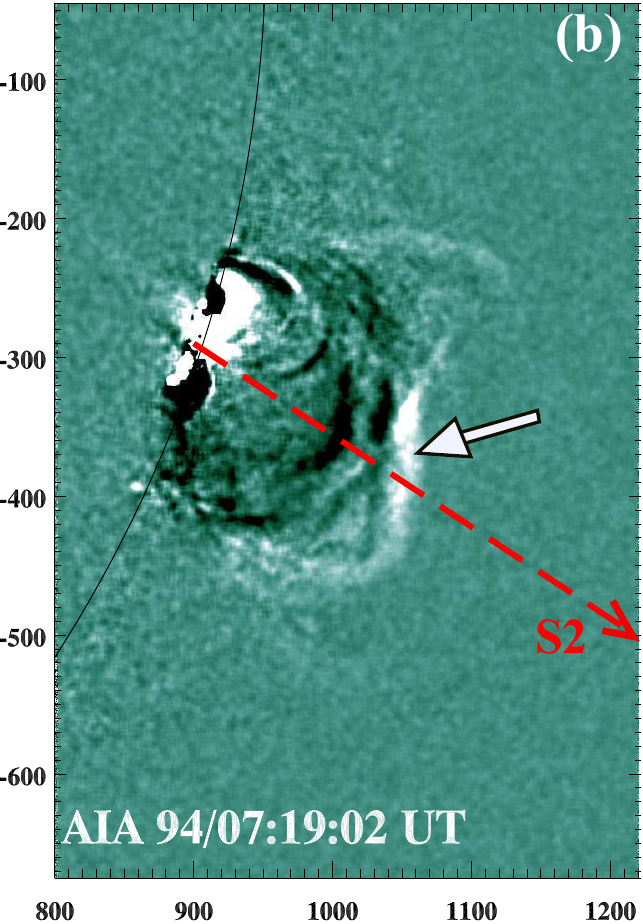}

\includegraphics[width=6cm]{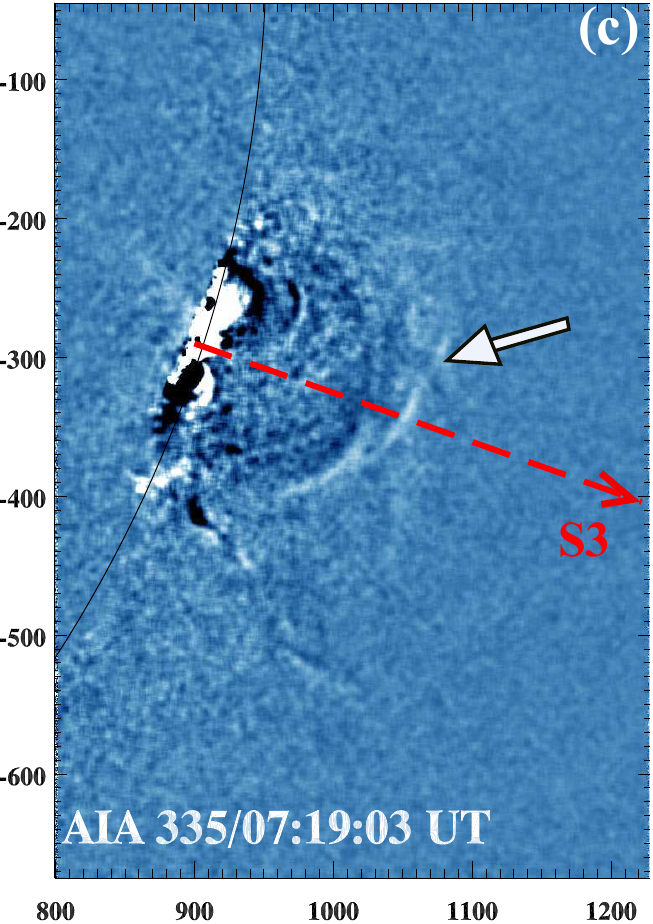}
\includegraphics[width=5.9cm]{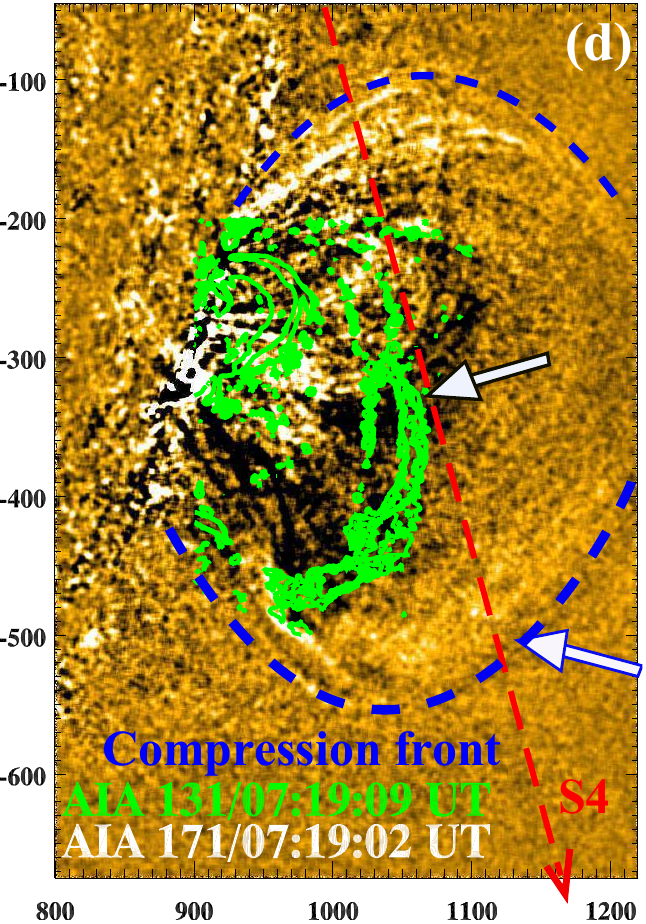}
}
\caption{SDO/AIA running difference images in 131, 94, 335, and 171 \AA~ channels showing the flux rope eruption. S1, S2, S3, and S4 are the slice cuts used for the stack plots. The flux rope contours (green) of AIA 131 \AA~ are overlaid on the AIA 171 \AA~ image (panel d). The blue dotted curve in AIA 171 \AA~ indicates the cool compression front of the CME. The temporal evolution of the flux rope and associated flare in 131, 94, 335, and 171 \AA~ can be found in a movie available in the online edition. }
\label{aia_rd}
\end{figure*}
\begin{figure*}
\centering{
\includegraphics[width=8.6cm]{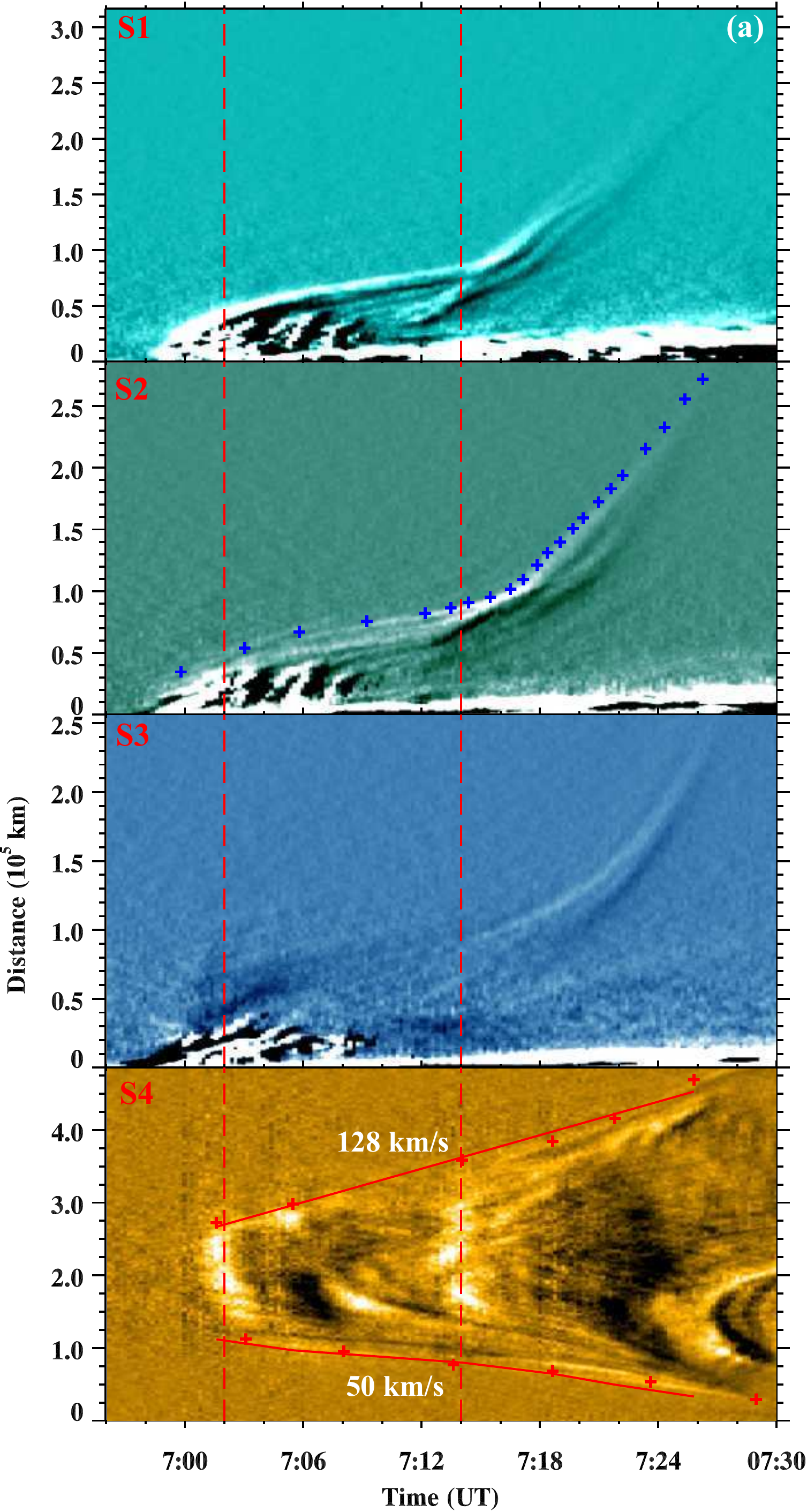}
\includegraphics[width=8.4cm]{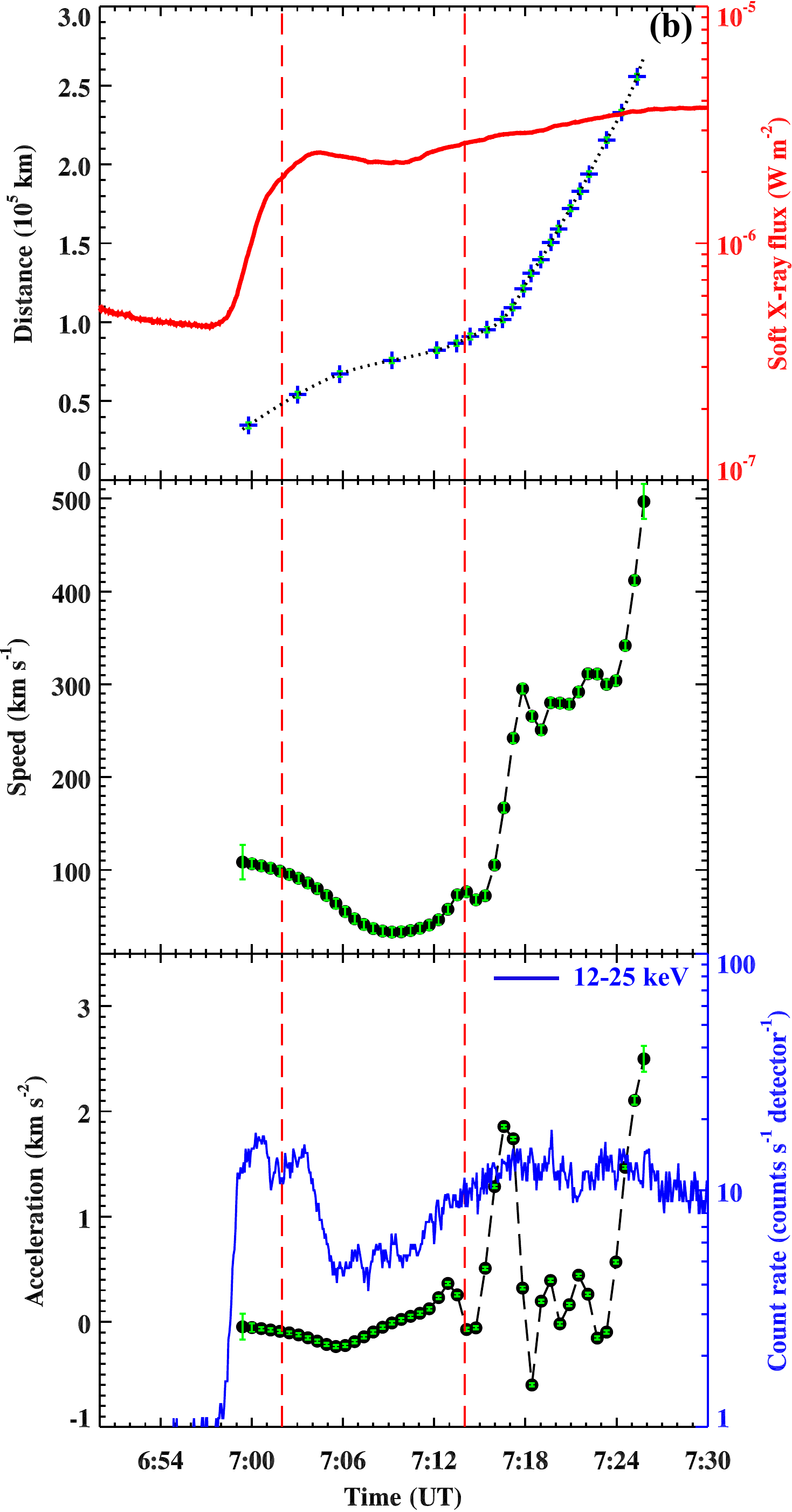}
}
\caption{(a) Space-time plots along the slices S1, S2, S3, and S4 using the AIA 131, 94, 335, and 171 \AA~ images. (b) Top: Height-time profile of the flux rope and GOES soft X-ray flux (red) in 1-8 \AA~ channel. Middle: Speed profile derived by the height-time measurements of the flux rope. Bottom: Acceleration profile of the flux rope plotted with RHESSI hard X-ray flux (blue) in 12-25 keV energy channel. Two vertical dotted lines indicate the timing of the 171 \AA~ brightenings and hard X-ray bursts.}
\label{stack}
\end{figure*}
\begin{figure*}
\centering{
\includegraphics[width=7.5cm]{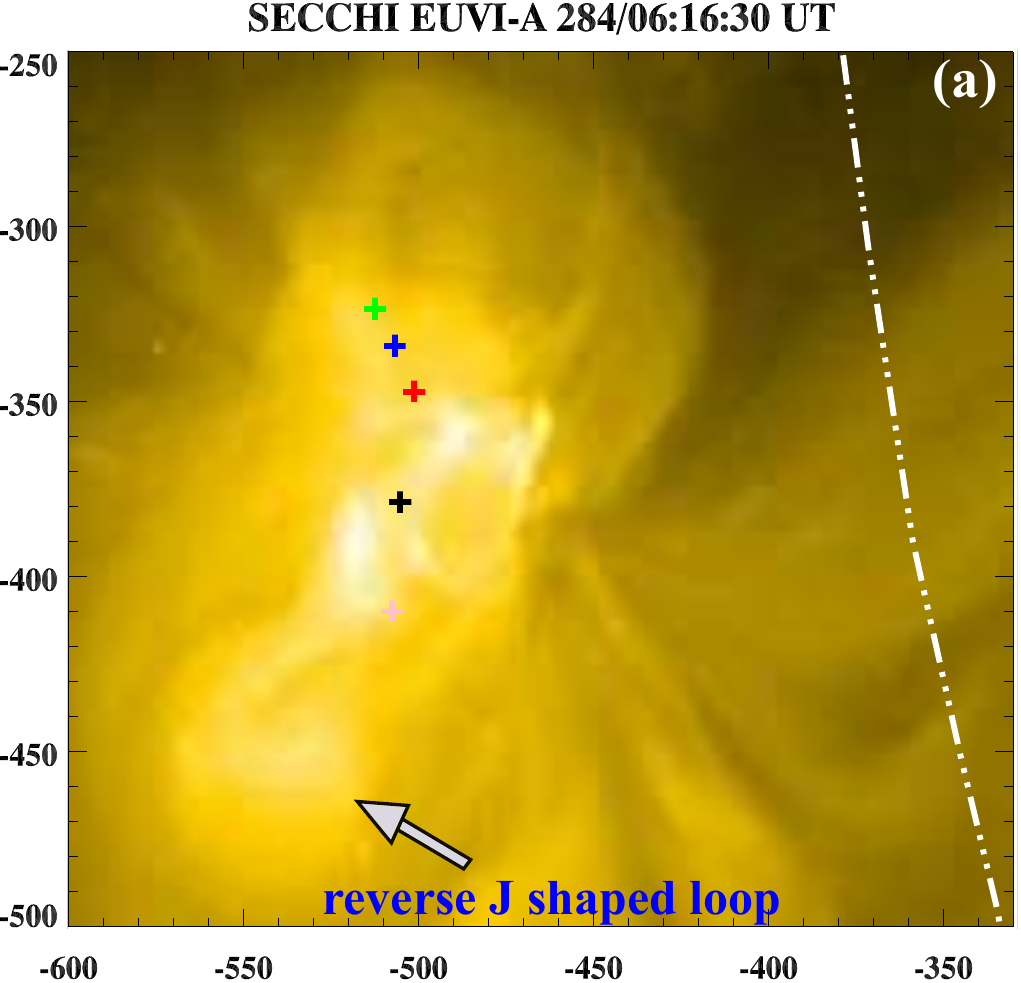}
\includegraphics[width=6.2cm]{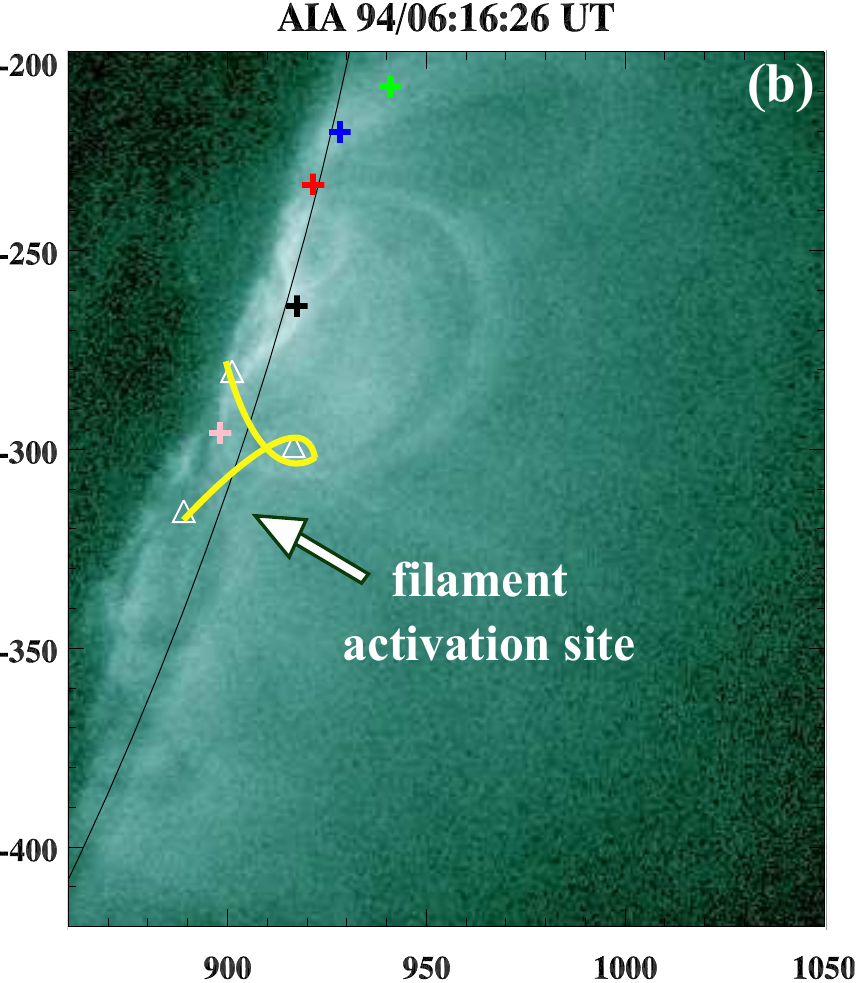}

\includegraphics[width=7.5cm]{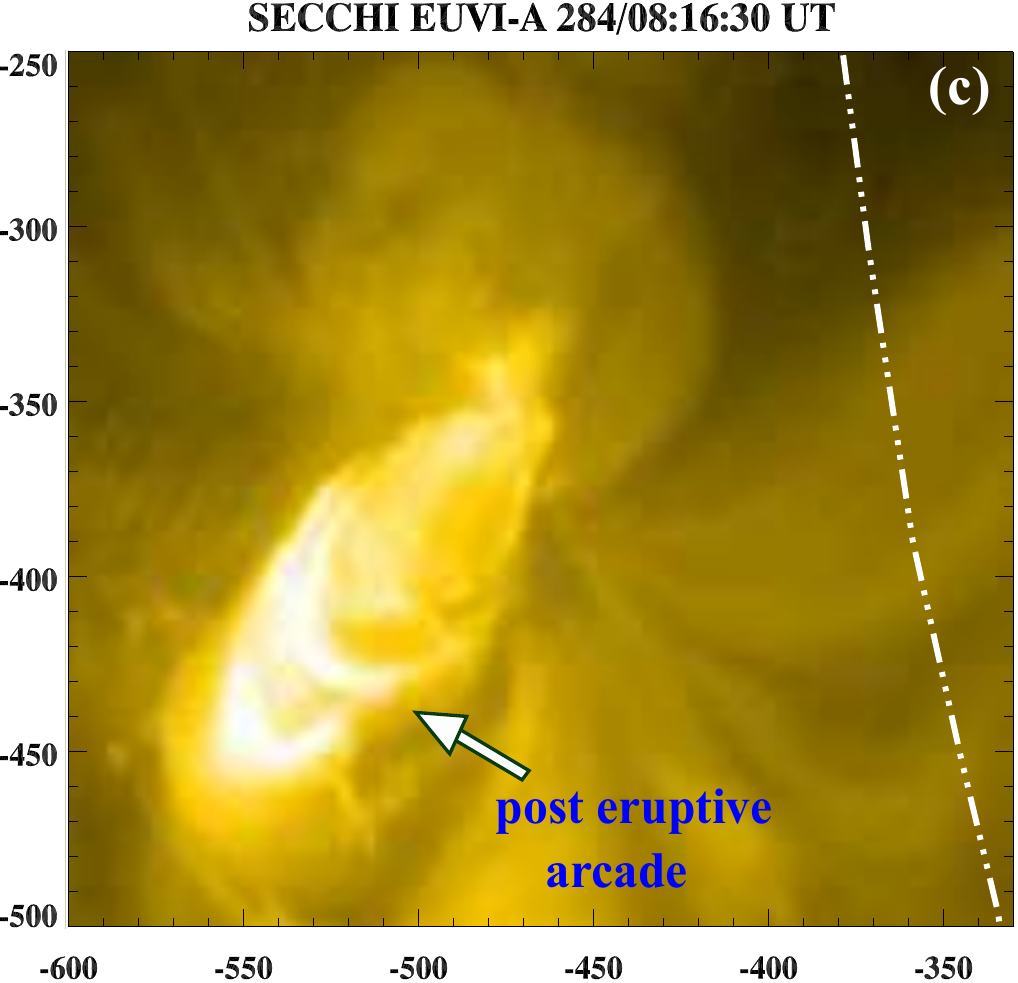}
\includegraphics[width=6.2cm]{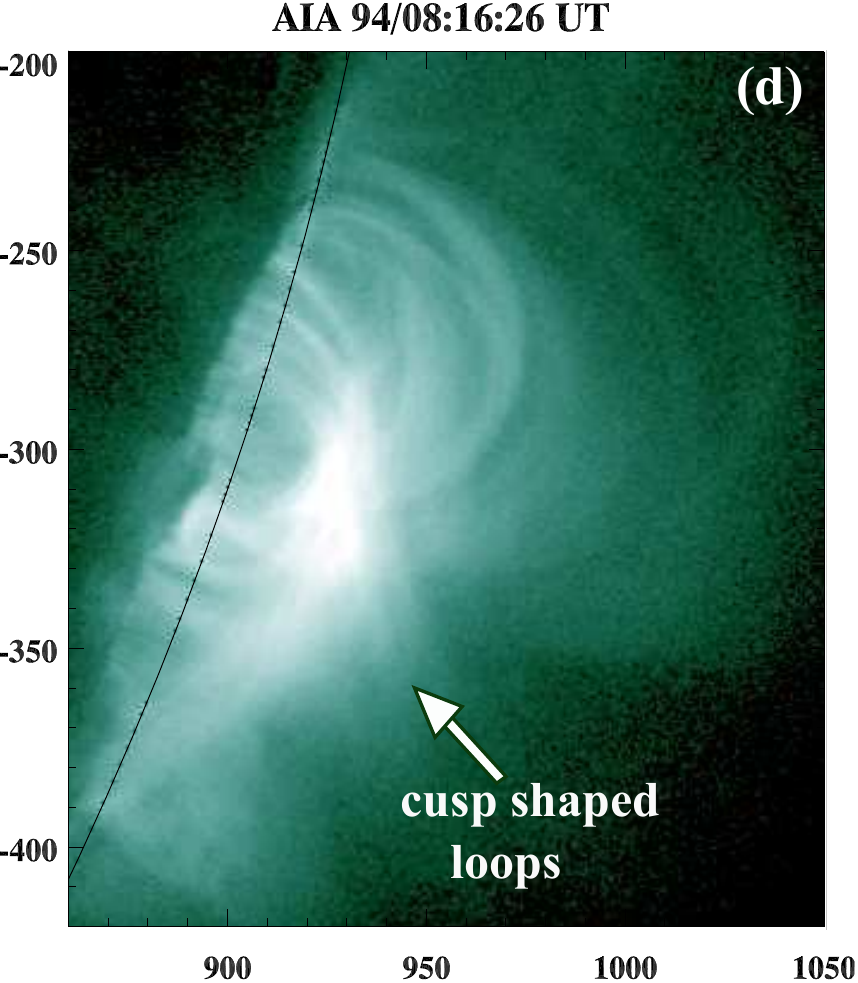}
}
\caption{STEREO SECCHI EUVI-A images showing the magnetic field configuration of the active region in 284 \AA~ channel before (a) and during (c) the flare. For comparison, the approximate positions of the magnetic structures are marked by `+' symbols. The position of the filament activation is drawn by using AIA 131 \AA~ image at 07:00 UT. The limb position from the AIA image is over-plotted on SECCHI images (white dotted lines). AIA 94 \AA~ images (b and d) of the eruption site.}
\label{st}
\end{figure*}
\begin{figure*}
\centering{
\includegraphics[width=7.0cm]{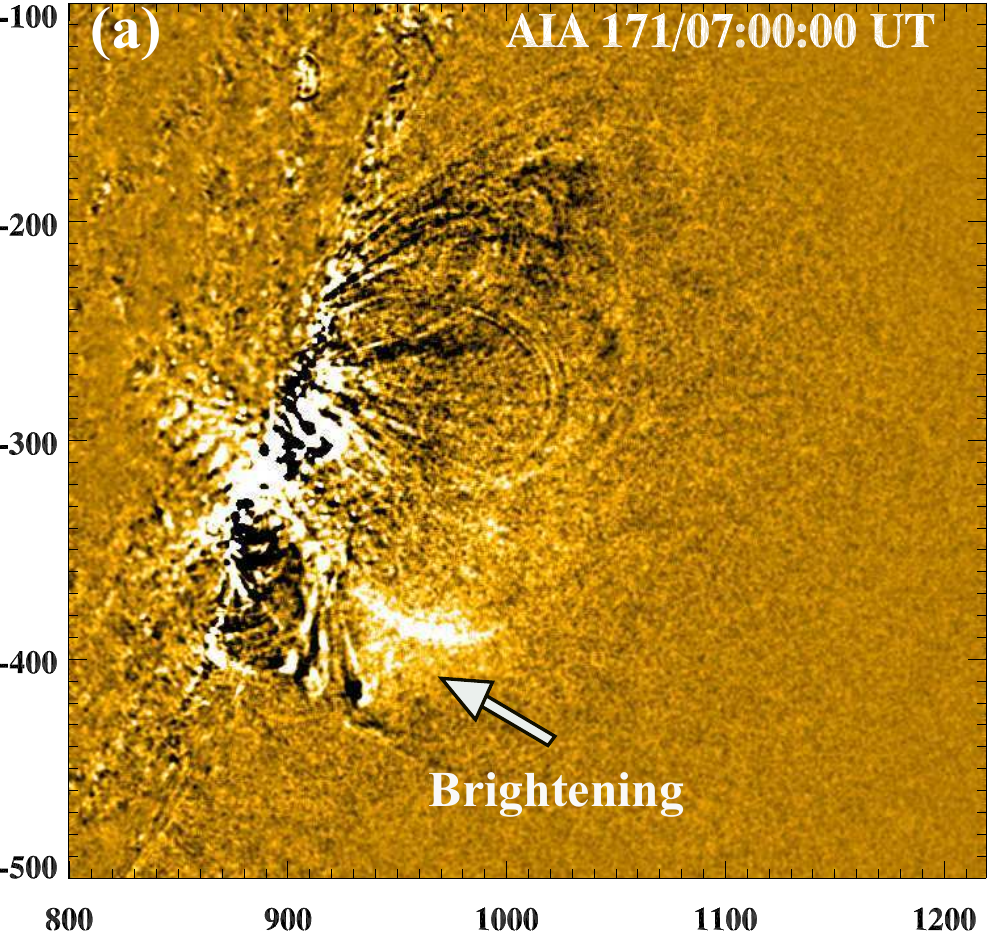}
\includegraphics[width=7.0cm]{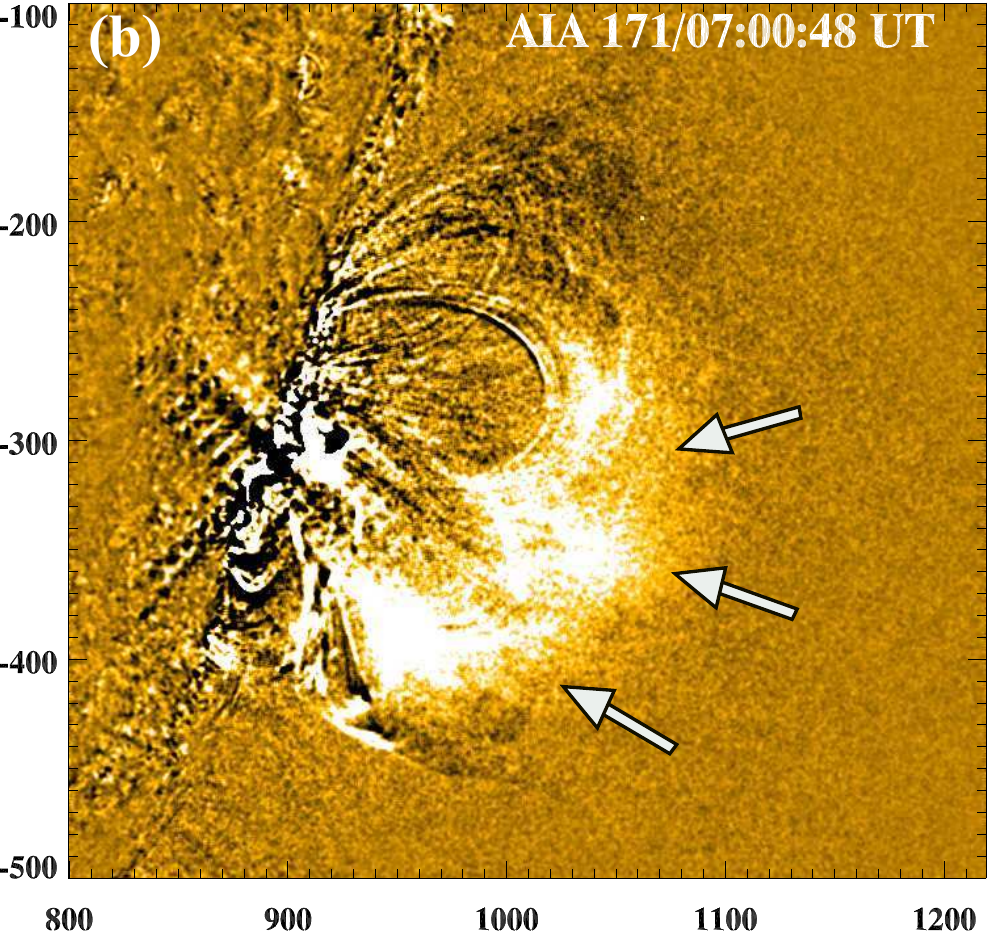}
\includegraphics[width=7.0cm]{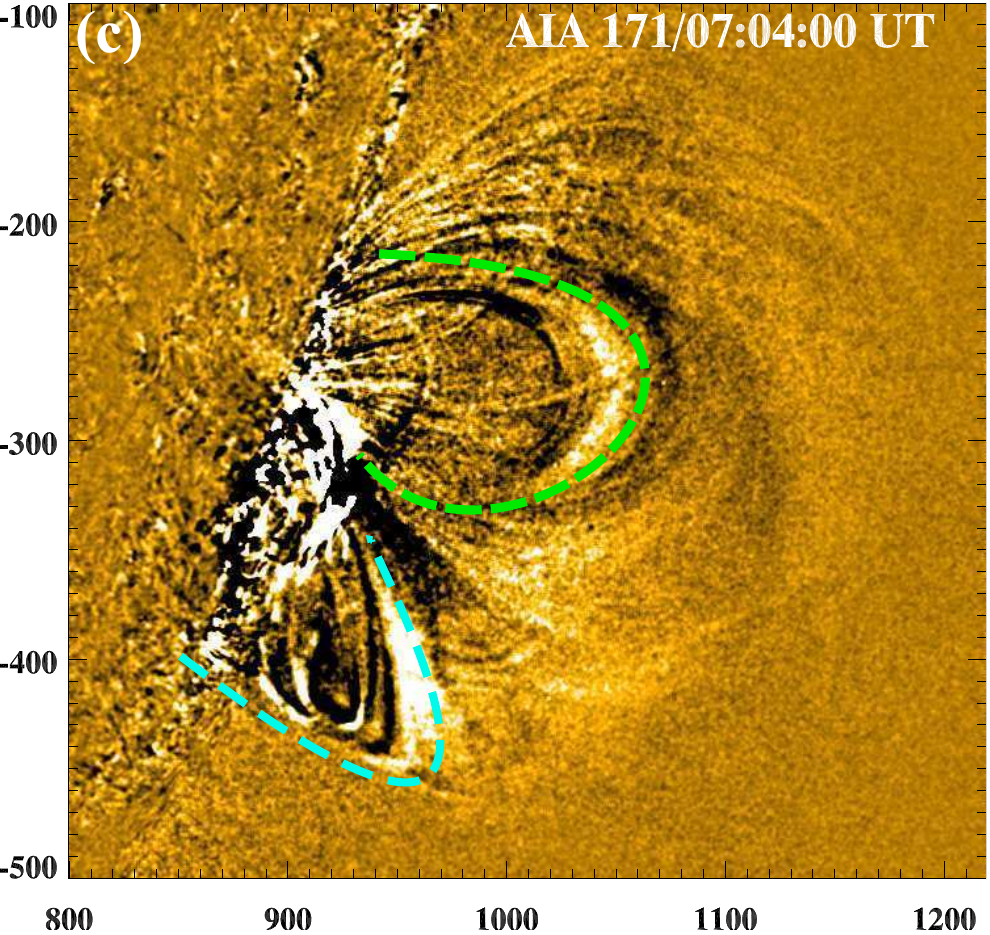}
\includegraphics[width=7.0cm]{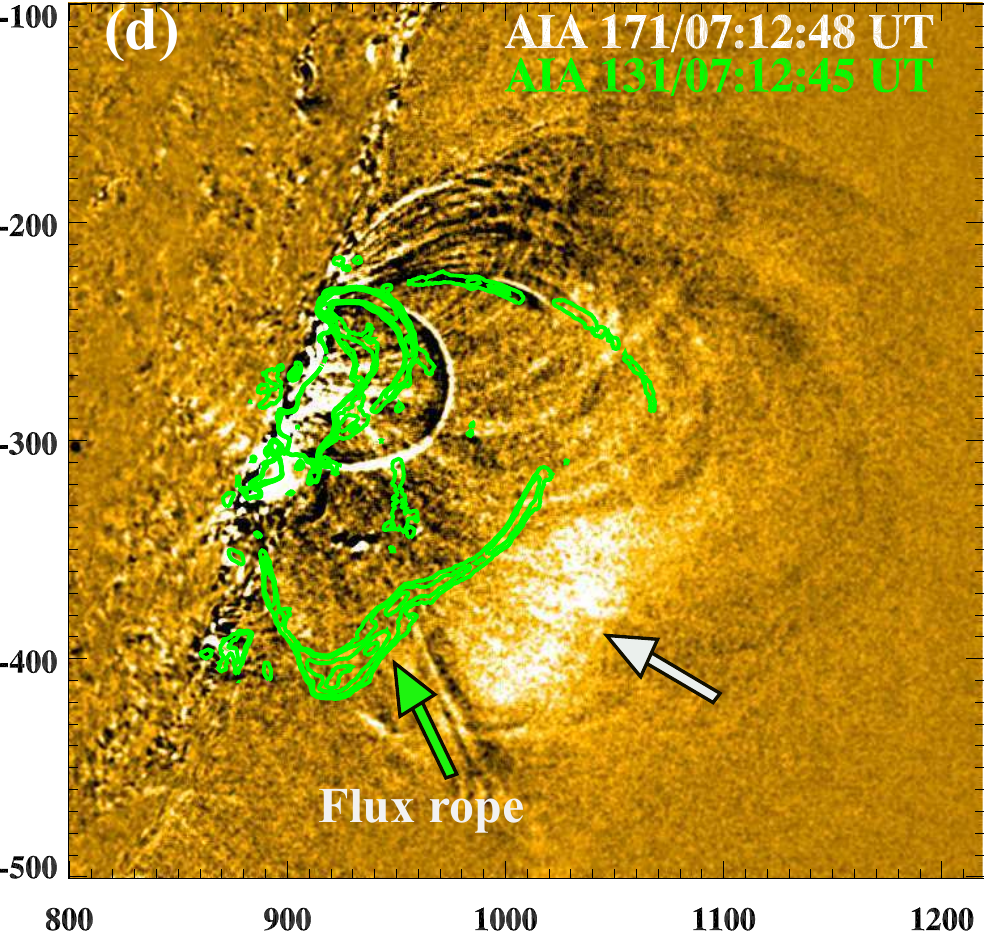}
}
\caption{(a-c) AIA 171 \AA~ running difference images showing the transient EUV brightenings (marked by arrows) above the flare site. Two underlying loop systems are marked by the dotted curves. (d) 171 \AA~ running difference image overlaid by 131 \AA~ running difference image contours (green).
}
\label{aia171_rd}
\end{figure*}


\subsection{Flux rope formation and eruption}
We used the AIA 94 \AA~ base difference and AIA 131 \AA~ running difference images (Figure \ref{aia131_94}) to investigate the hot plasma structures associated with the kinked filament and C3.9 flare. We chose one minute time difference to construct these images. The flux rope was observed only in these two hot channels, which suggests the high temperature of the flux rope (i.e, 6.3-16 MK) \citep{cheng2012,hannah2013}. Figure \ref{aia131_94}a-b show the activation of mini-filament along with the formation of a underlying flare loop near its southern leg during 6:56-6:57 UT. As the filament reached the height of about 18 Mm, we notice the first appearance of a mini flux rope structure (red dotted curve) above the filament during magnetic reconnection at 06:58 UT. 

Underlying hot loop near the southern leg of the filament is shown by the blue dotted curve.
Furthermore, Figure \ref{aia131_94}e-h display the dynamics of the flux rope (above the filament) during the slow rising of the filament. We observe appearance of a sheared loop \#1 at 06:59:21 UT (Figure \ref{aia131_94}f) during the reconnection process above the filament, and an another loop \#2 at 07:00:09 UT. The signatures of particle acceleration along these loops are observed as footpoint brightenings in the simultaneous AIA 1700 image (Figure \ref{hessi}d). In addition, the appearance of loop \#3 at 07:00:57 UT also indicates the reconnection above the filament. Simultaneously, we observed the lateral expansion of the flux rope along the southward direction. The process of the flux rope slow rising and expansion continued for $\sim$15-20 minutes (Figure \ref{aia131_94} i-k). The bottom panels show the larger field of view and the size of the upper panels is marked by the blue dotted rectangular box (Figure \ref{aia131_94}i). The twisted flux rope structure was best observed in AIA 131 \AA~ during 7:13-7:15 UT. The flare loop formation below the flux rope is evident at 07:15:09 UT (Figure \ref{aia131_94}k). The large-scale flux rope expansion and formation of the underlying loops are illustrated by dotted curves (Figure \ref{aia131_94}l).

We used soft X-ray images of the active region recorded by Hinode X-ray Telescope (XRT, \citealt{golub2007}). The field view of each image is 384$\times$384 pixels with a resolution of 1.02 arcsec pixel$^{-1}$. The images are captured by Ti-poly filter containing broad temperature response with a flat peak at 10$^{6.9}$ K. 
Figure \ref{sxt} displays the selected images of the eruption site shortly prior to and during the flux rope eruption. These images are coaligned with the AIA images.  Figure \ref{sxt}a shows the pre-existing arcade loop systems in the active region at 06:54:46 UT. We see the formation of a bright underlying loop (marked by arrow in Figure \ref{sxt}b) with the activation of mini-filament similar to the AIA 94 $\AA$ images. 
We could notice appearance of the flux rope structure above the kinked filament (yellow dotted curve), and formation of the underlying flare loop at the southern leg of the filament at 07:00:07 UT (Figure \ref{sxt}c). The flux rope structure is much clear in these images in comparison to the AIA images. This looks like a small-scale flux rope appeared above the kinked filament during the magnetic reconnection. We include simultaneous 131 $\AA$ image (Figure \ref{sxt}d) to compare the structure in X-ray and EUV. The structure observed above the kinked filament in 131 $\AA$ is consistent with the X-ray structure. Note that both images (EUV and X-ray) are sensitive to $\sim$10 MK plasma temperature. This structure expands slowly and probably reconnecting with the ambient fields within next 10-12 minutes, therefore forming a large-scale structure. The bottom row shows the evolution of the flux rope in the low corona in the larger field of view than the upper row. Figure \ref{sxt}e and f shows a clear twisted flux rope expanding laterally in the southward direction.
 
Initially at about 07:00 UT, we can not rule out the structure (above the filament) to be an arcade of loops, or a mini flux rope. However, later this structure was linked to a large-scale flux rope structure (observed in X-ray and EUV) after a series of magnetic reconnection above the mini-filament (within the arcade loops). Therefore, it is also possible that the initial structure above the filament could be a part of the flux rope observed later during the magnetic reconnection.  
 The observed structure (after 07:10 UT) most likely is a magnetic flux rope, because (i) it is observed only in the AIA hot channels (AIA 131 and 94 \AA), whose morphology and temperature (T$\sim$8 MK) are quite similar to the flux ropes reported in the recent case studies \citep{zhang2012,pats2013,cheng2014}. (ii) The structure is aligned and appeared above the neutral line (along the flare ribbons observed in the AIA 1600 \AA) and the orientation of the structure is across the post flare arcade (cusp) loops formed after the eruption (see Figure \ref{st}d). Furthermore, Figure \ref{sxt}h displays the AIA 171 \AA~ reverse color image (at 09:15 UT) during the decay phase of the flare (after flux rope eruption). Generally, the orientation of the flux rope can be derived from the post-flare loops \citep{yurchyshyn2008}. The flux rope lies along the neutral line (marked by the dotted line), which is in agreement with the flux rope elongation observed in the soft X-ray images.

On the basis of our observational findings, we draw a schematic cartoon to explain the different stages of the eruption.  The bottom panel of Figure \ref{sxt} shows (i) reconnection at the southern footpoint of the filament and its subsequent counterclockwise motion, (ii) breaking of the southern leg and formation of a hot underlying flare loop (blue). Reconnection above the filament results in the formation of hot loops, and appearance of the flux rope structure, (iii) lateral and radial expansion of the flux rope leading to the formation of a compression front around it (observed in the 171 \AA~ channel). The counterclockwise rotation of the filament reveals the left-handed twist.

To study the morphology the flux rope in different AIA channels,  we display the co-temporal AIA running difference images in 131, 94, 335 and 171 \AA~ channels at 07:19 UT (Figure \ref{aia_rd}). The AIA 131 \AA~ image shows a clear twisted structure. The flux rope was not initially visible in 335 \AA~ during its formation process, but later some part of the flux rope (north direction) was observed during its acceleration phase (Figure \ref{aia_rd}c). The expansion of the flux rope results in the formation of a cool compression front surrounding the flux rope (green contours), which was observed in the AIA 171 \AA~ channel (Figure \ref{aia_rd}d). The temporal evolution of the flux rope in four AIA channels is shown in the animation attached to Figure \ref{aia_rd}.

For the kinematics of the flux rope, we used slice cuts S1, S2, S3, and S4 in AIA 131, 94, 335, and 171 \AA~ channels respectively. The stack plot is shown in Figure \ref{stack}a. The slices S1 and S2 show the slow rising phase of the flux rope at $\sim$07:00 UT and an impulsive acceleration at $\sim$07:14 UT. The slice S3 shows only the acceleration phase of the flux rope in the 335 \AA. The location of the slice S3 is different from S1 and S2. It is likely that the initial temperature of the flux rope is high enough to be visible only in the 131 and 94 \AA~ channels as it is formed during the magnetic reconnection. Later, it might be cool down to a temperature of about 2.5 MK so that a part of it was observed in 335 \AA~ during the acceleration phase. This indicates the existence of multitemperature plasma within the flux rope. Alternatively, the reconnection of the flux rope with the preexisting field lines may produce some portion of the flux rope to be visible in 335 \AA.
 The lateral expansion of the CME was observed in AIA 171 \AA~, which shows a cool compression front around the flux rope. From the linear fit, the estimated lateral speed of the CME was $\sim$50 km s$^{-1}$ (north) and $\sim$128 km s$^{-1}$ (south). We also noticed EUV disturbances or brightenings (at 07:02 and 07:14 UT) and contraction/closing of the 171 \AA~ loop (at 07:28 UT) in the 171 \AA~ stack plot. These may be additional evidences for magnetic reconnection above the filament and below the newly formed flux rope.

The top panel of Figure \ref{stack}b displays height-time plot of the flux rope tracked in the stack plot of slice S2 (AIA 94 \AA). We used a cubic spline smoothing scheme to reduce the fluctuations in data points (marked by +) \citep{vrsnak2007}. We assume four pixel (2.4$\arcsec$) error in the height estimation. The middle panel shows the speed profile of the flux rope, which is computed from the height-time measurements using a numerical differentiation with three-point Lagrangian interpolation \citep{zhang2004}. The bottom panel displays the acceleration profile and evolution of the hard X-ray flux in 12-25 keV (blue). The uncertainty in the speed and acceleration is mainly due to the error in the height measurement. The initial speed of the flux-rope was $\sim$100 km s$^{-1}$ during the slow rising process at about 07:00 UT. It reduces to about 30-40 km s$^{-1}$ during 7:02-07:10 UT, and again started to increase at $\sim$7:12 UT. We notice an impulsive increase in the speed up to $\sim$300 km s$^{-1}$ at $\sim$7:16 UT. During the whole process, the flux rope speed varies from 40 to 500 km s$^{-1}$.
The flux rope does not show significant acceleration during the first hard X-ray burst at about 07:00 UT. Further, we notice an acceleration of $\sim$2 km s$^{-2}$ at 7:16 UT, which coincides well with the rise in the hard X-ray flux during 07:08-07:24 UT.

STEREO (Solar TErrestrial RElations Observatory)  SECCHI-A (ahead) \citep{wulser2004,howard2008} observed the same AR near the disc center. We utilized EUVI-A 284 $\AA~$ images to see the magnetic field topolgy of the active region. The size of the image is 2048$\times$2048 pixels ($\sim$1.6 arcsec pixel$^{-1}$) covering a field of view out to 1.7 R$_{\odot}$. The peak temperature response for the 284 $\AA~$ channel is 2.2 MK \citep{asc2008}. Fortunately, we could find two images of the eruption site, before and after the flux rope eruption. 
Figure \ref{st} shows the simultaneous images in EUVI-A 284 $\AA~$ and AIA 94 $\AA$ channels at 06:16 and 08:16 UT. Figure \ref{st}a shows a reverse `J' shaped structure in the AR. We used scc\_measure routine to compare the magnetic structures of the AR in 284 and 94 $\AA$. The approximate positions of the chosen structures are marked by `+' symbols with different colors in both the images. The location of the kinked filament is overplotted in Figure \ref{st}b. The comparison of both images indicates that the filament was located at the magnetic separatrix and below the reverse `J' shaped loop. Figure \ref{st}c shows the post eruptive arcade system formed at the site of filament activation and  `J' shaped loop. The cusp shaped loops were observed in AIA 94 (and 131) \AA~ images (Figure \ref{st}d). We also observed the supra-arcade downflows (e.g., \citealt{innes2014}) above the cusp shaped loops, which is an indirect evidence of magnetic reconnection below the flux rope during the eruption.
The combined magnetic field configuration (i.e., northern loops from SXT/AIA 94 \AA~ and southern sheared field lines from SECCHI 284 \AA) from two point observations supports the presence of quasi-separatrix layers (QSLs) \citep{demoulin1996,demoulin1997}, where the magnetic reconnection takes place above the kinked filament and cusp loops are formed during the decay phase of the flare.

\subsection{EUV disturbances/brightenings}
To see the coronal responses of the magnetic reconnection during the flux rope appearance, we generate AIA 171 \AA~ running difference images. Figure \ref{aia171_rd} displays some of the selected images during the flare. We observed multiple propagating brightenings in the corona associated with the flux rope formation during the magnetic reconnection. The first EUV brightening or pulse was observed during the first hard X-ray burst at about 07:00 UT (Figures \ref{aia171_rd} a and b). The location of brightening was above the mini-filament, which closely matched the coronal hard X-ray sources (sources B, C, and D in Figure \ref{hessi}).   
The second brightening was observed during the second hard X-ray burst at about 07:12 UT close to the flux rope (Figure \ref{aia171_rd}d). The more details about the spatial location and evolution of the EUV waves can be found in the AIA four channels composite movie (attached to Figure \ref{aia_rd}). The timing and location of these short duration brightenings suggest that they are closely associated with the magnetic reconnection and may be interpreted as a multiple EUV wavefronts \citep{liu2011}. The formation of southward closed loop systems (marked by dotted curve) are resulted by magnetic reconnection (Figure \ref{aia171_rd}c). These underlying closed loops are also observed in the AIA 94 \AA~ channel after the second hard X-ray burst (Figure \ref{aia131_94}l). 

\begin{figure*}
\centering{
\includegraphics[width=7cm]{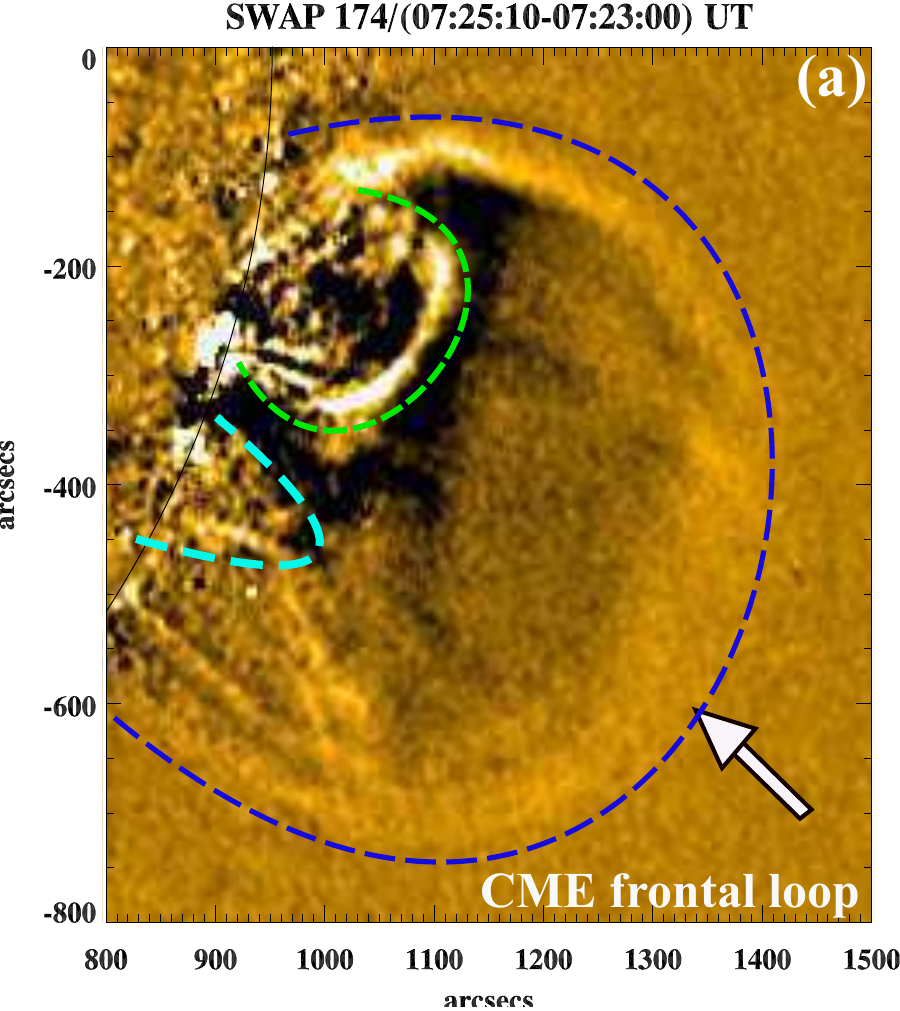}
\includegraphics[width=5.5cm]{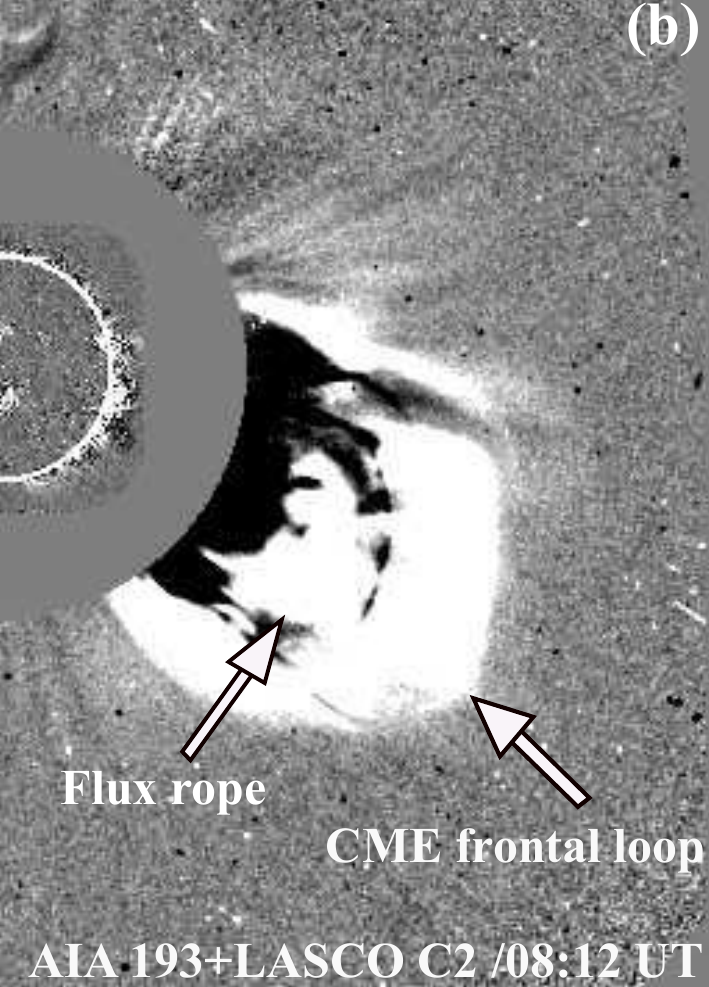}
\includegraphics[width=5.5cm]{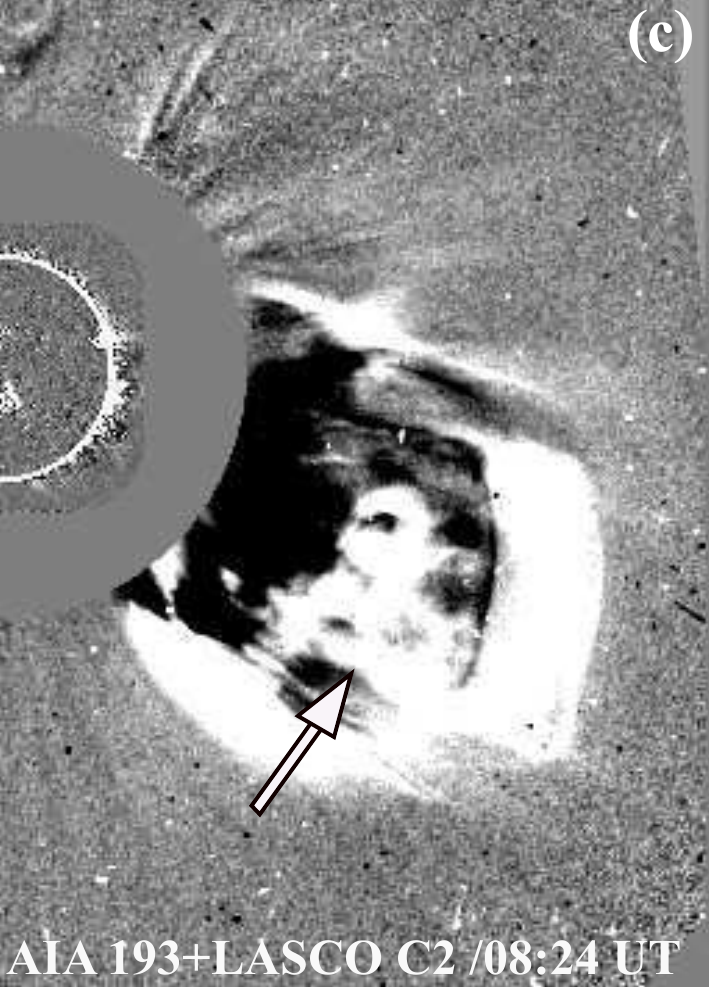}
}
\caption{(a) SWAP 174 \AA~ running difference image showing the compression front of the CME (blue dotted curve) and underlying loop systems (dotted curves). (b-c) AIA 193 \AA~ and LASCO C2 composite running difference images. The flux rope and `CME frontal loop' are shown by arrows.}
\label{lasco}
\end{figure*}
\begin{figure}
\centering{
\includegraphics[width=9cm]{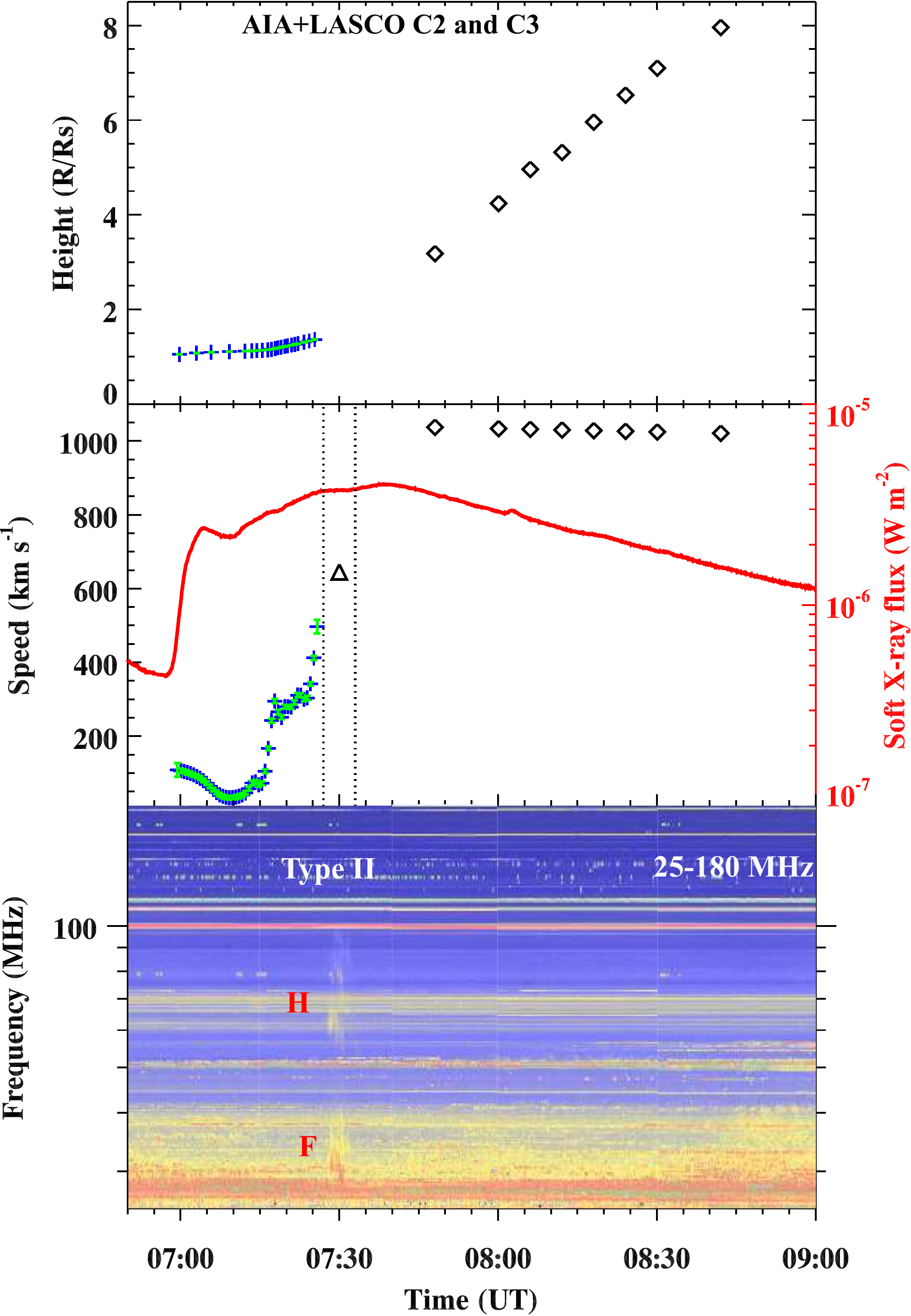}
}
\caption{Top: Height-time profile of the flux rope and CME frontal loop (diamond symbol) from AIA 94 \AA~ and LASCO C2/C3 respectively. Middle: Speed of the Flux rope and CME frontal loop, plotted with GOES soft X-ray profile (red curve). Triangle symbol (between two vertical dotted lines) shows the speed of the type II radio burst excitor. Bottom: Dynamic radio spectrum (25-180 MHz) from Learmonth radio observatory. `F' and `H' represent the fundamental and second harmonic emissions of the type II radio burst.}
\label{kin2}
\end{figure}

\subsection{CME}
To see the evolution of the CME in the larger field of view than AIA, we used the SWAP (Sun Watcher using Active Pixel System Detector and Image Processing) 174 \AA~ running difference image (7:25:10-07:23:00 UT) shown in Figure \ref{lasco}a. The SWAP telescope is a compact EUV imager on board the PROBA2 (PRoject for Onboard Autonomy) micro-satellite, which provides the coronal images at a temperature of $\sim$0.8 MK \citep{berghmans2006,halain2010}. The CME frontal loop shown by blue curve is clearly observed, which agrees well with the AIA 171 \AA~ images of the low corona (Figure \ref{aia_rd}d). Note that we do not observe the flux rope in cool channels. Two sets of underlying loops are shown by dotted curves indicating the reorganization of coronal magnetic field by magnetic reconnection. 

The white light running difference images from SOHO/LASCO C2 \citep{brueckner1995,yashiro2004} reveal the flux rope structure behind the CME frontal loop, which is shown by arrows in Figure \ref{lasco}b and c. It seems to cover the cavity region of the CME.
The dark cavities (surrounded by bright core, i.e., prominence) are often observed in the three part structure of CMEs, and are explained by the flux rope model \citep{illing1985,illing1986, bak2013}. In our case, the filament was failed to erupt and the cavity part is covered by the flux rope.

The top panel of Figure \ref{kin2} shows the composite height-time profile of the flux rope (observed in AIA 94 \AA~) and frontal loop from LASCO C2 and C3. The speed of the flux rope is derived from the three point Lagrangian interpolation method (see Figure \ref{stack}). The speed of the CME frontal loop is estimated from a second order polynomial function fitted to the height-time measurements. The average speed of the CME in LASCO field of view is $\sim$992 km s$^{-1}$, and an acceleration of $\sim$-4.98 m s$^{-2}$. The bottom panel displays the dynamic radio spectrum in 25-180 MHz observed from Learmonth radio observatory, Australia. A type II radio burst (25-102 MHz) was observed during the impulsive acceleration of the flux rope between 07:27 UT-07:33 UT. The timing and formation height of the type II source indicates the presence of the piston driven shock ahead of the CME front. From the frequency drift rate, the speed of the type II excitor was $\sim$645 km s$^{-1}$, which should be close to the flux rope speed at the formation height of type II burst. Interestingly, no type III burst was observed during the activation of kinked filament at $\sim$07:00 UT. This suggests that there is no opening of the field lines into the interplanetary medium. It is also possible that the presence of the flux rope above the mini-filament did not allow the particles to escape outward.
 It should be noted that first impulsive rise in SXR flux  ($\sim$07:00 UT) is not synchronized with the CME speed (i.e., impulsive acceleration) as shown by \citet{zhang2006}. As discussed in the first section, it is related to the magnetic reconnection at the filament leg and apex. However, the impulsive acceleration of the flux rope $\sim$07:12 UT onwards is synchronized with the gradual rise of the SXR flux, which could be related to the successive magnetic reconnection below the flux rope. In addition, the tether cutting reconnection below the flux rope usually increases the magnetic pressure by adding the poloidal flux that enhances the hoop force of the flux rope \citep{vrsnak2004,zhang2006}.   


\section{Summary and Discussion}
We presented multiwavelength study of a flux rope eruption associated with failed eruption of the kinked mini-filament and C3.9 flare. The main results of this study are summarized as follows:

(1) The activation and slow rising motion ($\sim$40 km s$^{-1}$) of the kinked mini-filament was associated with the formation of a hot loop near its southern leg. The first magnetic reconnection possibly occur at or above the southern leg of the filament that causes the formation of a hot loop and breaking of the southern leg. RHESSI hard X-ray images (12-25 keV) clearly show the two footpoints and a looptop source coinciding with the underlying hot loop. 

After breaking the leg, the filament twist is converted into the writhe as seen clearly in the SOT H$\alpha$ images.
The filament showed unwinding motion of the apex and northern leg, in the counterclockwise direction (left handed twist) and fell back to the solar surface within $\sim$10 minutes. The morphology of the filament and its rotation support the occurrence of kink instability and the relaxation of twist during the eruption \citep{torok2005,kliem2012}.

(2) Prior to the flare onset, we observed a mini-filament in the AIA 304 \AA~ channel. The flux rope appeared during the second magnetic reconnection above the apex of the mini-filament. We have sufficient evidences to support this argument; (i) appearance of the RHESSI hard X-ray source (12-25 keV, source B) above the filament apex during 06:59:40-07:00 UT, suggesting the particle acceleration site above the filament, (ii) at the same time, we see the appearance of a mini flux rope (SXT, AIA 94 and 131 \AA) and loops \#1 and \#2 associated with the footpoint brightenings toward north of the filament (AIA 1700 \AA, kernels K1 and K2), and (iii) generation of the 171 \AA~ EUV propagating brightening/disturbances above the filament, which might be associated with the energy release during the magnetic reconnection. All these evidences suggest that the magnetic reconnection took place above the filament.

(3) However, presence of the filament before the eruption confirms the existence of the flux rope (supporting the filament) prior to the eruption. Additionally, the reverse J shaped loop system in the STEREO EUVI 284 \AA~ images also suggest the presence of sheared field lines at the eruption site, which were transformed into post-eruptive (cusp-shaped) arcade loops. It should be noted that there are two flux rope, one supporting the filament that becomes kink unstable and another one formed above the filament in the sheared arcade loops (observed in the AIA hot channels).
 The magnetic reconnection above the filament may heat the flux rope more than about 6 MK temperature and makes it to be visible in the SXT images and AIA hot channels (131 and 94 \AA). The flux rope expands slowly, and developed into a large-scale structure possibly by magnetic reconnection with the surrounding magnetic fields within 15-20 minutes. 
 \citet{cheng2014} also reported a similar flux rope structure exhibiting the twisted/helical threads in hot AIA channels for a different eruptive event. \citet{li2013} studied the four homologous flux rope structures, appeared during the magnetic reconnection in AIA hot channels. The fourth flux rope (in their observation) was associated with a CME and had a filament underneath showing counterclockwise rotation and failed eruption.
In our case, we consider the mini-filament associated with kink instability (including magnetic reconnection at its footpoint) causing the onset of the eruption (i.e., by magnetic reconnection above it) and appearance of a large scale flux rope above it.

(4) The speed of the flux rope varies in between 30-500 km s$^{-1}$. The flux rope does not show significant acceleration (i.e., slow rise phase) during the first energy release at about 07:00 UT, but an impulsive acceleration of $\sim$2 km s$^{-2}$ later at 07:14 UT associated with the hard X-ray flux rise and an EUV pulse in 171 \AA. Moreover, the soft X-ray flux also rises after $\sim$7:12 UT. These evidences suggest the ongoing reconnection in the solar corona resulting the acceleration of a rather big flux rope and associated particle acceleration from the current sheet underneath the flux rope. In addition, the decrease of the overlying field strength with height also determines the acceleration of the flux rope in the corona by torus instability \citep{torok2005,kliem2006, aulanier2010,olmedo2010}. However, in our case, magnetic reconnection below the flux rope and onset of the torus instability at about 1$\times$$10^{5}$ km height jointly may cause the impulsive acceleration of the flux rope after 07:12 UT.
     
Some of the observational features of this event match the prominence eruption studied by \citet{kumar2012}. They observed breaking of the northern leg of the prominence followed by an unwinding motion (anticlockwise) caused by the reconnection below the prominence followed by the formation of a helix. However, they observed neither the hard X-ray source (12-25 keV) nor the hot flux rope appearance above the prominence. Probably due to absence of reconnection above the prominence that could not heat the flux rope to be appear in the hot channel. In addition, the prominence successfully erupted and became the core of the CME. Note that here we observed breaking of the southern leg and unwinding motion (anticlockwise) in a similar way but in a small-scale filament including magnetic reconnection above it, which causes the appearance of the flux rope and failed eruption of the filament. 

We do not observe the hard X-ray source in between the crossing legs of the filament, which may be an evidence of particle acceleration site at the current sheet formed in between the crossing legs by kink instability \citep{torok2005,alex2006,cho2009,kliem2010,joshi2013}. As suggested in the two flux rope model of \citet{gilbert2001}, the magnetic reconnection above the filament (within the flux rope) may cause the failed eruption of the filament, that result in the formation of an inverse polarity flux rope (new one) above it (middle panel of their Figure 3). However, kink instability is not discussed in their model.
 Very likely the reconnection above the filament (in our case) suppresses it downward resulting in the failed eruption \citep{gilbert2001,gilbert2007}. In addition, our case study demonstrates multiple reconnection above the filament (hard X-ray flux and EUV pulses) that may also serve the sheared magnetic flux to form a large-scale twisted flux rope.

In conclusion, we have reported a unique event of the flux rope eruption above a kinked mini-filament. The magnetic reconnection above the kinked filament is rare and this type of event has not been observed earlier. This study indicates the role of multiple reconnection and kink instability in the trigger of the eruption. Future studies of the similar events using high-resolution multiwavelength data sets will shed more light on the detailed processes involved in such eruptive phenomena related to the flux rope formation and eruption. 

\begin{acknowledgements}
 We thank the referee for valuable comments and suggestions to improve the manuscript.
SDO is a mission for NASA's Living With a Star (LWS) program. RHESSI is 
a NASA Small Explorer. Hinode is a Japanese mission developed and launched by ISAS/JAXA, with NAOJ as
domestic partner and NASA and STFC (UK) as international partners. It is
operated by these agencies in co-operation with ESA and the NSC (Norway). SWAP is a project of the Centre Spatial de Liege and the Royal Observatory of Belgium funded by the Belgian Federal Science Policy Office (BELSPO). SOHO is a project of international cooperation between ESA and NASA. This work was supported by the ``Development of Korea Space Weather Center" of KASI
and the KASI basic research funds.
 
\end{acknowledgements}
\bibliographystyle{aa}
\bibliography{reference}

\end{document}